\documentclass[12pt]{article}
\usepackage[greek,british]{babel}
\def\AnswerYes{y}
\def\pdflatex{y}                  
\def\ShowLineNumberVersion{n}     
\def\ShowLabelsVersion{n}         
\def\ShowChangesVersion{n}        
\def\ShowAnnotationsVersion{n}    
\def\ShowFigures{y}               
\def\feynVersion{n}               
\def\MakeArXivLinksActive{y}      
\ifx\pdflatex\AnswerYes
   \usepackage[update,prepend]{epstopdf} 
   \usepackage{grffile} 
\fi
\ifx\pdflatex\AnswerYes
    \usepackage{color}
\else
    \usepackage[dvips]{color}
\fi
\ifx\pdflatex\AnswerYes
    \usepackage{thumbpdf}    %
\else
    \usepackage[dvips]{thumbpdf}
\fi
\ifx\pdflatex\AnswerYes
    \usepackage[ 
        final,
        breaklinks=true,
        colorlinks=true,           
        pdfborder={0 0 1},    
        citecolor={blue},linkcolor={blue},urlcolor={red},
        citebordercolor={1 0 0},linkbordercolor={0 0 1},urlbordercolor={1 0 0},
        pdfpagemode=UseOutlines,
        bookmarks=true,bookmarksopenlevel=4
        ]{hyperref}         
\else
    \usepackage[dvips,              
        final,
        breaklinks=true,
        colorlinks=true,      
        pdfborder={0 0 1},    
        citecolor={blue},linkcolor={blue},urlcolor={red},
        citebordercolor={1 0 0},linkbordercolor={0 0 1},urlbordercolor={1 0 0},
        pdfpagemode=UseOutlines,
        bookmarks=true,bookmarksopenlevel=4
         ]{hyperref}         
\fi
\ifx\MakeArXivLinksActive\AnswerYes
   \usepackage{xparse}
   \NewDocumentCommand{\arxiv} %
   {r [: u{ [} u{]]} }{[\href{http://arxiv.org/abs/#2}{arXiv:#2}~[#3]]}
   \NewDocumentCommand{\arxivold} {r[]}{[\href{http://arxiv.org/abs/#1}{#1}]}
   \NewDocumentCommand{\arXiv} %
   {r [: u{ [} u{]]} }{[\href{http://arxiv.org/abs/#2}{arXiv:#2}~[#3]]}
   \NewDocumentCommand{\arXivold} {r[]}{[\href{http://arxiv.org/abs/#1}{#1}]}
\else
   \newcommand{\arxiv}[1][]{[#1]}
   \newcommand{\arxivold}[1][]{[#1]}
   \newcommand{\arXiv}[1][]{[#1]}
   \newcommand{\arXivold}[1][]{[#1]}
\fi
\usepackage{doi} 
\usepackage{tikz}
\usepackage{calc}
\newlength{\ORCIDidheight}
\newlength{\ORCIDidunit}
\definecolor{ORCIDgreen}{HTML}{A6CE39}
\newcommand{\ORCIDid}{%
  \settoheight{\ORCIDidheight}{AXg}%
  \setlength{\ORCIDidunit}{1.2pt * \ratio{\ORCIDidheight}{256 pt}}%
  \raisebox{0.5\depth}{\parbox{\ORCIDidheight}{%
    \begin{tikzpicture}[x=\ORCIDidunit, y=\ORCIDidunit, inner sep=0pt, outer sep=0pt]%
      \fill[ORCIDgreen] (128,128) circle (128);
      \fill[white] (70,177) rectangle (86,70);
      \fill[white] (78,200) circle (10);
      \fill[white] (109,177) -- (150,177) %
      .. controls (190,177) and (208,149)%
      .. (208,123)%
      .. controls (208,96) and (186,70)%
      .. (150,70)%
      -- (109,70)%
      -- (109,177) -- cycle%
      (124,84)%
      -- (150,84)%
      .. controls (186,84) and (192,110)%
      .. (192,123)%
      .. controls (192,145) and (178,163)%
      .. (150,163) -- (124,163)%
      -- (124,84) -- cycle;
  \end{tikzpicture}%
  }}%
}
\usepackage{etoolbox}
\DeclareRobustCommand\orcidlinkX[3]{\href{https://orcid.org/#2}{%
\ifstrempty{#1}{}{#1\,}\ORCIDid\,\ifstrempty{#3}{}{\,#3}}}

\newcommand{\orcidlink}[1]{\orcidlinkX{}{#1}{}}
\usepackage[UKenglish]{isodate} 
\usepackage[numbers,sort&compress]{natbib}
\ifx\ShowFigures\AnswerYes
   \usepackage{graphicx}          
\else
   \usepackage[draft]{graphicx}   
   \renewcommand{\includegraphics}[2][]{\fbox{#2}}
\fi
\graphicspath{{figures/}{./}}

\usepackage{multirow}                   
\usepackage{amssymb}
\usepackage{amsmath}
\usepackage{marvosym} 
\usepackage{bbm} 
\usepackage{textcomp}
\usepackage{xspace}                    
\xspaceaddexceptions{[]}
\usepackage{dcolumn}                    
\usepackage{pbox}

\usepackage{bm}                        

\usepackage{cancel}	                  

\usepackage{rotating}

\usepackage{stfloats}
\fnbelowfloat 

\usepackage{floatpag}           
\floatpagestyle{plain}          
\usepackage{slashed}
\ifx\feynVersion\AnswerYes
   \usepackage{feynmp-auto} 
\fi

\usepackage{chngcntr} 
\counterwithin*{equation}{section} 

\ifx\ShowLabelsVersion\AnswerYes
   \usepackage[color]{showkeys} 
   \definecolor{refkey}{gray}{.3}
   \definecolor{labelkey}{gray}{.3}
   
\fi
\ifx\ShowAnnotationsVersion\AnswerYes
   \newcommand{\comment}[1]{{\scriptsize\sffamily\bfseries{#1}}}
   \newcommand{\margin}[1]{\marginpar{\scriptsize\sffamily\bfseries{#1}}}
   \newcommand{\drafty}{\textbf{Draft version \today} \hfill}
\else
   \newcommand{\comment}[1]{}
   \newcommand{\margin}[1]{}
   \newcommand{\drafty}{}
\fi
\ifx\ShowLineNumberVersion\AnswerYes
   \usepackage{lineno}
   \linenumbers
\fi
\ifx\ShowChangesVersion\AnswerYes
   \usepackage{ulem}
    
   \newcommand{\delete}[1]{\sout{#1}}            
   \renewcommand{\emph}[1]{\textit{#1}}           
\else

   \newcommand{\sout}[1]{}
   \newcommand{\xout}[1]{}

   \newcommand{\delete}[1]{}
\fi

\newcommand{\absatz}{\vspace{2ex}\noindent}

\newcommand{\red}[1]{#1}

\newcommand{\cf}{\textit{cf.}\xspace}
\newcommand{\eg}{\textit{e.g.}\xspace}
\newcommand{\etal}{\textit{et al.}\xspace}
\newcommand{\etc}{\textit{etc.}\xspace}
\newcommand{\ie}{\textit{i.e.}\xspace}

\newcommand{\contradict}{\red{\Lightning}}
\newcommand{\non}{\nonumber}

\newcommand{\half}{\frac{1}{2}}
\newcommand{\e}{\mathrm{e}}
\newcommand{\ii}{\mathrm{i}}
\newcommand{\dd}{\mathrm{d}}

\newcommand{\deint}[2]{\dd^{#1}\;\!\! #2\;}

\newcommand{\vectorwithspace}[1]{\vec{#1}\mkern2mu\vphantom{#1}}
\newcommand{\vect}[1]{\vectorwithspace{#1}}

\newcommand{\ev}{\vectorwithspace{e}}

\newcommand{\lv}{\vectorwithspace{l}}

\newcommand{\qv}{\vectorwithspace{q}}

\renewcommand{\Re}{\mathrm{Re}}
\renewcommand{\Im}{\mathrm{Im}}

\newcommand{\arccot}{\ensuremath{\mathrm{arccot}}}

\newcommand{\kcotdelta}{\ensuremath{\mathrm{k}\!\cot\!\delta}}
\newcommand{\cotdelta}{\ensuremath{\cot\!\delta}}

\newcommand{\Folgt}{\Longrightarrow}

\newcommand{\mpi}{\ensuremath{m_\pi}}

\newcommand{\fpi}{\ensuremath{f_\pi}}
\newcommand{\gA}{\ensuremath{g_A}}
\newcommand{\MeV}{\ensuremath{\mathrm{MeV}}}
\newcommand{\fm}{\ensuremath{\mathrm{fm}}}

\newcommand{\ChiEFT}{\foreignlanguage{greek}{q}EFT\xspace}
\newcommand{\ChiEFTPP}{\foreignlanguage{greek}{q}EFT(p\foreignlanguage{greek}{p})\xspace}
\newcommand{\ChiEFTPPUE}{\foreignlanguage{greek}{q}EFT(p\foreignlanguage{greek}{p})$_\text{UE}$\xspace}

\newcommand{\NoPion}{\foreignlanguage{greek}{p}\hspace*{-0.48em}/}
\newcommand{\EFTNoPion}{EFT(\NoPion)\xspace}

\newcommand{\LambdaNN}{\ensuremath{\overline{\Lambda}_{\N\N}}}
\newcommand{\QNoPion}{\ensuremath{Q_{\text{\NoPion}}}}
\newcommand{\OPE}{OPE\xspace}

\newcommand{\NXLO}[1]{N\ensuremath{{}^{#1}}LO\xspace}

\newcommand{\wave}[3]{\ensuremath{{}^{#1}\mathrm{#2}_{#3}}\xspace}
\newcommand{\oneS}{\wave{1}{S}{0}}

\newcommand{\threeS}{\wave{3}{S}{1}}
\newcommand{\threeD}{\wave{3}{D}{1}}
\newcommand{\threeSD}{\wave{3}{SD}{1}}

\newcommand{\SWsigma}{\mathrm{SW}\!_\sigma}

\newcommand{\fourHe}{\ensuremath{{}^4}He\xspace}

\newcommand{\p}{\ensuremath{\mathrm{p}}}
\newcommand{\n}{\ensuremath{\mathrm{n}}}
\newcommand{\N}{\ensuremath{\mathrm{N}}}
\newcommand{\ptyp}{\ensuremath{p_\mathrm{typ}}}
\newcommand{\kfit}{\ensuremath{k_\mathrm{fit}}}

\newcommand{\aoneS}{\ensuremath{a(\oneS)}}
\newcommand{\roneS}{\ensuremath{r(\oneS)}}
\newcommand{\athreeS}{\ensuremath{a(\threeS)}}
\newcommand{\rthreeS}{\ensuremath{r(\threeS)}}

\newcommand{\SU}{\ensuremath{\mathrm{SU}}}

\newcommand{\MN}{\ensuremath{M}}

\newcommand{\whitey}{\phantom{0}}
\newcommand{\whit}{\phantom{.}}

\newcommand{\calE}{\ensuremath{\mathcal{E}}}

\newcommand{\calO}{\ensuremath{\mathcal{O}}} 
 
\newcommand{\calS}{\ensuremath{\mathcal{S}}}

 \newcommand{\rmD}{\ensuremath{\mathrm{D}}}

 \newcommand{\rmP}{\ensuremath{\mathrm{P}}}
 
\newcommand{\rmS}{\ensuremath{\mathrm{S}}}

\newcommand{\mytitle}[1]{\begin{center}\LARGE{\textbf{#1}}\end{center}}
\newcommand{\myauthor}[1]{\textbf{#1}}
\newcommand{\myaddress}[1]{\textit{#1}}
\newcommand{\mypreprint}[1]{\begin{flushright}#1\end{flushright}}

\begin{document}
%

\begin{titlepage}
  \setcounter{page}{0} \mypreprint{
    \drafty
    13th October 2024\\Revised versions 20th October 2024,
    8th May 2025, 10th July 2025\\
    Final version 15th July 2025 accepted by Europ.~Phys.~J.~\textbf{A}
  }
  
  
  \mytitle{On Two Nucleons Near Unitarity\\[0.2ex] with Perturbative Pions}


\begin{center}
  \myauthor{Yu-Ping
    Teng\orcidlink{0009-0007-8663-2998}$^{a,b}$}\footnote{Email:
    yteng@uwm.edu}
  \emph{and} 
  \myauthor{Harald W.\
    Grie\3hammer\orcidlink{0000-0002-9953-6512}$^{a}$}\footnote{Email:
    hgrie@gwu.edu (corresponding author)}
  
  \vspace*{0.5cm}
  
  \myaddress{$^a$ Institute for Nuclear Studies, Department of Physics, \\The
    George Washington University, Washington DC 20052, USA}

 \vspace*{0.2cm}
 
  \myaddress{$^b$ Department of Physics, University of Wisconsin-Milwaukee, Milwaukee WI 53211, USA}


\end{center}


\begin{abstract}
  We explore the impact of perturbative pions on the Unitarity Expansion in
  the two-nucleon $\rmS$-waves of Chiral Effective Field Theory at
  next-to-next-to leading order (\NXLO{2}).  Pion exchange explicitly breaks
  the nontrivial fixed point's universality, \ie~invariance of $\rmS$ waves
  under both conformal and Wigner's combined $\SU(4)$ spin-isospin
  transformations. On the other hand, Unitarity explicitly breaks chiral
  symmetry. The two seem incompatible in their respective exact-symmetry
  limits.
  \ChiEFT with Perturbative Pions in the Unitarity Expansion resolves the
  apparent conflict in the Unitarity Window (phase shifts
  $45^\circ\lesssim\delta(k)\lesssim135^\circ$), \ie~around momenta
  $k\approx\mpi$ most relevant for low-energy nuclear systems. Its only LO
  scale is the scattering momentum; NLO adds only scattering length, effective
  range and non-iterated one-pion exchange (\OPE); and \NXLO{2} only
  once-iterated \OPE.
  Agreement in the \oneS channel is very good. Apparently large discrepancies
  in the \threeS channel even at $k\approx100\;\MeV$ are remedied by taking at
  \NXLO{2} only the central part of \OPE. In contradistinction to the tensor
  part, it is identical in the \oneS and \threeS channels.
  Both channels then match empirical phase shifts and pole parameters well
  within mutually consistent quantitative theory uncertainty estimates. Pionic
  effects are small, even for $k\gtrsim\mpi$.
  Empirical breakdown scales are consistent with
  $\LambdaNN=\frac{16\pi\fpi^2}{\gA^2M}\approx300\;\MeV$, where iterated
  \OPE is not suppressed.
  We therefore conjecture: Both conformal and Wigner symmetry in the
  Unitarity Expansion show \emph{persistence}, \ie~the footprint of both
  combined dominates even for $k\gtrsim\mpi$ and is more relevant than chiral
  symmetry, so that the tensor/Wigner-$\SU(4)$ symmetry-breaking part of \OPE
  does not enter before \NXLO{3}. We also discuss the potential relevance of
  entanglement and possible resolution of a conflict with the strength
  of the tensor interaction in the large-$N_C$ expansion.
\end{abstract}

\vspace*{\fill} 
\noindent
\pbox[t]{\linewidth}{Suggested\\Keywords} \hfill
\begin{minipage}[t]{13.9cm} 
    Chiral Perturbation Theory, Chiral Effective Field Theory, perturbative
    pions, Unitarity, Universality, scale/conformal invariance, Wigner-$\SU(4)$
    spin-isospin symmetry, chiral symmetry, two-nucleon scattering,
    entanglement, Large-$N_C$.
  \end{minipage}
  
  \vspace*{0.5cm}
\end{titlepage}

\setcounter{footnote}{0}

\newpage

%
\section{Introduction}
\label{sec:introduction}

Why are the observables of few-nucleon systems dominated by anomalous scales?
The deuteron is an exceptionally shallow bound state in the \threeS channel of
$\N\N$ scattering, with a binding momentum of $45\;\MeV$ set by the inverse of
the scattering length $a(\threeS)\approx5\;\fm$. Likewise the \oneS channel
has a virtual bound state at a binding momentum of $-7\;\MeV$ from
$a(\oneS)\approx-24\;\fm$. These scales differ markedly from the natural
low-momentum QCD scales of Nuclear Physics, the pion mass
$\mpi\approx140\;\MeV$ and $\frac{1}{\mpi}\approx1.4\;\fm$. The combined
evidence of lattice computations and chiral extrapolations suggest that this
fine tuning holds in QCD only in a small window around the physical value of
the pion mass~\cite{Epelbaum:2002gb, Braaten:2003eu, Beane:2008dv}; \cf~\cite{BaSc:2025yhy}.

Effective Field Theories (EFTs) of Nuclear Physics do not offer an
explanation. They simply impose an ordering scheme whose leading-order (LO)
iterates a $\N\N$ interaction infinitely often. In the usually employed
version of Chiral Effective Field Theory (\ChiEFT), the one ``with
Nonperturbative Pions'' (see \eg~\cite{Hammer:2019poc, Machleidt:2024bwl} for
recent reviews), LO iterates one-pion exchange (\OPE) and contact terms, tuned
to reproduce the shallow bound states. Perfectly valid and consistent versions
of \ChiEFT exist in which LO is perturbative and the binding momentum (energy)
of light nuclei is set by the scale $\mpi$ ($\frac{\mpi^2}{\MN}$, with $\MN$
the nucleon mass). But these are not realised in Nature.

And yet, one may recover the intrinsic momentum scales of $\rmS$ waves by
expanding about scale zero, namely about Unitarity.  This presentation
explores how such an expansion emerges in \ChiEFT, \ie~with pions, despite an
apparent incompatibility with chiral symmetry, the cornerstone of \ChiEFT,
and finally argues that its highly symmetric limit is key for the wider
question. The rest of the Introduction sets the stage: the Unitarity Expansion
and its symmetries in $\N\N$ systems; issues with pions as explicit degrees of
freedom; the Conjecture we will infer from our results with perturbative
pions; and the organisation of this article.

\absatz
Let us first address the importance of Unitarity in general terms. 
The centre-of-mass two-body scattering amplitude at relative momentum $k$ is
\begin{equation}
    \label{eq:amplitude}
    A(k)=\frac{4\pi}{M}\;\frac{1}{\kcotdelta(k)-\ii k}\;\;.
\end{equation}
The term ``$-\ii k$'' ensures Unitarity of the $S$ matrix, \ie~probability
conservation. All information about the interaction is encoded in
$\kcotdelta(k)$. This invites two expansions.

In the \textbf{Born Corridor}, $|\kcotdelta| \gtrsim |\ii k|$, the phase shift
is ``small'', $|\delta(k)|\lesssim45^\circ$, so contributions from
interactions are small and can be treated perturbatively (Born approximation,
no bound state). Their details enter at LO and are therefore crucial:
\begin{equation}
    A(k)\Big|_\mathrm{Born}=\frac{4\pi}{M}\;\frac{1}{\kcotdelta(k)}
    \left[1+\frac{\ii}{\cotdelta(k)}+\frac{\ii^2}{\cot^2\delta(k)}
      +\calO(\cot^{-3}\delta)\right]\;\;.
\end{equation} 
This is the expansion about the ``trivial/Gau\3ian'' fixed point of zero interactions.

On the other hand, in the \textbf{Unitarity Window}
$|\kcotdelta|\lesssim|\ii k|$ about the Unitarity Point $\cotdelta=0$, the
phase shift is ``large'',
$45^\circ\lesssim|\delta(k)|\lesssim135^\circ$. Albeit interactions are so
strong that they must be treated non-perturbatively at leading order by
resumming an infinite iteration, it is Unitarity which dominates the amplitude:
\begin{equation}
  \label{eq:unitarity-amp}
  A(k)\Big|_\mathrm{Uni}=\frac{4\pi}{M}\;\frac{1}{-\ii k}
  \left[1+\frac{\cotdelta(k)}{\ii}
    +\frac{\cot^2\delta(k)}{\ii^2}+\calO(\cot^3\delta)\right]\;\;.
\end{equation}
This seems paradoxical, but the interactions are actually so strong that their
details do not matter as much as the simple fact that they are very strong --
so strong indeed that probability conservation limits their impact on
observables. For example, the cross section is saturated,
$\sigma=\frac{4\pi}{k^2}[1-|\calO(\cotdelta)|]$. Moreover, an anomalously
shallow bound state emerges naturally since the amplitude's pole is at zero,
$k_\mathrm{pole}=0+\calO(\cotdelta)$. Since the leading amplitude has no
intrinsic scale, all dimensionless $\N\N$ observables are zero or infinite at
LO, while all dimensionful quantities (like the cross section) are homogeneous
functions of $k$. Their exponents are set by dimensional analysis, and their
dimensionless coefficients are independent of details of the interactions.
This expansion in the Unitarity Window is hence one about the
``Unitarity/non-trivial'' fixed point of ``maximally-strong'' interactions,
and Unitarity naturally implies Universality: 
Systems share the same behaviour, independent of interaction details at short
distances. This applies when they are so close to the same Unitarity fixed
point that differences between their interactions can be treated in
perturbation. Universality classes differ by different sets of
symmetries. That is important for imposing a dominant symmetry in
sect.~\ref{sec:inferences}.

Universality, in turn, has been proposed as key to the emergence of simple,
unifying patterns in complex systems like the nuclear
chart~\cite{Kievsky:2015dtk, Konig:2016utl, Kolck:2017zzf, Kievsky:2018xsl,
  vanKolck:2019qea, Konig:2019xxk}. Expanding about it quantitatively
reproduces key observables like the binding of \fourHe and its shallow
first excitation. Such indications exist also for heavier systems and
nuclear matter~\cite{Tews:2016jhi, Gattobigio:2019omi, Kievsky:2020sni,
  Georgoudis:2020sbg, Georgoudis:2022hzk, Contessi:2023pxk};
see~\cite{vanKolck:2020plz, Kievsky:2021ghz} for reviews.

\begin{figure}[!b]
    \begin{center}
      \includegraphics[width=0.5\linewidth]{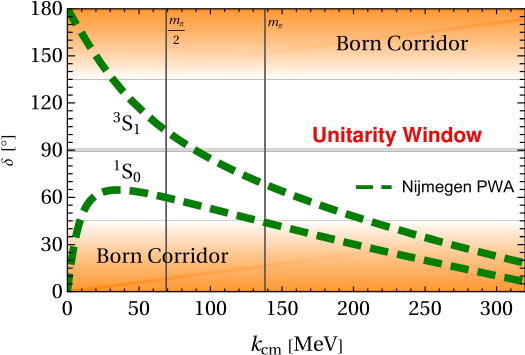}
      \caption{(Colour on-line) Born Corridor (shaded) and Unitarity Window
        (clear) of $\N\N$ \oneS and \threeS phase shifts in the Nijmegen
        PWA~\cite{Stoks:1993tb}. Marked are also the scales of the first and
        second non-analyticity in $\kcotdelta(k)$, from the branch points
        $k=\pm\ii\frac{\mpi}{2}$ and $\pm\ii\mpi$ of non- and once-iterated \OPE.}
    \label{fig:unitaritywindow}
    \end{center}
\end{figure}

The transition between Born Corridors and Unitarity Window is of course
gradual rather than abrupt. One expects computations of observables still to
be reliable to some degree as one intrudes into the other. The borders should
therefore not be taken literally but are fuzzily located around phase shifts
of about $45^\circ$ and $135^\circ$ ($|\cotdelta|\approx1$); see
fig.~\ref{fig:unitaritywindow}.

\absatz This leads to the natural question: Is the Unitarity Window relevant
in $\N\N$? According to Partial-Wave Analyses (PWAs)~\cite{Stoks:1993tb,
  NavarroPerez:2013usk, NavarroPerez:2013mvd,
  RuizArriola:2019nnv}\footnote{While we use the Nijmegen PWA, uncertainties
  are minuscule in these channels. The \oneS and \threeSD phase shifts and
  mixing angles reported by the Granada group differ by less that $0.6^\circ$
  up to $k=350\;\MeV$.}, only the \oneS and \threeS channels contain
anomalously shallow (real/virtual) bound states. Only their phase shifts are
clearly inside it at momenta relevant for low-energy properties of systems of
nucleons, namely $k\lesssim300\;\MeV$ (lab energies $\lesssim200\;\MeV$). From
fig.~\ref{fig:unitaritywindow}, that happens for
$35\;\MeV\lesssim k\lesssim200\;\MeV$. The estimate is conservative, and the
boundaries are again to be understood as fuzzy. 

In none of the other channels does the magnitude of the phase shift exceed
even $25^\circ$ ($|\cotdelta|\approx2$) for $k\lesssim300\;\MeV$. The ERE
mandates that $\kcotdelta_l(k\to0)\propto k^{-2l}\xrightarrow{k\to0}\infty$ is
at low momenta well outside the Unitarity Window in partial waves with orbital
angular momentum\footnote{A more careful $K$-matrix analysis of partial-wave
  mixing does not change this conclusion~\cite{TengMSthesis}.}  $l\ge1$. The
$\N\N$ system knows no physical mechanism to overcome that at higher energies
and push phase shifts into the Unitarity Window. \threeS is part of a coupled
channel, but the magnitudes of \threeSD mixing angle and \threeD phase shift
do not exceed $25^\circ$ and may be amenable to treatment in perturbation. The
\wave{3}{P}{0,2}, \wave{3,1}{P}{1} and \wave{3,1}{D}{2} channels hardly reach
$25^\circ$ ($|\cotdelta|\gtrsim2$), either. The degree to which they must be
treated nonperturbatively is evolving~\cite{Wu:2018lai, Kaplan:2019znu,
  Peng:2020nyz}, and we are for the purpose of this presentation agnostic
about the issue. All others have even smaller phase shifts
($|\delta(k)|\lesssim10^\circ$, $|\cotdelta|\gtrsim5$) and can easily be
described in perturbation~\cite{Kaiser:1997mw, Beane:2001bc, Birse:2005um,
  Birse:2009my}. Therefore, $l\ge1$ partial waves are not considered here --
not because they would not be important, but simply because the premises of
the Unitarity Expansion do not apply. Unless explicitly mentioned, statements,
results or conclusions in this article hold only for such a theory of $\N\N$
$\rmS$ waves, in the r\'egime where it applies.

Unitarity is so attractive because its two-nucleon state is highly
symmetric. At the fixed point, nonrelativistic field theories are
automatically not only scale-invariant but invariant under the larger
Schr\"odinger group of nonrelativistic conformal
transformations~\cite{Mehen:1999nd, Nishida:2007pj}. The ``Un-nuclear
Physics''~\cite{Hammer:2021zxb} this inspired is however not our focus. In
systems of two spin-$\half$ particles, \oneS and \threeS amplitudes are
identical at Unitarity, so that the symmetry group is augmented by invariance
under Wigner's combined $\SU(4)$ spin-isospin transformations~\cite{Wigner,
  Hund, Mehen:1999qs}. Different scattering-length and effective-range
corrections are included perturbatively; see~\cite{Epelbaum:2001fm,
  Riska:2002vn} for numerical evidence in phenomenological $\N\N$
potentials. The tightest-bound light nuclei, \fourHe, $^{12}$C and $^{16}$O,
are all near-perfect spin-isospin singlets; \cf~Similarity Renormalisation
Group studies in ref.~\cite{Timoteo:2011tt, RuizArriola:2013kdu}. Recently, Li
Muli \etal~\cite{LiMuli:2025zro} saw it dominate the $\beta$-decay of light
nuclei. In QCD's large-$N_C$ expansion, the leading non-tensor part of the
$\N\N$ interaction is automatically Wigner-$\SU(4)$ symmetric in even partial
waves~\cite{Kaplan:1995yg, Kaplan:1996rk}; \cf~\cite{CalleCordon:2008cz,
  RuizArriola:2016vap} for a concise summary. [Sect.~\ref{sec:largeNc} returns
to the interplay with large-$N_C$.]

While it would be an anthropomorphism to surmise that Nature prefers
expansions about configurations with very high degrees of symmetry, theorists
certainly do: Since Noether's theorem~\cite{Noether} relating continuous
symmetries to conserved quantities, employing symmetry principles to construct
systematic theories have become a cornerstone of modern Physics. As
$\frac{1}{a(\oneS)}\ne\frac{1}{a(\threeS)}$ and neither is zero, both
conformal invariance and Wigner-$\SU(4)$ symmetry are weakly broken in Nature
for $|\kcotdelta|<1$. Therefore, a \textbf{Unitarity Expansion} about the
nontrivial renormalisation-group fixed point with the symmetry group discussed
above should be useful~\cite{Birse:1998dk, Barford:2002je, Birse:2005um,
  Birse:2009my}.

In the language of Information Theory, it is the high degree of symmetry at
Unitarity which may imply that the existence of anomalous scattering lengths
is important, while their values are demoted to be less consequential, on par
with other information (like effective ranges). The result is a compression of
informational content by identifying relevant and filtering out irrelevant
information.

Wigner-$\SU(4)$ symmetry plays also a prominent r\^ole in ``Pionless EFT''
(\EFTNoPion). Its $3\N$ Counter Term (CT) needed to renormalise LO is
automatically spin-isospin invariant, with its Low-Energy-Coefficient (LEC)
fixed by a $3\N$ datum~\cite{Bedaque:1999ve}. Other examples include
classifying parity-violating and conserving $3\N$
interactions~\cite{Griesshammer:2010nd, Vanasse:2011nd, Lin:2022yaf,
  Lin:2024bor}; and binding patterns in nuclear systems~\cite{Birse:2012ih,
  Lu:2018bat, Lee:2020esp, Izosimov:2020thi, Liu:2025say, Niu:2025uxk}, like
the $^{12}$C ground and Hoyle
states~\cite{Shen:2021kqr}. 

However, the first non-analyticity of $\kcotdelta$ marks the formal breakdown
scale of \EFTNoPion, \ie~its mathematical radius of convergence. It comes from
the left-hand branch point $k=\pm\ii\frac{\mpi}{2}$ of $\N\N$ \OPE, the
longest-range nonlocal exchange interaction. The associated scale
$k\sim\frac{\mpi}{2}$ lies well inside the Unitarity Window, as
fig.~\ref{fig:unitaritywindow} shows. It also appears to be right around where
both partial waves are close to perfect Unitarity, $\delta=90^\circ$.
And yet, \EFTNoPion is of all EFTs the only one for which the Unitarity
Expansion has thus far been explored~\cite{Konig:2016utl, Kolck:2017zzf,
  Konig:2019xxk, Tews:2016jhi, vanKolck:2019qea, vanKolck:2020plz,
  Contessi:2023pxk}. That 
it appears to converge much better in practical applications, is
well-known. For example, a recent combination of Bayesian order-by-order
convergence analyses of $\n\p$ and $\n\mathrm{d}$ scattering put its breakdown
scale with $68\%$ degree of belief at
$[0.8; 1.2]\;\mpi$~\cite{Ekstrom:2024dqr, Ekstrom:2025ncs}. Even that makes it
cover barely more than half of the Unitarity Window.
Such tension makes one ask how the Unitarity Expansion emerges with pionic
degrees of freedom, \ie~in \ChiEFT which should extend to higher momenta and
automatically embed the non-analyticities~\cite{perspectives2}.

\absatz However, the question is not simply how big the window is in which the
Unitarity Expansion of \ChiEFT is relevant, \ie~both converges and describes
data efficiently. More fundamental is \emph{how} the key aspects of Unitarity
and Universality emerge with pions. In \ChiEFT, conformal and Wigner-$\SU(4)$
invariance are not manifest in the chiral Lagrangean but hidden. They are not
imposed when one constructs the EFT~\cite{Griesshammer:2021zzz}, in
contradistinction to chiral, Lorentz and other symmetries. Rather, they are
accidentally ``discovered'' when parameters are determined by data. They are
\emph{emergent phenomena}. Certainly, neither is manifest in \OPE~[$\ev_q$:
unit vector of momentum transfer $\qv$; $\vec{\sigma}_i,\vec{\tau}_i$: nucleon
$i$'s spin, isospin]:
\begin{equation}
  \label{eq:OPE}
  \begin{split}
  V_\mathrm{OPE}&=-\frac{\gA^2}{12\fpi^2}\;\frac{\qv^2}{\qv^2+\mpi^2}
  \bigg[
    \left(\vect{\sigma}_1\cdot\vect{\sigma}_2\right)+
  [3\left(\vect{\sigma}_1\cdot\ev_q\right)
    \left(\vect{\sigma}_2\cdot\ev_q\right)
    -\left(\vect{\sigma}_1\cdot\vect{\sigma}_2\right)]\bigg]
  \left(\vect{\tau}_1\cdot\vect{\tau}_2\right)\\[0.5ex]
  &=:\left(\vect{\sigma}_1\cdot\vect{\sigma}_2\right)
    \left(\vect{\tau}_1\cdot\vect{\tau}_2\right) V_C+
    [3\left(\vect{\sigma}_1\cdot\ev_q\right)
    \left(\vect{\sigma}_2\cdot\ev_q\right)
    -\left(\vect{\sigma}_1\cdot\vect{\sigma}_2\right)]
  \left(\vect{\tau}_1\cdot\vect{\tau}_2\right) V_T\;\;.
  \end{split}
\end{equation}
The pion decay constant $\fpi$ and $\mpi$ break scale (and hence conformal)
invariance explicitly because they carry mass dimensions, and the spin-isospin
structure of the tensor part, $V_T$, induces $\rmS\leftrightarrow\rmD$ and
$\rmD\to\rmD$ transitions. These are only possible in \threeS, and not in
\oneS where $V_T$ is of course identically zero. Such \threeSD mixing
therefore manifestly breaks the Wigner-$\SU(4)$ symmetry between the
$\mathrm{S}$ waves. In a possibly slight abuse of language, this is what we
call ``Wigner-$\SU(4)$ symmetry breaking''; see also
sect.~\ref{sec:WignerKSW}.

On the other hand, the spin-isospin structure of the central part, $V_C$, is
identical in these channels,
$(\vect{\sigma}_1\cdot\vect{\sigma}_2)(\vect{\tau}_1\cdot\vect{\tau}_2)=-3$,
and can in a similarly slight abuse of language be called ``Wigner-invariant''
(projected onto the \oneS-\threeS system).  It is also the \emph{only}
contribution in the \oneS channel, where the tensor piece is of course
identically zero. Thus, only amplitudes without $V_T$ contributions are
unchanged under Wigner-$\SU(4)$ transformations in the subspace of the
\oneS-\threeS channels; \cf~more formal discussion in
sect.~\ref{sec:WignerKSW}. Consequently, it is not self-understood how
explicit pionic degrees of freedom are reconciled with the symmetries of the
Unitarity Expansion which they appear to break rather strongly, while
concurrently extending the range to the whole Unitarity Window, including
$k\gtrsim\mpi$.

The Unitarity Expansion with pions sets thus up as a apparent clash of
fundamental but apparently incompatible symmetries. On the one side is the
Unitarity Expansion whose weakly broken conformal and Wigner-$\SU(4)$
symmetries treat the \oneS and \threeS channels as fundamentally identical,
and do therefore not allow \threeSD mixing. On the other side, the weakly
broken chiral symmetry of the Goldstone mechanism dictates the form of the
\OPE of eq.~\eqref{eq:OPE}, \threeS mixing, and that the \threeS and \oneS
channels are fundamentally different as the tensor interaction in one is
absent in the other, but \emph{prima facie} explicitly breaks Conformal and
Wigner-$\SU(4)$ symmetry rather strongly. All these symmetries are
approximate, but neither scenario accommodates the idealisation from which the
other starts. Each symmetry appears to be ``hidden'' and thus ``accidental''
from the perspective of the other. 

It makes thus sense to investigate in detail how to reconcile the two
approaches, so that both become manifest in observables. To our
knowledge, that has not yet been attempted.

We therefore choose to investigate the transition from ``pionless'' to
``pionic'' EFT in \ChiEFT ``with Perturbative/KSW Pions'', proposed by Kaplan,
Savage and Wise~\cite{Kaplan:1998tg, Kaplan:1998we}. In it, LO consists still
only of iterated, momentum-independent contact interactions. Together with
corrections to these, \OPE without iteration enters at next-to-leading order
(NLO), and once-iterated \OPE at \NXLO{2}. Twice-iterated \OPE as well as
correlated two-pion exchange is relegated to higher orders. At its breakdown
scale $\LambdaNN$, \OPE becomes non-perturbative~\cite{Kaplan:1998tg,
  Kaplan:1998we, Fleming:1999ee}, and its dimensionless expansion parameter at
typical momenta $\ptyp\sim k,\mpi$ is
\begin{equation}
  \label{eq:Q}
  Q=\frac{\ptyp}{\LambdaNN}\;\;\mbox{, with }
  \LambdaNN=\frac{16\pi\fpi^2}{\gA^2M}\approx300\;\MeV\;\;.
\end{equation}
Birse~\cite{Birse:2005um} pointed out that this scale depends on the total
angular momentum of the partial wave, but we reiterated that we are only
interested in $\rmS$ waves. From here on, we rephrase factors in OPE in terms
of this scale, \eg~$\frac{\gA^2}{12\fpi^2}=\frac{4\pi}{3M\,\LambdaNN}$ in
eq.~\eqref{eq:OPE}.
While numerically $\LambdaNN\approx2\mpi$ in the real world, it is not
explicitly dependent of $\mpi$ and hence remains close to that value in the
chiral limit. Fleming, Mehen and Stewart (FMS) found~\cite{Fleming:1999bs,
  Fleming:1999ee} that perturbative pions at \NXLO{2} fit \threeS phase shifts
rather poorly for $k\gtrsim140\;\MeV$, \ie~well below $\LambdaNN$, and found
order-by-order convergence elusive. And yet, this is the only \ChiEFT
generally accepted to be self-consistent and renormalisable order by order,
with a well-understood power counting~\cite{Beane:2001bc}, and thus regularly
employed for predictions in the absence of data, \eg~in
beyond-the-standard-model processes~\cite{Oosterhof:2019dlo,
  Oosterhof:2021lvt, OosterhofPhD}. Modifications of this perturbative-pion
version with better $\rmS$-wave convergence have also been
explored~\cite{Kaplan:1999qa, SanchezSanchez:2017tws, Beane:2008bt,
  Kaplan:2019znu}.

\absatz We thus propose the \textbf{\ChiEFT with Perturbative Pions in the
  Unitarity Expansion (\ChiEFTPPUE)}. Its \NXLO{2} amplitudes are based on the
work by Rupak and Shoresh (RS) in the \oneS channel~\cite{Rupak:1999aa} and by
FMS in \threeSD~\cite{Fleming:1999bs, Fleming:1999ee}.  We will argue that it
has a wider range of both convergence and agreement with PWAs than \EFTNoPion,
if the following holds:

\newcommand{\Conjecture}{\hspace*{\fill}
  \pbox{0.955\linewidth}{\absatz\textbf{Conjecture:} The symmetries of the
    Unitarity Limit are broken weakly in Nuclear Physics. Their footprint
    shows \emph{persistence}, \ie~their impact dominates observables at
    momentum scales $\ptyp\sim\mpi$ and beyond, and is there more relevant
    than chiral symmetry.  In particular, the tensor/Wigner-$\SU(4)$
    symmetry-breaking part of one-pion exchange in the $\N\N$ \threeS channel
    is super-perturbative, \ie~does not enter before \NXLO{3}.}}

\Conjecture

\absatz Co-author Teng's MSc thesis gave first results~\cite{TengMSthesis}.
Recently, we discussed findings and Conjecture at the ECT* workshop \emph{The
  Nuclear Interaction: Post-Modern Developments}, the \emph{11th International
  Workshop on Chiral Dynamics (CD2024)}, programme \emph{INT-24-3 Quantum Few-
  and Many-Body Systems in Universal Regimes} and workshop \emph{INT-25-92W
  Chiral EFT: New Perspectives}~\cite{talk1, talk2, talk3,
  talk4}. Ref.~\cite{Griesshammer:2025els} contains a digest and addenda.

\absatz This article is organised as follows. Appendices detail important
aspects which for the sake of flow are only summarised in the main text.
Section~\ref{sec:amplitudes} is devoted to methodology, with details relegated
to app.~\ref{app:amplitudes}. We briefly review the \EFTNoPion amplitudes to
\NXLO{2} in the Unitarity Expansion (sect.~\ref{sec:amplitudes-pionless}), and
our choice to extract phase shifts, pole positions and residues from the
amplitudes via $\kcotdelta$ (sect.~\ref{sec:amplitudestophaseshiftsandpoles}),
with details in apps.~\ref{app:amplitudestophaseshifts}
and~\ref{app:amplitudestopoles}. Section~\ref{sec:amplitudesKSW} discusses the
\ChiEFT amplitudes with Perturbative Pions in the Unitarity Expansion at
$\calO(Q^1)$ (\NXLO{2}; sects.~\ref{sec:LOamplitudes}
to~\ref{sec:N2LOamplitudes}); with app.~\ref{app:limits} on their $k\to0$
limit. It reviews the power counting (sect.~\ref{sec:PC}) and group theory of
Wigner-$\SU(4)$ symmetry, and identifies \OPE's transformation properties
(sect.~\ref{sec:WignerKSW}), followed by specifying renormalisation point and
parameters (sect.~\ref{sec:pathway}).
Section~\ref{sec:results} on the results starts with the zero-momentum and
pole properties. Our central results are sects.~\ref{sec:results1S0} on the
\oneS channel, and~\ref{sec:results3S1} on the \threeS channel both with and
without its Wigner-$\SU(4)$ breaking parts. A detailed accounting of the
robustness of our results and theory uncertainties is relegated to
apps.~\ref{app:uncertainties} and~\ref{app:fitpoints} but summarised in
sect.~\ref{sec:consistency}: order-by-order convergence, expansion parameters
inside the Unitarity Window, convergence to the PWA, complementing phase shift
extractions, and variation of the fit point.
After such fact-oriented investigations, sect.~\ref{sec:interpretation} turns
to ideas inspired by them. It develops the Conjecture
(sect.~\ref{sec:inferences}) and speculates about the impact of the nontrivial
fixed point on chiral symmetry (sect.~\ref{sec:fixedpoint}), as well as about
the interplay of Wigner symmetry with quantum-mechanical entanglement
(sect.~\ref{sec:entanglement}), and with QCD's large-$N_C$ limit
(sect.~\ref{sec:largeNc}).
After the customary summary, sect.~\ref{sec:conclusions} outlines steps to
test if the symmetries of the Unitarity Limit persist and dominate well into a
momentum r\'egime in which pions must be accounted for.

\section{Amplitudes and Observables}
\label{sec:amplitudes}

\subsection{Amplitudes in Pionless EFT}
\label{sec:amplitudes-pionless}

To set up \ChiEFTPPUE, we first consider the Unitarity Expansion at very low
energies. That is the r\'egime of \EFTNoPion, which is at \NXLO{2} identical
to the Effective Rage Expansion (ERE)~\cite{Schwinger, Chew, Barker,
  Bethe}. We will use it in sect.~\ref{sec:results} to check how important
pions actually are. In addition, some subtleties on extracting observables in
app.~\ref{app:amplitudestophaseshifts} become more transparent.

The function $\kcotdelta(k)$ which parametrises the interaction part proceeds
for the expansion about $k=0$ in powers of $k^2$ for
$k\lesssim\frac{\mpi}{2}$, \ie~below non-analyticities from non-iterated \OPE:
\begin{equation}    
    \label{eq:effrange}
    \kcotdelta(k)=-\frac{1}{a}+\frac{r}{2}\;k^2+
    \sum\limits_{n=2}^\infty v_n\;k^{2n}\;\;,
\end{equation}
with $a$ the scattering length, $r$ the effective range, $v_n$ the shape
parameters\footnote{Other conventions include $v_n=-P_n\;r^{2n-1}$ with
  dimensionless $P_n$ and factors of $2$, $4$ and $n!$.}. From
fig.~\ref{fig:unitaritywindow}, phase shifts are inside the Unitarity Window
even at these low momenta, so we choose:
\begin{equation}
  \label{eq:kcotdeltapionless}
  \kcotdelta_{0,-1}^\text{\NoPion}=0\;\;,\;\;
  \kcotdelta_{0,0}^\text{\NoPion}=-\frac{1}{a}+\frac{r}{2}\;k^2\;\;,\;\;
  \kcotdelta_{0,1}^\text{\NoPion}=0\;\;.
\end{equation}
The first subscript denotes the $\rmS$ channel ($l=0$); the second the order
in the expansion parameter $\QNoPion$ of \EFTNoPion. Inserting into the
Unitarity Expansion of eq.~\eqref{eq:unitarity-amp}, one finds at $\calO(\QNoPion^1)$
(\NXLO{2}):
\begin{equation}
\label{eq:amplitudes-pionless}
  A_{-1\text{\NoPion}}^{(\rmS)}(k)=\frac{4\pi\ii}{M}\;\frac{1}{k}\;\;,\;\;
  A_{0\text{\NoPion}}^{(\rmS)}(k)=-\frac{4\pi}{Mk}\;\left(
    \frac{1}{ka}-\frac{kr}{2}\right)\;\;,\;\;
  A_{1\text{\NoPion}}^{(\rmS)}(k)=\frac{[A_{0\text{\NoPion}}^{(\rmS)}(k)]^2}
  {A_{-1\text{\NoPion}}^{(\rmS)}(k)}\;\;.
\end{equation}
Therefore, the Unitarity Expansion in \EFTNoPion proceeds in powers of and
applies for $Q\sim\frac{1}{ka},\frac{rk}{2}\ll1$. This also implies
$\frac{r}{2a}\sim \QNoPion^2\ll1$, \ie~$r$ is of natural size for anomalously large
$a$. It breaks down for momenta which are either small,
$k\lesssim\frac{1}{a}$, or large, $k\gtrsim \frac{2}{r}$, relative to the ERE
scales. The dimensionless expansion parameters at, for example,
$k\approx\frac{\mpi}{2}$ are $\frac{1}{ka(\oneS)}\approx0.1$,
$\frac{kr(\oneS)}{2}\approx0.5$, $\frac{1}{ka(\threeS)}\approx0.5$ and
$\frac{kr(\threeS)}{2}\approx0.3$. Since none of these are particularly small,
convergence of observables must be assessed carefully; see
sect.~\ref{sec:consistency}. The two NLO contributions are of relative size
$|\frac{ar}{2}\;k^2|\approx4$ for $k\approx\frac{\mpi}{2}$ in \oneS and
$\approx0.6$ in \threeS. So we follow~\cite{Konig:2016utl} to define the
Unitarity Expansion such that $\calO(\QNoPion^0)$ (NLO) includes both the scattering
length and effective range. At higher $k$, $a$ could be considered \NXLO{2}
since $r$ dominates, but we strive for an expansion applicable throughout the
Unitarity Window.

No new Physics enters in \EFTNoPion at \NXLO{2} since the amplitude is
entirely determined by the LO and NLO results. The first additional datum
comes then from the shape parameter $v_2$ at $\calO(\QNoPion^2)$ (\NXLO{3}),
\ie one order higher than considered here. Additional information from $v_n$
enters at $\calO(\QNoPion^{2n-2})$ (\NXLO{2n-1}). No new information enters at
odd powers of $Q$ (\NXLO{2n}). Since pionic effects enter at each order beyond
LO, that changes in \ChiEFTPPUE whose power counting is described in
sect.~\ref{sec:PC}.

\subsection{From Amplitudes to Phase Shifts and Pole Positions}
\label{sec:amplitudestophaseshiftsandpoles}

We follow the canonical Stapp-Ypsilanti-Metropolis (SYM/''bar'') phase-shift
parametrisation~\cite{SYM}, also used by the Nijmegen~\cite{Stoks:1993tb} and
Granada~\cite{NavarroPerez:2013usk, NavarroPerez:2013mvd, RuizArriola:2019nnv}
PWAs. Phase shifts and bound-state properties ($S$-matrix poles and residues)
are then found by an order-by-order expansion of the amplitude in ``strict''
perturbation as defined in Mathematical Perturbation Theory~\cite{Bellman,
  BenderOrszag, Murdock, Holmes}; see also~\cite{Rupak:1999aa, Bedaque:1999vb}
and a technical summary~\cite{hgrienotes}. This preserves all symmetries at
each order independently, including $S$-matrix unitarity. We will mostly use
the expansion of $\kcotdelta$, as this is the fundamental variable of the
Unitarity Expansion. The path from amplitudes via $\kcotdelta(k)$,
eqs.~(\ref{eq:LOkcotdeltaSS}-\ref{eq:N2LOkcotdeltaSS}), to phase shifts and
pole parameters is described in apps.~\ref{app:amplitudestophaseshifts} and
app.~\ref{app:amplitudestopoles}, respectively. There, we also comment on
approach's advantage outside the Unitarity Window. We show in
app.~\ref{app:uncertainties-delta} that inside the Unitarity Window,
extraction variants are consistent within expected higher-order uncertainties.

\subsection{Amplitudes in \texorpdfstring{\ChiEFT}{ChiEFT} with Perturbative
  Pions about Unitarity}
\label{sec:amplitudesKSW}

\subsubsection{Power Counting}
\label{sec:PC}

Guided by the discussion around eq.~\eqref{eq:Q} in the Introduction and
sect.~\ref{sec:amplitudes-pionless}, we count powers of $Q$ in \ChiEFTPPUE as
follows. In an EFT with resummed LO in the 2-body system, LO must for
consistency on general grounds be counted as $\calO(Q^{-1})$; see
\eg~\cite{Griesshammer:2020fwr, Griesshammer:2021zzz}.  The Unitarity Window
of the $\N\N$ system with perturbative \OPE dictates then
$\frac{1}{|a|}\ll\ptyp\ll\LambdaNN$. For simplicity, we equate the two
dimensionless expansion parameters into
\begin{equation}
  \label{eq:Qunitarity}
  Q:=\frac{\ptyp=(k,\mpi)}{ \LambdaNN}\approx\frac{1}{a\,\ptyp}\to Q\;\;,
\end{equation}
since they are numerically close for $k\approx\mpi$ and physical values of the
$\N\N$ scattering lengths. We also set $r\sim\mpi^{-1}$,
\ie~$\frac{kr}{2}\sim Q$. All this implies
$\frac{1}{a\LambdaNN}\sim\frac{r}{2a}\sim Q^2$ and
$\frac{r\LambdaNN}{2}\sim Q^0$. That is not inconsistent with the real-world
numerical values. With this,
LO is Unitarity; NLO ($\calO(Q^0)$) adds non-iterated \OPE, finite scattering
length and nonzero effective range (\cf~sect.~\ref{sec:amplitudes-pionless});
and \NXLO{2} ($\calO(Q^1)$) once-iterated \OPE and corrections in $a,r$. Neither
is zero.

\subsubsection{Wigner-$\SU(4)$ Symmetry and Its Breaking in
  \texorpdfstring{\ChiEFTPPUE}{ChiEFT(pp)UE}}
\label{sec:WignerKSW}

We first briefly review the group theory of Wigner-$\SU(4)$ spin-isospin
transformations, following~\cite{CalleCordon:2008cz}.  Its second Casimir
operator $C_2^\mathrm{Wigner}:=\frac{15+\sigma+\tau+\sigma\tau}{2}$ has
eigenvalue $5$ for super-sextuplets ($l$ even), and $9$ for super-decuplets
($l$ odd) (defining $\sigma:=\vect{\sigma}_1\cdot\vect{\sigma}_2$ \etc). As it
commutes with the orbital angular momentum operator, we decide to
classify super-multiplets in addition by a superscript $l$ for
$\vect{l}^2=l(l+1)$.

The \oneS-\threeS channels form an (antisymmetric) super-sextuplet Irrep
$[\mathbf{6}_A]^{l=0}$ ($(S,I)=(0,1)\vee(1,0)$). In general, a super-sextuplet
$[\mathbf{6}_A]^l$ applies to all partial waves with even orbital angular
momentum $l$. The next is $[\mathbf{6}_A]^{2}$, the spin-singlet
isospin-triplet \wave{1}{D}{2} combined with the Wigner-symmetric components
of the spin-triplet-isospin-singlets \wave{3}{D}{2} and \wave{3}{DF}{3}, as
well as \threeD which mixes with the \oneS-\threeSD Irrep; see below. Odd
partial waves pair to (symmetric) super-decuplet Irreps $[\mathbf{10}_S]^l$,
$(S,I)=(0,0)\vee(1,1)$. The lowest-$l$ representation combines the
spin-singlet-isospin-singlet \wave{1}{P}{1} and the Wigner-symmetric
components of the spin-triplet isospin-triplets
\wave{3}{P}{0}-\wave{3}{P}{1}-\wave{3}{PF}{2}.
[One could also classify Wigner-supermultiplets by total angular momentum
$j$. In mixed channels like \threeSD, the supermultiplet axes are then
slightly tilted off the $l=j\pm1$ state. However, PWAs show that $\N\N$
phase-shift mixing angles are very small, $|\epsilon|<5^\circ$, \ie~pure
$l=j\pm1$ states can be treated as $[\mathbf{W}]^l=[\mathbf{W}]^j$ at LO, with
corrections in perturbation; \cf~also~\cite[sect.~V.C]{CalleCordon:2008cz}.]

The origin of Wigner-$\SU(4)$ breaking is thus twofold. The central operator
$\vect{\sigma}_1\cdot\vect{\sigma}_2$ splits members of each super-multiplet
$[\mathbf{W}]^l$ at fixed $l$ by spin $S=0$ (eigenvalue $-1$) or $1$
(eigenvalue $3$), \eg~by different effective-range parameters $a$ and $r$ in
the two channels. However, it does not mix super-multiplets with different $l$
since it commutes with both $C_2^\mathrm{Wigner}$ and $\lv^2$. On the other hand, \OPE's tensor
operator in eq.~\eqref{eq:OPE} does not commute with $\vect{l}^2$ (but of
course with $\vect{j}^2$), yet it does commute with $C_2^\mathrm{Wigner}$. It
lifts therefore the degeneracy both within and between super-multiplets
$[\mathbf{6}_A]^{l}$ and $[\mathbf{10}_S]^{l}$ for different $l$ (where
sextets do not mix with decuplets because parity and $j$ must be
conserved). This justifies the term ``Wigner-breaking''. It does not affect
$S=0$ states, but mixes exclusively the $S=1$ state in the same
super-multiplet with one in another super-multiplet with the same $j$ but
different $l=j\pm1$, like the $j=1$ states \threeS in $[\mathbf{6}_A]^{0}$ and
\threeD in $[\mathbf{6}_A]^{2}$. The operator of \OPE's central part,
$V_C\propto\sigma\tau$, also commutes with the Casimir operator. It acts on
all members of super-sextuplets like \oneS-\threeS by returning the same state
with eigenvalue $-3$ and is thus there proportional to the unit operator. That
justifies the name ``Wigner-$\SU(4)$ symmetric'' contribution, albeit some may
prefer ``Wigner-$\SU(4)$ covariant''. In a super-decuplet
$[\mathbf{10}_S]^{l}$ like \wave{1}{P}{1}-\wave{3}{P}{0,1,2}, the latter is
more appropriate since it returns different values: $9$ for $S=I=0$ and $1$
for $S=I=1$. The impact of Wigner-symmetry there is worth
exploring~\cite{CalleCordon:2008cz}. As the Unitarity Expansion is
inapplicable in all partial waves except for the \oneS-\threeS super-sextet,
we continue with the attributions ``symmetric'' and ``breaking''.

We now discuss at which order and how Wigner-breaking terms enter in
\ChiEFTPPUE. At NLO, \OPE is inserted once. Symbolically, one projects
$V_\mathrm{OPE}$ (with contact terms for renormalisation) onto the \oneS and
\threeS channels:
\begin{equation}
  \begin{split}
    A_{0}(\oneS)&=\langle\oneS|V_C|\oneS\rangle=:A_{0}^{\rmS}\\
    A_{0}(\threeS)&=
    \begin{pmatrix}|\threeS\rangle\\0\end{pmatrix}^\dagger
    \begin{pmatrix}V_C&\sqrt{8}\;V_T\\\sqrt{8}\;V_T&V_C-2V_T\end{pmatrix}
    \begin{pmatrix}|\threeS\rangle\\0\end{pmatrix}
    =\langle\threeS|V_C|\threeS\rangle=A_{0}^{\rmS}
  \end{split}
\end{equation}
The LO amplitudes and states $|\oneS\rangle$ and
$|\threeS\rangle$ are $\rmS$-waves only and Wigner-$\SU(4)$ invariant,
$\sigma\tau|\threeS\rangle=\sigma\tau|\oneS\rangle$. The pion part at
NLO is therefore necessarily Wigner-$\SU(4)$ symmetric. \OPE-induced
differences between the partial waves only enter at \NXLO{2}:
\begin{equation}
  \label{eq:N2LOWigner}
  \begin{split}
    A_{1}(\oneS)&=\langle\oneS|V_CGV_C|\oneS\rangle=:A_{1\mathrm{sym}}^{(\rmS)}\\
    A_{1}(\threeS)&=
    \begin{pmatrix}|\threeS\rangle\\0\end{pmatrix}^\dagger
    \begin{pmatrix}V_C&\sqrt{8}\;V_T\\\sqrt{8}\;V_T&V_C-2V_T\end{pmatrix}G
    \begin{pmatrix}V_C&\sqrt{8}\;V_T\\\sqrt{8}\;V_T&V_C-2V_T\end{pmatrix}
    \begin{pmatrix}|\threeS\rangle\\0\end{pmatrix}\\[1ex]
    &=A_{1}(\oneS)+ \langle\threeS|8V_TGV_T|\threeS\rangle=:
    A_{1\mathrm{sym}}^{(\rmS)}+A_{1\mathrm{break}}^{(\rmS)}
  \end{split}
\end{equation}
NLO is $A_{0}^{\rmS}$; the symmetric (breaking) \NXLO{2} part is
$A_{1\mathrm{sym}}^{\rmS}$ ($A_{1\mathrm{break}}^{\rmS}$).  The ``free'' LO
two-nucleon propagator $G$ is Wigner-symmetric at Unitarity. Wigner-breaking comes exclusively from once iterating \OPE via
$V_T$ and induces a $\rmS\to\rmD\to\rmS$ intermediate state.
A $\rmD\rmD$ matrix element does not appear before twice-iterated \OPE,
\ie~\NXLO{3}. Extension to other channels is
straightforward~\cite{future1}. It is natural to assume that the amount of
breaking at higher orders grows exponentially with the number of $V_T$
insertions.

We reiterate that this reasoning only applies to the pion-part of the
interaction and the corresponding CTs which absorb regulator-dependence. In
our case, their ``finite parts'' are set by the empirical scattering
lengths and effective ranges. These break Wigner-$\SU(4)$ symmetry explicitly
but weakly. For a meaningful distinction between Wigner-symmetric and breaking
contributions, it must of course be renormalisation-group invariant. That is
fulfilled by construction if the regulator preserves Wigner-$\SU(4)$ symmetry,
like in dimensional regularisation with Power-Divergence
Subtraction~\cite{Kaplan:1998tg, Kaplan:1998we} which we use.

\subsubsection{Pathway and Parameters}
\label{sec:pathway}


\begin{figure}[!b]
\begin{center}
     \includegraphics[width=\linewidth]{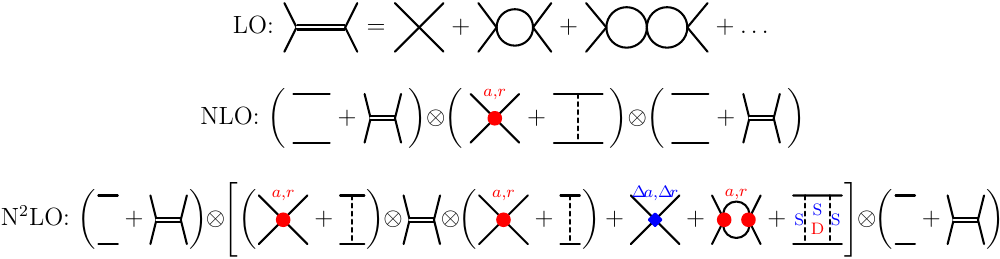}
     \caption{(Colour on-line) \ChiEFTPPUE at LO (top); NLO (middle) with CTs
       (red circle) fixed to reproduce scattering length $a$ and effective
       range $r$; \NXLO{2} with CTs (blue diamonds) fixed so that $a$ and $r$
       do not change from the NLO value. Last \NXLO{2} term in square
       brackets: once-iterated \OPE, with orbital angular momentum of the
       intermediate state as indicated.}
\label{fig:amplitudes}
\end{center}
\end{figure}

The $\N\N$ amplitudes in the \oneS and coupled \threeSD channels with
perturbative (KSW) pions were derived for finite scattering length by Kaplan,
Savage and Wise~\cite{Kaplan:1998tg, Kaplan:1998we} to NLO, and extended to
\NXLO{2} by RS (\oneS)~\cite{Rupak:1999aa}, and then by FMS (\threeS,
\oneS)~\cite{Fleming:1999bs, Fleming:1999ee} (both \oneS results
agree). Figure~\ref{fig:amplitudes} shows the contributions at each order. The
last line includes those diagrams with $\mathrm{SD}$-mixing interactions, all
of which enter only at \NXLO{2} as just discussed. Those which enter only in
the phase shift of the \threeD channel and of the $\mathrm{SD}$ mixing angle
are discussed in an MSc thesis~\cite{TengMSthesis} and will be subject of a
future publication~\cite{future1}. In order to confirm correct coding, we
reproduced the FMS results with their parameter values. This is nontrivial
since some are highly sensitive to the exact numbers used\footnote{We are
  grateful for Iain Stewart's help in numerous conversations, both
  electronically and in person.}.

We then expanded the amplitudes about Unitarity to find the LECs at its
natural renormalisation point, $\kfit=0$.  Section~\ref{sec:consistency}
summarises discussions of different $\kfit\ne0$ in app.~\ref{app:fitpoints}.

FMS find their LO LECs (equivalent to our fixing $a$) from the pole positions,
and their dimensionless NLO and \NXLO{2} LECs $\zeta_1$ to $\zeta_4$ from
weighted least-squares fits to the Nijmegen PWA which do not change the LO
pole positions. We did confirm FMS' finding that further coefficients
$\zeta_5,\zeta_6$ of higher-order contributions do not solve the poor
convergence at \NXLO{2} and discard them here from the start; \cf~discussion
in sect.~\ref{sec:difference-sym-vs-break}.

For clarity, we regrouped the amplitudes into Wigner-$\SU(4)$ symmetric and
breaking parts, each sorted by powers of the dimensionless ratio
$\frac{k}{\mpi}$. The results appear considerably shorter that those of
FMS~\cite{Fleming:1999ee} for three reasons. First, they are rewritten using
the ERE parameters $a$, $r$ in favour of FMS' $\zeta_1$ to $\zeta_4$. Second,
the Unitarity Expansion leads to a few simplifications. Third, we found some
more economical ways to rewrite certain terms.

On a technical note, we do not differentiate between $\N\N$ CTs which
explicitly break chiral symmetry by quark (and hence pion) mass
dependence. Staring at NLO, a $\N\N$ LEC $D_2\mpi^2$
enters~\cite{Kaplan:1998sz} with the same operator structure as the CT whose
LEC is determined by the scattering length. Here, we do not vary the pion
mass, and the amplitudes we use are regulator-independent, so such effects
cannot be disentangled.

We use the average nucleon and pion masses $\MN=938.91897\;\MeV$,
$\mpi:=\frac{2\mpi^\pm+\mpi^0}{3}=138.037\;\MeV$, the axial pion-nucleon
coupling $\gA=1.267$, pion decay constant\footnote{KSW and FMS use
  $f=\sqrt{2}\fpi$~\cite{Kaplan:1998tg, Kaplan:1998we, Fleming:1999bs,
    Fleming:1999ee}.} $\fpi=92.42\;\MeV$, and $197.327\;\fm\,\MeV=1$. The
position of cuts and poles in amplitudes induced by pionic effects becomes
manifest when one replaces $\mpi^2\to\mpi^2-\ii\epsilon$ with
$\epsilon\searrow0$.

Generic scattering lengths and effective range are denoted by $a$ and $r$ and
replaced by the \oneS or \threeS values as appropriate. We choose those
reported by the Granada group~\cite{RuizArriola:2019nnv}:
$\aoneS=-23.735(6)\;\fm$, $\roneS=2.673(9)\;\fm$; and
$\athreeS=5.435(2)\;\fm$, $\rthreeS=1.852(2)\;\fm$. The slightly different
results and parameter values in the MSc thesis~\cite{TengMSthesis} used the
Nijmegen group's 1995 analysis~\cite{Stoks:1993tb, deSwart:1995ui}:
$\aoneS=-23.714\;\fm$, $\roneS=2.73\;\fm$; $\athreeS=5.420(1)\;\fm$,
$\rthreeS=1.753(2)\;\fm$. However, induced differences even at
$k\approx300\;\MeV$ are smaller than $2.7^\circ$ and hardly exceed the line
widths.

\subsubsection{Amplitude at $\calO(Q^{-1})$ (Leading Order)}
\label{sec:LOamplitudes}

The LO contribution at Unitarity is identical in the \threeSD and \oneS
channel and therefore obviously both conformally and Wigner-$\SU(4)$
invariant. Only the \rmS-wave component is nonzero and of course identical to
the pionless result of eq.~\eqref{eq:amplitudes-pionless}:
\begin{equation}
  \label{eq:LOamplitude}
  A_{-1}^{(\rmS)}(k)=\frac{4\pi\ii}{M}\;\frac{1}{k}\;\;.
\end{equation}

\subsubsection{Amplitude at $\calO(Q^0)$ (Next-To-Leading Order)}
\label{sec:NLOamplitudes}

Now, scattering length, effective range and pions enter:
\begin{equation}
    \label{eq:NLOamplitude} 
    A_{0}^{(\rmS)}(k)=
    -\frac{4\pi}{M}\left[\frac{1}{k}\left(\frac{1}{ka}-\frac{kr}{2}\right)
    +\frac{1}{\LambdaNN}\left(1-\frac{\mpi^2}{4k^2}\;
      \ln[1+\frac{4k^2}{\mpi^2}]\right)\right]\;\;.
\end{equation}
This, including its \OPE part, is indeed suppressed against LO in the
Unitarity Window ($\frac{1}{ka},\frac{k}{\LambdaNN}<1$), consistent with the
power counting.

The first term is the NLO amplitude of the pionless version,
eq.~\eqref{eq:amplitudes-pionless}. Differences of scattering lengths and
effective ranges induce the \emph{only} Wigner-$\SU(4)$ breaking effects at
this order. While the pionic potential of eq.~\eqref{eq:OPE} breaks conformal
and Wigner symmetry, \OPE at NLO naturally accommodates Wigner-$\SU(4)$
symmetry as its projections onto the \oneS and \threeS channels are identical,
as anticipated in sect.~\ref{sec:WignerKSW}. It is also the only non-analytic
contribution, with just one branch point at $k=\pm\ii\frac{\mpi}{2}$.

As $k\to0$, the non-pionic part is exactly the \EFTNoPion result of
eq.~\eqref{eq:amplitudes-pionless}. Corrections come exclusively from pions
and vanish in dimensionless units as $\frac{2k^3}{\LambdaNN\mpi^2}$; see
eq.~\eqref{eq:NLOlimit} in app.~\ref{app:limits}.  Thus, they contribute
substantially only for $k\gtrsim100\;\MeV$: Pion contributions violate scale
invariance only weakly for low momenta, and not at all for zero momenta. This
is expected since all scale-breaking at $k=0$ is subsumed into the different
scattering lengths and effective ranges in \oneS and \threeS. The
$\calO(\gA^2)$ contribution in eq.~\eqref{eq:NLOamplitude} constitutes only
the long-range part of \OPE (without iteration), after CTs absorb divergences
to reproduce the empirical $a,r$. The phrase ``pion contribution'' is
understood in that sense: the pionic long-distance (non-analytic) part after
renormalisation within the chosen renormalisation scheme (Power Divergence
Subtraction in dimensional regularisation\cite{Kaplan:1998tg, Kaplan:1998we})
and renormalisation condition ($\kfit=0$ with empirical $a$ and $r$).

\subsubsection{Amplitude at $\calO(Q^1)$ (Next-To-Next-To-Leading Order)}
\label{sec:N2LOamplitudes}

No new low-energy scattering parameters enter. As argued in the Introduction
and sect.~\ref{sec:WignerKSW}, the Wigner-$\SU(4)$ symmetric part is identical to the contribution in the \oneS channel:
\begin{align}
    A_{1}^{(\oneS)}(k)\equiv& A_{1\mathrm{sym}}^{(\rmS)}(k)=
    \frac{\left[A_{0}^{(\rmS)}(k)\right]^2}{A_{-1}^{(\rmS)}(k)}
    +\frac{8\pi}{M\,\LambdaNN}
        \bigg\{\frac{4}{3a\mpi}-\frac{\mpi}{k}\left(\frac{1}{ka}-\frac{kr}{2}
  \right)
    \label{eq:N2LOamplitudeWignersym}\\&
  -
  \frac{M}{4\pi}A_{0}^{(\rmS)}(k)\;\frac{\mpi^2}{2k}\;\arctan[\frac{2k}{\mpi}]
  -\frac{\mpi}{\LambdaNN}
  \bigg[\frac{1}{12}+\left(\frac{\mpi^2}{4k^2}-\frac{1}{3}\right)\ln2
  -G_\pi(\frac{k}{\mpi})\bigg]\bigg\}\;\;.\non
\end{align}
The first term combines the \NXLO{2} \EFTNoPion amplitude of
eq.~\eqref{eq:amplitudes-pionless} with insertions of the pionic NLO pieces;
see eq.~\eqref{eq:NLOamplitude} and first term in square brackets in
fig.~\ref{fig:amplitudes}. The second term is a correction from (non-iterated)
\OPE and CTs to keep $a$ and $r$ at \NXLO{2} fixed to the NLO values (\NXLO{2}
diagrams labelled $a,r$ and $\Delta\!\,a,\Delta\!\,r$ in
fig.~\ref{fig:amplitudes}). The $\calO(\LambdaNN^{-2})$ contributions of the
last line encode the long-range part of once-iterated \OPE via
$\rmS\to\rmS\to\rmS$ only (last \NXLO{2} diagram in square brackets of
fig.~\ref{fig:amplitudes} and eq.~\eqref{eq:N2LOWigner}). The amplitude has a
branch point from non-iterated \OPE at $k=\pm\ii\frac{\mpi}{2}$, and in
addition from once-iterated \OPE at $k=\pm\ii\,\mpi$. The latter comes from
the last term of the dimensionless function
\begin{equation}
    \label{eq:auxfunction}
    G_\pi(x):=\frac{1}{8x^3}\left\{\arctan[2x]\ln[1+4x^2]-
      \Im\Big[\mathrm{Li}_2[\frac{2\ii x+1}
      {2\ii x-1}]+2\mathrm{Li}_2[\frac{1}{2\ii x-1}]\Big]\right\}
\end{equation}
since Euler's Dilogarithm (Spence function)
$\mathrm{Li}_2(z)=-\int\limits_0^z\deint{}{t}\frac{\ln[1-t]}{t}$ has a branch
point at $z=1$~\cite[sect.~25.12(i)]{NIST}. Furthermore,
$G_\pi(x\to0)=1-\frac{10x^2}{3}+\calO(x^4)$ is finite as $k\to0$, and
monotonically falling towards its chiral limit
$G_\pi(x\to\infty)=\frac{\pi}{8}\;\frac{\ln[2x]}{x^3}+\calO(x^{-4})$.

The non-pionic ($(\LambdaNN)^0$) part reduces to the pionless version,
eq.~\eqref{eq:amplitudes-pionless}, \ie~$a$ and $r$ are unchanged from
NLO. The pionic contribution to $\kcotdelta$ vanishes faster than $k^2$,
namely in dimensionless units as
$\frac{4k^3}{\LambdaNN\mpi}(\frac{1}{a\mpi^2},r)$ dominated by the size of
$r\approx\frac{1}{\mpi}$. With $a\,\mpi\gg r\,\mpi\approx1$, the scale for
once-iterated \OPE is of the same magnitude,
$\frac{k}{\LambdaNN}\,\frac{k^2}{\mpi^2}$. Both are hence again small for
$k\lesssim\mpi$.  $A_{0}^{(\rmS)}(k\to0)$ is given in
eq.~\eqref{eq:N2LOlimitWigner} of app.~\ref{app:limits}.

A Wigner-$\SU(4)$ breaking part enters in the \threeS amplitude:
\begin{align}
    \label{eq:N2LOamplitude3S1}
    A_{1}^{(\threeS)}(k)=&A_{1\mathrm{sym}}^{(\rmS)}(k)+A_{1\mathrm{break}}^{(\rmS)}(k)\;\;,
  \\
  A_{1\mathrm{break}}^{(\rmS)}(k)=&\frac{16\pi\mpi}{M\,\LambdaNN^2}
           \bigg\{\overbrace{\frac{571-352\ln2}{210}}^{=1.5572\dots}
                                    -\left(1+\frac{3\mpi^2}{2k^2}
    +\frac{9\mpi^4}{16k^4}\right)G_\pi(\frac{k}{\mpi})
    \non\\&+\frac{2\mpi^2}{5k^2}(\ln4-1)+
    \frac{3\mpi^4}{16k^4}
    -\frac{3}{2}\left[\left(\frac{k}{\mpi}+\frac{\mpi}{k}\right)
  -\left(\frac{\mpi^3}{8k^3}+\frac{3\mpi^5}{16k^5}\right)\right]
  \arctan[\frac{k}{\mpi}]\non\\&
  +\frac{3}{16}\left(\frac{\mpi^4}{k^4}+\frac{3\mpi^6}{4k^6}\right)
  \ln[\frac{16(k^2+\mpi^2)}{4k^2+\mpi^2}]\bigg\}
  -\frac{\left[A_{0}^{(\rmS\rmD)}(k)\right]^2}{A_{-1}^{(\rmS)}}\;\;.
    \label{eq:N2LOamplitudeWignerbreak}
\end{align}
Since $A_{1\mathrm{break}}^{(\rmS)}$ is $\calO(\LambdaNN^{-2})$, it comes exclusively
from the transition $\rmS\to\rmD\to\rmS$ of once-iterated \OPE; see
fig.~\ref{fig:amplitudes} and $\langle\threeS|8V_TGV_T|\threeS\rangle$ of
eq.~\eqref{eq:N2LOWigner}. In curly brackets is the part which is not simply
twice the non-iterated OPE between on-shell nucleons. It has branch points
again at $k=\pm\ii\frac{\mpi}{2},\;\pm\ii\,\mpi$, with the latter from
$\arctan\frac{k}{\mpi}$, $\ln[k^2+\mpi^2]$ and $G_\pi$. The very last term is
the on-shell piece, \ie~the NLO $\rmS\rmD$-mixing amplitude-squared:
\begin{equation}
    \label{eq:NLOamplitudeSD}
    A_{0}^{(\rmS\rmD)}(k)=-\frac{\ii\pi\sqrt{2}\,\mpi^2}{M\,\LambdaNN\,k^2}
    \left[\frac{3\mpi}{2k}-\left(1+\frac{3\mpi^2}{4k^2}\right)
      \arctan[\frac{2k}{\mpi}]\right]\;\;.
\end{equation}
However, that term cancels the same contribution in the phase shift,
eq.~\eqref{eq:N2LOphaseshiftSS}, and in $\kcotdelta$,
eq.~\eqref{eq:N2LOkcotdeltaSS}, irrespective of whether $A_{-1}^{(\rmS)}$ is
at Unitarity or not. Since $A_{0}^{(\rmS\rmD)}(k=0)=0$, it does not contribute
to shifting the pole, either; \cf~eq.~\eqref{eq:N2LOpole}. The LO $\rmS\rmD$
amplitude is zero; neither its \NXLO{2} nor any $\rmD\rmD$ amplitude enters for
$\rmS$ waves at this order~\cite{TengMSthesis, future1}.

According to eq.~\eqref{eq:N2LOlimitbreak} of app.~\ref{app:limits},
$\kcotdelta(k\to0)$ vanishes again as
$\frac{k}{\LambdaNN}\,\frac{k^2}{\mpi^2}$ ($a\,\mpi\gg r\,\mpi\approx1$), so
that $a$ and $r$ are not shifted from the NLO values.

We see that all \NXLO{2} \OPE contributions are in the Unitarity Window
suppressed by powers of $Q\sim\frac{1}{ka}\sim\frac{k,\mpi}{\LambdaNN}<1$
against LO, and against both pionic and non-pionic NLO parts separately. This
confirms $Q$ and $\LambdaNN$ of eq.~\eqref{eq:Qunitarity} in $\rmS$-waves for
the Unitarity Expansion of 
\ChiEFT with Perturbative Pions.

\section{Results and Analysis}
\label{sec:results}

All observables include Bayesian truncation uncertainties at \NXLO{2},
estimated semi-quanti\-tatively as $68\%$ degree-of-belief (DoB) intervals via
the ``$\max$'' criterion~\cite[sect.~4.4]{Griesshammer:2012we} with reasonable
priors following~\cite{Cacciari:2011ze, Furnstahl:2015rha,
  Griesshammer:2015ahu}. As posteriors are non-Gau\3ian, the width of
$95\%$ DoB intervals is about $2.7$ times $68\%$ DoB. These and other
theory uncertainties are discussed in sect.~\ref{sec:consistency}, and
apps.~\ref{app:uncertainties},~\ref{app:fitpoints}. The centre-of-mass
momentum $k$ is the primary variable.

\subsection{Zero-Momentum Limit and Pole Parameters}
\label{sec:resultspoles}

As $k\to0$, NLO and \NXLO{2} contribute to $\kcotdelta$ only
via even powers of $k$, as required by analyticity below
$k=|\pm\ii\frac{\mpi}{2}|$~\cite{Schwinger, Chew, Barker, Bethe}; see
app.~\ref{app:limits}.
Since the renormalisation point is ERE at $k=0$ with Unitarity,
$\frac{1}{a}=0$, LO is $\kcotdelta_{0,-1}=0$. NLO is
$\kcotdelta_{0,0}=\left(-\frac{1}{a}+\frac{r}{2}k^2\right)+\calO(k^4)$;
eq.~\eqref{eq:NLOlimit}. \NXLO{2} vanishes, $\kcotdelta_{0,1}=0+\calO(k^4)$;
eqs.~(\ref{eq:N2LOlimitWigner}/\ref{eq:N2LOlimitbreak}). Therefore, pole
positions and residues in \ChiEFTPPUE, table~\ref{eq:polevalues}, are
unchanged from ERE/\EFTNoPion, eqs.~(\ref{eq:pole}/\ref{eq:residue}) in
app.~\ref{app:amplitudestopoles}. We use the Granada group's empirical values
(with negligible uncertainties)~\cite{RuizArriola:2019nnv}: in the \threeS
channel, derived from the deuteron's binding energy and asymptotic
normalisation ($\gamma=\sqrt{M B_2}$, $Z=\frac{A_S^2}{2\gamma}$); fr the \oneS
channel, inferred from the scattering length, effective range and shape
parameter as described in~\cite{Griesshammer:2004pe}.  From
app.~\ref{app:uncertainties-Bayes}, the $68\%$ DoB interval for theory
truncation uncertainties is $\pm\frac{\ii}{a}\frac{r^3}{2a^3}$ for the pole
position and $\pm\frac{4r^3}{a^3}$ for its residue (times about $2.7$ for
$95\%$ DoBs). Within these, comparison to the empirical values is favourable.

\begin{table}[!h]
  \centering\footnotesize
  \setlength{\tabcolsep}{4.5pt}
  \begin{tabular}{|l||ll|ll|}
    \hline
    \rule[-1.5ex]{0ex}{4ex}&\multicolumn{2}{c|}{\oneS}
    &\multicolumn{2}{c|}{\threeS}\\[-0.5ex]
    &\hfill$\ii\gamma\;[\MeV]$\hspace*{\fill}&\hfill$Z$\hspace*{\fill}
    &\hfill$\ii\gamma\;[\MeV]$\hspace*{\fill}&\hfill$Z$\hspace*{\fill}\\
    \hline\hline
    \rule[-1.5ex]{0ex}{4ex}
    NLO &$\:-\whitey8.314\,\ii$&$\:0.887$&$\:+36.3\,\ii$&$\:1.34$\\
    \rule[-1.5ex]{0ex}{4ex}
    \NXLO{2}&$[-\whitey7.898\pm0.006]\ii$&$[0.906\whitey\pm0.006]$
               &$[+44.6\whitey\whitey\whitey\pm0.7]\ii$
               &$[1.52\whitey\whitey\pm0.16]$\\
    \rule[-1.5ex]{0ex}{4ex}
    empirical  &$[-\whitey7.892\pm0.040]\ii$&$[0.9034\pm0.0005]$
               &$[+45.7023\pm0.0001]\ii$&$[1.6889\pm0.0031]$\\\hline
  \end{tabular}
  \caption{Pole position and residue in PWA and \ChiEFTPPUE at
    \NXLO{2} (Bayesian uncertainties from app.~\ref{app:uncertainties-Bayes})
    and NLO (no errors).
    \label{eq:polevalues}}
\end{table}


\subsection{The Spin Singlet, Isospin Triplet: \oneS Phase Shifts}
\label{sec:results1S0}


\begin{figure}[!h]
\begin{center}
  \includegraphics[height=0.56\textheight]
  {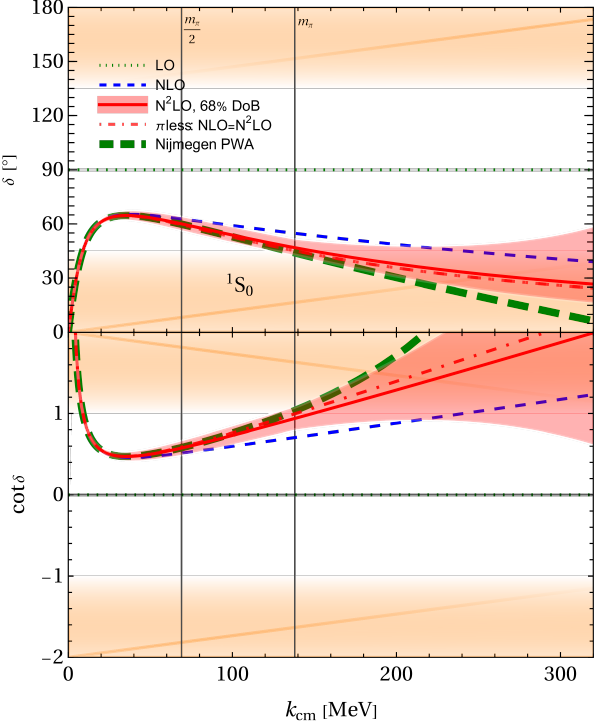}
  \caption{(Colour on-line) Phase shift (top) and $\cotdelta(k)$ (bottom) in
    the \oneS channel as a function of $k$, compared to the Nijmegen
    PWA~\cite{Stoks:1993tb} (thick green dashed). Green dotted: LO; blue
    dashed: NLO; red solid: \NXLO{2} with $68\%$ degree-of-belief (DoB)
    interval based on the Bayesian truncation uncertainty under the
    assumptions of sect.~\ref{app:uncertainties-Bayes}; red dash-dotted:
    NLO$=$\NXLO{2} in \EFTNoPion. Shaded areas indicate the ``Born Corridors''
    of fig.~\ref{fig:unitaritywindow}, \ie~regions in which the Unitarity
    Expansion can \emph{a priori} not be expected to hold. Marked is also the
    scale of the first and second non-analyticity in $\kcotdelta(k)$, from the
    first two branch points $k=\pm\ii\frac{\mpi}{2},\pm\ii\mpi$ of \OPE.}
\label{fig:results1S0}
\end{center}
\end{figure}

The pionic contribution to the \oneS channel consists only of the
Wigner-$\SU(4)$ invariant part. By inspection, its phase shift (top of
fig.~\ref{fig:results1S0}) converges well order-by-order until
$k\to\LambdaNN\approx300\;\MeV$ approaches the putative breakdown scale. This
point lies actually well outside the Unitarity Window (non-shaded area), but
even there, the difference between \NXLO{2} and PWA is only as large as the
difference between NLO and \NXLO{2}. Remarkably, \NXLO{2} is nearly
indistinguishable from \EFTNoPion at NLO$=$\NXLO{2}. [Recall that the \NXLO{2}
contribution of \EFTNoPion is zero; see eqs.~\eqref{eq:kcotdeltapionless}
and~\eqref{eq:EREcoeffs}.] Thus, explicit pionic degrees of freedom appear to
have a minuscule impact even close to or beyond the breakdown scale of
\EFTNoPion, which is at least the scale of the  first branch point $k\gtrsim\frac{\mpi}{2}$.
The bottom of fig.~\ref{fig:results1S0} also displays $\cotdelta$ since the
Introduction identified the ``physical part'' of the amplitude, $\cotdelta$ in
eq.~\eqref{eq:unitarity-amp}, as the primary variable in the Unitarity
Expansion. Included in both figures are the \NXLO{2} intervals in which
higher-order corrections should lie with $68\%$ DoB, based on a Bayesian
order-by-order convergence analysis under assumptions detailed in
sect.~\ref{app:uncertainties-Bayes}.  Section~\ref{sec:consistency} will
argue that this provides also a reasonable estimate of the smallest reasonable
range of theory uncertainties combining several assessments.

The excellent agreement between PWA, \EFTNoPion and \ChiEFTPPUE at low $k$ is
a result of our fit to the ERE parameters. At high momenta, differences to the
original results by RS, and by FMS~\cite{Rupak:1999aa,
  Fleming:1999ee} are small. While our \NXLO{2} is above the PWA, theirs lies
a bit below. We attribute this to two aspects: KSW fit at low $k$ to the PWA,
while we use the ERE parameters; and we find phase shifts via $\kcotdelta$,
while they determine $\delta$ directly using
eqs.~(\ref{eq:LOphaseshift}-\ref{eq:N2LOphaseshiftSS}); see
apps.~\ref{app:uncertainties-delta}
and~\ref{app:amplitudestophaseshifts}. That we expand about Unitarity while
they use a finite $a$ at LO, has little impact since $\frac{1}{ka}\ll1$ for
large $k$.

In this channel, \ChiEFTPPUE appears to 
accomplish the goal of a self-consistent theory with pions in all of the
Unitarity Window, up to its breakdown scale $\LambdaNN$.

\subsection{The Spin Triplet, Isospin Singlet: \threeS Phase Shifts}
\label{sec:results3S1}

\begin{figure}[!t]
\begin{center}
  \includegraphics[height=0.56\textheight]
  {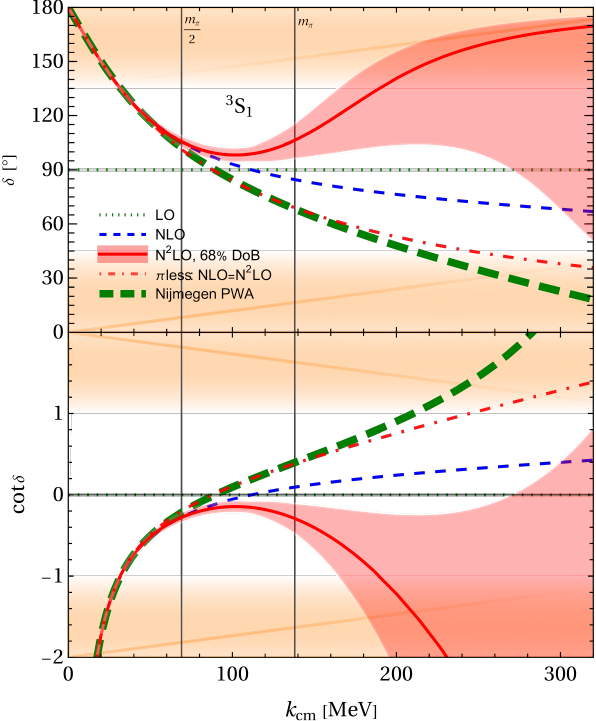}
  \caption{(Colour on-line) Phase shift (top) and $\cotdelta(k)$ (bottom) in
    the \threeS channel, compared to the Nijmegen PWA~\cite{Stoks:1993tb}
    (thick green dashed). Details as in fig.~\ref{fig:results1S0}.}
\label{fig:results3S1}
\end{center}
\end{figure}

The result with Wigner-breaking \OPE pieces at $\calO(Q^1)$ (\NXLO{2}) in
fig.~\ref{fig:results3S1} is catastrophic, as FMS already
noticed~\cite{Fleming:1999ee}. While NLO is by eye reasonable, \NXLO{2}
deviates dramatically from the PWA just above the \OPE branch-point scale of
$\frac{\mpi}{2}$. This is the more puzzling as \EFTNoPion agrees well with the
PWA even at $k\gtrsim100\;\MeV$. With pions, the difference between NLO and
\NXLO{2} becomes around $k\approx\mpi$ as large as between LO and NLO, and
larger than the deviation from the PWA. Not even the sign of $\cotdelta$ is
correct. The breakdown is hardly gradual but sudden, with no hint at NLO of
the unnaturally large \NXLO{2} curvature around $100\;\MeV$. This is the more
concerning as phase shifts are there nearly at the centre of the Unitarity
Window. \NXLO{2} pions have an outsized and wrong impact for
$k\gtrsim100\;\MeV$.

That most of the Unitarity Window is outside the radius of convergence and
hardly extends beyond the scale $\frac{\mpi}{2}$, is unsatisfactory for a
``pionful theory''.

\begin{figure}[!t]
\begin{center}
  \includegraphics[height=0.56\textheight]
  {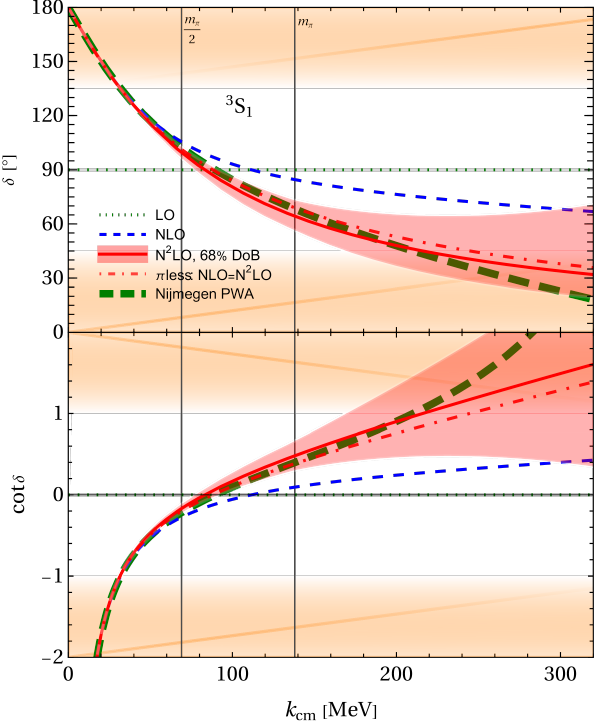}
  \caption{(Colour on-line) Phase shift (top) and $\cotdelta(k)$ (bottom) in
    the Wigner-symmetric \threeS channel, compared to the
    Nijmegen PWA~\cite{Stoks:1993tb} (thick green dashed). Details as in
    fig.~\ref{fig:results1S0}.}
\label{fig:results3S1Wigner}
\end{center}
\end{figure}

What is the origin of the stark discrepancy of both convergence and reasonable
range of applicability between the \oneS and \threeS channels? The former is
exclusively Wigner-$\SU(4)$ invariant; the latter has a symmetry-breaking
component which according to sect.~\ref{sec:WignerKSW} comes exclusively from
the tensor ($\rmS\to\rmD\to\rmS$) part of once-iterated \OPE. Mindful of the
importance of this symmetry around Unitarity discussed in the Introduction,
fig.~\ref{fig:results3S1Wigner} shows the \threeS channel without the
Wigner-$\SU(4)$ symmetry-breaking term of
eq.~\eqref{eq:N2LOamplitudeWignerbreak}.

The qualitative and quantitative improvement is obvious. Similar to \oneS, the
phase shift converges now order-by-order even as
$k\to\LambdaNN\approx300\;\MeV$ just outside the Unitarity Window. The
difference between \NXLO{2} and PWA is wholly within the Bayesian $68\%$ DoB
band of app.~\ref{app:uncertainties-Bayes}, even better than in the \oneS
channel and quite a bit smaller than the shift from NLO to \NXLO{2}. Since
\ChiEFTPPUE at \NXLO{2} and \EFTNoPion at NLO$=$\NXLO{2} are nearly
indistinguishable, pionic degrees of freedom have again a minuscule impact
beyond $k\gtrsim\frac{\mpi}{2}$. Even before the much more robust discussion
in sect.~\ref{sec:consistency}, one can thus place the empirical breakdown
scale of \ChiEFTPPUE soundly around $k\gtrsim250\dots300\;\MeV$, which aligns
with the formal breakdown scale $\LambdaNN$ at which \OPE needs to be iterated
fully.

We conclude that the Wigner-$\SU(4)$ invariant version of \ChiEFTPPUE appears
well-suited to overcome the formal inapplicability of \EFTNoPion above the
scale of the first branch point. In both the \oneS and \threeS channels, it is
a self-consistent, convergent theory with pions in all of the Unitarity Window
-- and potentially somewhat beyond -- with a common breakdown scale of
$k\gtrsim250\;\MeV\approx\LambdaNN$, plus good agreement with PWAs even at
high momenta.

\subsection{Wigner-$\SU(4)$ Symmetric vs.~Breaking Amplitude in \threeS}
\label{sec:difference-sym-vs-break}

\begin{figure}[!t]
\begin{center}
  \includegraphics[height=0.56\textheight]
  {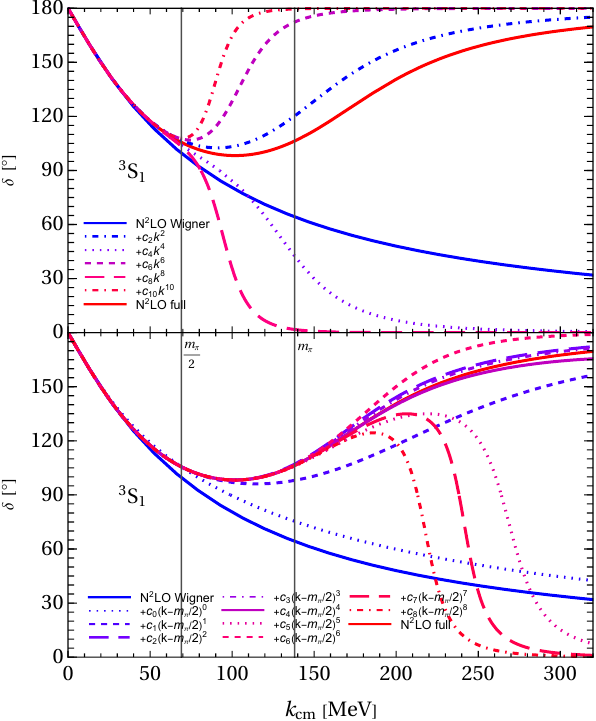}
  \caption{(Colour on-line) Interpolation between the Wigner-symmetric and
    full \threeS phase shift as expansion in powers of $k$ about $k=0$ (top)
    and $\frac{\mpi}{2}$ (bottom); see text for details.}
\label{fig:3S1expansion}
\end{center}
\end{figure}

Why do versions with and without Wigner-$\SU(4)$ symmetry-breaking terms lead
to such different results? We are unable to cast blame on any combination of
terms in $A_{1\mathrm{break}}^{(\rmS)}(k)$,
eq.~\eqref{eq:N2LOamplitudeWignerbreak} and thus turn to a low-energy
expansion for further insight. In contradistinction to \EFTNoPion, the pionic
contributions to all amplitudes in \ChiEFT with Perturbative Pions lead to
nonzero coefficients of the ERE, and hence to predictions of all shape
parameters $v_n$ in~eq.~\eqref{eq:effrange}; see eqs.~\eqref{eq:NLOlimit},
\eqref{eq:N2LOlimitWigner} and \eqref{eq:N2LOlimitbreak} in
app.~\ref{app:limits}.  Cohen and Hansen~\cite{Cohen:1998jr, Cohen:1999iaa,
  Cohen:1999ds} argued that PWA and FMS results are incompatible due to a
``pattern of gross violation'': The coefficients $v_n$ in \ChiEFTPP follow the
pattern of the na\"ive $v_n\sim\mpi^{-2n+1}\sim r^{2n+1}$ \emph{a-priori}
estimate in \ChiEFT and \EFTNoPion but are by a factor of $5$ to $10$ larger
than the PWA values, even after updates~\cite{RuizArriola:2019nnv} to the
latter. Their scales seem in the real world rather set by $\LambdaNN$ or an
even higher scale. Interestingly, Thim recently showed that this issue does
not exist when pions are included nonperturbatively and found instead very
good agreement with phenomenology within respective
uncertainties~\cite{Thim:2024jdv}.

In \ChiEFTPPUE, the phase shifts at NLO and \NXLO{2} as well as \EFTNoPion all
describe the PWA analyses of both the \oneS and \threeS channel very well for
$k\lesssim70\;\MeV$, albeit $v_n\equiv0$ in \NXLO{2} \EFTNoPion; see
figs.~\ref{fig:results1S0} to~\ref{fig:results3S1Wigner}.  We find a generic
scaling as $v_n\sim\frac{1}{\LambdaNN^m}$ with $m-1$ the number of OPE
iterations (entering at \NXLO{m+1}), times combinations of powers of
$\mpi,a,r$ obscured by both quite large and quite small numerical
coefficients; see \eg~eqs.~(\ref{eq:NLOlimit}-\ref{eq:N2LOlimitbreak}) in
app.~\ref{app:limits}.  This suggests that the \emph{string} of $v_n$
predictions which is individually too large in \ChiEFT, is highly correlated
and produces \emph{in toto} large cancellations between different $v_n$.
Indeed, FMS~\cite{Fleming:1999ee} already saw that all pion contributions at
\NXLO{2} are for $k\to0$ \emph{non-analytic} in the quark mass ($\mpi^2$) and
can hence \emph{never} be fully absorbed into higher-order \ChiEFT
LECs\footnote{We thank U.~van Kolck to bring this to our attention.}; see
app.~\ref{app:limits}. Cancellations may be forced in some momentum window but
will fail overall. This generalises easily: The low-momentum limit of \ChiEFT
with Perturbative Pions leads at odd orders $Q^{2n-1}$ (\NXLO{2n}) to
non-analyticities because no contact term determined by an effective-range
enters. This observation does not necessarily conflict with the \ChiEFT power
counting since, as discussed above, the contributions to \emph{all} $v_n$ are
nonzero separately at \emph{each} order of $Q$. The $v_n$ may be vastly
changing between orders, but their combined effect on the low-momentum phase
shifts (observables) does not. Eventually, a CT at $\calO(Q^{2n-2})$
(\NXLO{2n-1}) can be used to exactly match $v_n$ to the empirical value,
eliminating the issue up to that order and ERE coefficient altogether. Still,
the non-analyticities in $\mpi^2$ will persist.

This can explain that the \oneS and Wigner-symmetric \threeS results agree
with PWA despite discrepant predictions for the $v_n$, but is it also key to
the breakdown of the Wigner-breaking version? When one expands the
once-iterated, symmetry-breaking \OPE of
eq.~\eqref{eq:N2LOamplitudeWignerbreak} for $k\to0$, the dominant part is
$\propto k^4$ with a purely numerical coefficient of $0.2827\dots$;
eq.~\eqref{eq:N2LOlimitbreak}. That is about a factor $7$ bigger than the
corresponding symmetric coefficient $0.0408\dots$ from
eq.~\ref{eq:N2LOamplitudeWignersym}; see eq.~\eqref{eq:N2LOlimitWigner}. The
next terms, $\propto k^6$ \etc, show a similar pattern. This could lead to
subtle cancellations at low momenta which are overwhelmed at high $k$.

The top graph of fig.~\ref{fig:3S1expansion} compares therefore the full
amplitude to one in which one adds an expansion of
$A_{1\mathrm{break}}^{(\rmS)}(k)$ up to $\calO(k^{10})$ to the symmetric part,
$A_{1\mathrm{sym}}^{(\rmS)}(k)$. [The series proceeds in powers of $k^2$ since
the ERE is analytic~\cite{Schwinger, Chew, Barker, Bethe}.] If this converges,
then it should smoothly interpolate between the Wigner-symmetric and full
result as powers of $k^2$ are added.

However, large coefficients are clearly not the issue. Just as before, the
symmetry-breaking portion is very small for $k\lesssim\frac{\mpi}{2}$, the
scale of the \OPE branch point. Above it, the expansion does not converge at
all. Each new term is bigger and alternates sign.

We also expanded also around $\frac{\mpi}{2}$, where all non-negative integer
powers of $k$ contribute. As discussed in sects.~\ref{sec:NLOamplitudes}
and~\ref{sec:N2LOamplitudes}, contributions from this branch point appear at
NLO and in both the symmetric and breaking terms of \NXLO{2}. According to the
lower plot in fig.~\ref{fig:3S1expansion}, this series interpolates well
between symmetric and full amplitudes. We find that the region of convergence
extends, simply because the distance from the first branch point is increased
up to $k\approx\frac{\mpi}{2}+\frac{\mpi}{\sqrt{2}}\approx170\;\MeV$. Above
that, convergence is asymptotic: the best, and indeed very good, approximation
uses all terms up to and including $k^4$. We checked that Wigner-breaking
terms with branch point at $k=\pm\ii\;\mpi$ alone do not explain the big
differences. Rather, the issue seems in combining all symmetry-breaking terms.

\subsection{Consistency, Robustness and Uncertainties}
\label{sec:consistency}

We now summarise detailed studies of robustness, consistency and theory
truncation uncertainties which are relegated to the appendix.  According to
the ``democratic principle'', different but reasonable choices should agree up
to higher-order corrections. In concert, these thus test to which degree
various EFT assumptions are consistent with the outcomes. To encourage
discussion~\cite{Griesshammer:2021zzz}, we disclose our choices and why we
believe they are reasonable.

We start with \emph{a-priori} estimates based on the idea that the expansion
parameter $Q$ of \ChiEFTPPUE should be small enough so that \OPE is
perturbative (app.~\ref{app:uncertainties-LambdaNN}) and one is in the
Unitarity Window (app.~\ref{app:uncertainties-Born}). We find that these are
consistent with a ``Lepage''
analysis~\cite[chap.~2.3]{Landau}~\cite{Lepage:1997cs} of \emph{data-driven}
convergence to PWA results (app.~\ref{app:uncertainties-Lepage}), with
\emph{a-posteriori} order-by-order convergence via a simple Bayesian analysis
of the convergence pattern
(app.~\ref{app:uncertainties-Bayes})~\cite{Cacciari:2011ze, Furnstahl:2015rha,
  Griesshammer:2015ahu}, and with \emph{a-posteriori} comparison of different
phase shift extractions, directly or via $\kcotdelta$ inside the Unitarity
Window (app.~\ref{app:uncertainties-delta}).  Since all focus on complementing
aspects, their uncertainties should be combined, but it is not clear how, and
how uncorrelated these approaches are. 

Overall, all $5$ findings are consistent with the more qualitative
observations discussed with figs.~\ref{fig:results1S0}
to~\ref{fig:results3S1Wigner}: a breakdown scale of
$\LambdaNN\approx300\;\MeV$; a lower bound of the Unitarity Window of
$ k_\mathrm{low}\lesssim 8\;\MeV$ in \oneS and $35\;\MeV$ in \threeS; and an
expansion in $Q$ as in eq.~\eqref{eq:Qunitarity}. The Bayesian uncertainty
bands at \NXLO{2} appear to be a good estimate of truncation errors at about
the $68\%$ degree-of-belief (DoB) level, possibly extending to slightly
smaller phase shifts at given $k$. For these, we followed the \textsc{Buqeye}
collaboration to derive probability distributions and DoB intervals via the
``$\max$'' criterion~\cite[sect.~4.4]{Griesshammer:2012we} with a simple
choice of priors~\cite{Cacciari:2011ze, Furnstahl:2015rha,
  Griesshammer:2015ahu}: a uniform (flat) distribution for all known and
unknown coefficients of the EFT expansion of an observable up to some (unknown
but existing) maximum which follows a $\mathrm{log}$-uniform
distribution. Reasonable variations of the priors induce variations by
$\lesssim20\%$ in these posteriors~\cite{Furnstahl:2015rha}. For these priors,
the $95\%$ interval at \NXLO{2} is about $2.7$ times the $68\%$ DoB, \ie~not
twice like for a Gau\3ian distribution. The posteriors fall off much slower,
namely with inverse powers. It may be of note that we exploit that it is
natural in the Unitarity Expansion to set the scale by the Unitarity-ensuring
factor $-\ii k$ of the scattering amplitude,
eq.~\eqref{eq:amplitude}\footnote{We thank D.~R.~Phillips for this
  suggestion.}, and that uncertainties are constructed to vanish at the
renormalisation point (where quasi-exact PWA values must be reproduced). We
use the same Bayesian scheme for the uncertainty estimate of the pole
parameters of table~\ref{eq:polevalues}.

We caution that Bayesian uncertainty estimates of the factual
convergence pattern of observables complement, but do not replace, conceptual
arguments. If an expansion breaks down, for example because non-analyticities
limit its radius of convergence, then an order-by-order assessment may be able
to heuristically extend the range of applicability beyond that for some
processes. But there is no reason why this should hold in all cases. If an EFT
is not self-consistently renormalisable order-by-order, no statistical
analysis can validate it.

All tests show a clear failure of the expansion when Wigner-breaking terms
enter at \NXLO{2} in \threeS, usually around $k\approx[100\dots150]\;\MeV$.

We do not test insensitivity to varying the cutoff scale which probes
momenta beyond the EFT's range; see
\eg~\cite{Griesshammer:2015osb, Griesshammer:2021zzz}. The amplitudes are
analytical in the PDS scheme of dimensional
regularisation~\cite{Kaplan:1998tg, Kaplan:1998we}. That choice eliminates any
unphysical quantities.

Another assessment of EFT uncertainties varies the renormalisation point,
\ie~the input determining LECs. Such alternatives add another
``uncertainty band'' and test the stability of observables, expansion
parameter and breakdown scale: The Callan-Symanzik renormalisation group
equation's variation of the renormalisation condition needs to be fulfilled up
to higher-orders in the EFT expansion, wherever the EFT converges.
In \ChiEFT, a preferred ``Goldilocks corridor'' $\kfit\sim\ptyp\sim\mpi$
captures the Physics at the scales the EFT is designed for and helps avoid
potential fine-tuning of coefficients between different orders.

So far, we used $\kfit=0$ because that is the natural scale at Unitarity.  In
app.~\ref{app:fitpoints}, we add studies with the parameters determined at
points inside the Unitarity Window which are natural in \ChiEFTPPUE: the pole
position of the amplitude, or at $\kfit=\frac{\mpi}{2}$, or at $\mpi$, as the
branch-point scales of non- and once-iterated \OPE.  That $\frac{\mpi}{2}$ is
also in the region where both \oneS and \threeS are closest to Unitarity
($\delta\approx\frac{\pi}{2}$), makes it conceptually attractive. Remember
that \ChiEFTPPUE is constructed to explore precisely this very region
$\ptyp\sim\mpi$ around Unitarity.  We find that these reasonable choices of
renormalisation points at \NXLO{2} do not lead to statistically significant
changes of phase shifts or empirical breakdown scales.

\subsection{Partial-Wave Mixing}
\label{sec:partialwavemixing}

As discussed in sect.~\ref{sec:WignerKSW}, tensor \OPE induces \threeSD mixing
in the original Wigner super-multiplet \oneS-\threeS. What is its impact of
Unitarity, and to which extent are such Wigner-breaking terms demoted? Both
mixing angle $\varepsilon$ and \threeD-wave phase shift $\delta_2$ are small,
but \ChiEFT has no satisfactory explanation. By imposing Wigner-$\SU(4)$
symmetry at \NXLO{2}, the mixing angle and $\rmD$ wave are nonzero at most as
early as $\calO(Q^2)$ (\NXLO{3}), so $\varepsilon,\delta_2=0+\calO(Q^2)$. With
$Q\sim0.5$ at $k\approx\mpi$ or so and LO being $\calO(Q^{-1})$, their natural
magnitude can \emph{a priori} be estimated as $(0.5)^3\lesssim10^\circ$. The
actual values are $\varepsilon(\mpi)\approx2^\circ$
($\varepsilon(2\mpi)\approx3^\circ$), $\delta_2(\mpi)\approx-5^\circ$
($-15^\circ$). This is markedly smaller, but not entirely incompatible with
what is only intended as a rough estimate. It might therefore indicate that
$V_T$ enters only at even higher orders. The Unitarity Expansion might thus
naturally explain why both the \threeSD mixing angle and \threeD wave phase
shift are so small. First results are given in refs.~\cite{TengMSthesis,
  Griesshammer:2025els}, and more are forthcoming~\cite{future1}.

\section{Interpretation}
\label{sec:interpretation}

While we focused so far to establish facts, we now discuss ideas to which these
may lead.

\subsection{Inferences and Conjecture}
\label{sec:inferences}

In sects.~\ref{sec:results3S1} and~\ref{sec:difference-sym-vs-break}, we
presented evidence that \ChiEFTPPUE converges quite well in the \oneS and
\threeS channels at \NXLO{2} both order-by-order and to the PWA only if the
tensor/Wigner-breaking effects of once-iterated \OPE are demoted to \NXLO{3}
or higher. Traditional chiral power counting for perturbative pions adds them
at \NXLO{2}, but such phase shifts fail for $k\gtrsim100\;\MeV$.  Indeed, it
has been argued that the breakdown of \ChiEFT with Perturbative Pions occurs
at momenta so low that the tensor interaction is simply
irrelevant~\cite{Birse:2005um, Birse:2009my}. This is unsatisfactory since
\threeS is there actually close to Unitarity,
$\delta(\threeS,k\approx100\;\MeV)\approx85^\circ$. That would be the failure
of a description which wants to merge Unitarity and chiral Physics in all of
the Unitarity Window.  We therefore propose an alternative idea:

\Conjecture

\absatz Our study considered only the Wigner-doublet \oneS-\threeS channels
since only their phase shifts are well inside the Unitarity Window. We did
not address at which order the tensor/Wigner-$\SU(4)$ breaking part of
one-pion exchange (\OPE) enters -- except that it is at least \NXLO{3}.

This Conjecture is a first step to convert the emergent phenomena of what
appear to be accidental/hidden conformal and Wigner-$\SU(4)$ symmetries into
manifest but weakly broken symmetries in the \ChiEFT Lagrangean
itself. Demoting interactions based on renormalisation-group arguments is less
familiar than promoting. However, it is expected when one fails to
impose symmetries on an EFT but data fits find symmetry-breaking LECs which
appear anomalously small, or even zero if the symmetry is exact. For example,
QED with different couplings for left- and right-handed electrons would
display what one could call an emergent symmetry because the parity-violating
interactions are (nearly) zero -- and that becomes manifest by postulating
that the QED Lagrangean must be parity-conserving.

The Conjecture provides context to the long list of studies which see
Wigner-$\SU(4)$ symmetry manifest itself both with and without pionic degrees
of freedom in essential nuclear interactions, potentials, binding and matrix
elements well into the heavy-element region~\cite{Epelbaum:2001fm,
  Riska:2002vn, Timoteo:2011tt, RuizArriola:2013kdu, LiMuli:2025zro,
  Kaplan:1995yg, Kaplan:1996rk, CalleCordon:2008cz, RuizArriola:2016vap,
  Bedaque:1999ve, Griesshammer:2010nd, Vanasse:2011nd, Lin:2022yaf,
  Lin:2024bor, Birse:2012ih, Lu:2018bat, Lee:2020esp, Izosimov:2020thi,
  Liu:2025say, Niu:2025uxk, Shen:2021kqr}. It adds that the symmetry is not
isolated or accidental but a natural consequence of its exact and unavoidable
realisation at the Unitarity Fixed Point; see next subsection.

If the Conjecture is to be successful, then one must find a systematic
expansion guided by Wigner-$\SU(4)$ symmetry and Unitarity. Like any ordering
scheme, it must, \emph{before} the computation, provide reliable,
semi-quantitative assessments which Wigner-breaking pion interactions to
include at a given order. Ordinarily, correlated two-pion exchange starts for KSW pions at \NXLO{4} and has both Wigner-symmetric and breaking
pieces. Are the latter demoted? The power counting should also illuminate at
which momenta either Wigner-$\SU(4)$ symmetry or chiral symmetry
dominates. Preferably, it will also differentiate the Unitarity Window from
Born Corridors, \eg~by being small in one but large in the other. Before
considering two candidates: quantum-mechanical operator entanglement
(sect.~\ref{sec:entanglement}), and QCD's large-$N_C$ limit
(sect.~\ref{sec:largeNc}), we discuss the neighbourhood of the fixed point.

\subsection{Universality and Chiral Symmetry Close to the Fixed Point}
\label{sec:fixedpoint}

Central to the Conjecture is that it breaks chiral symmetry at \NXLO{2},
resolving the conflict of symmetries in favour of Unitarity. However, that the
pion's central and tensor interactions in few-nucleon systems enter at the
same chiral order and with the same strength, is a direct consequence of its
derivative coupling to the nucleon spin, rooted in chiral symmetry. The
Goldstone mechanism converting global chiral symmetry into local, weakly
interacting field excitations may eventually be too strong to overcome and
lead to the Conjecture' downfall. Why is this violation not that important?

\begin{figure}[!b]
\begin{center}
  \includegraphics[width=0.43\textwidth]
  {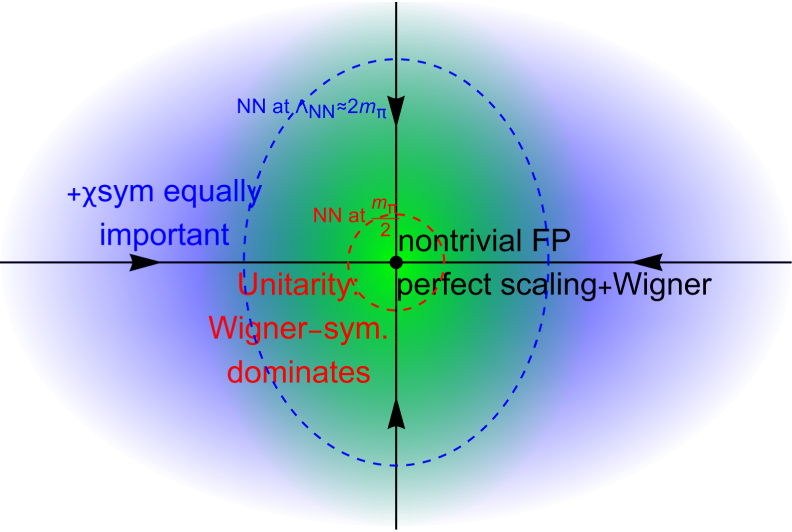}
  \caption{(Colour on-line) Illustration of symmetry-stacking around the
    Unitarity FP.}
\label{fig:fixedpoint}
\end{center}
\end{figure}

The renormalisation group around the Unitarity fixed point (FP)
in the two- and few-nucleon sector may provide some insight; see
fig.~\ref{fig:fixedpoint}.
The FP is non-Gau\3ian (LO is nonperturbative) and displays conformal
invariance. In Nuclear Physics, its universality class is defined by adding
invariance under Wigner-$\SU(4)$ transformations, so all such theories must
agree at Unitarity. One of these is \EFTNoPion, the EFT to which \ChiEFT
reduces very close to the FP (very low momenta). As chiral symmetry is unknown
in \EFTNoPion, it is subdominant in the immediate vicinity of the FP;
Wigner-$\SU(4)$ symmetry and its weak breaking dominate. If this were
incorrect, then \ChiEFT would belong to a different universality class than
\EFTNoPion. In this scenario, the Unitarity FP protects Wigner-$\SU(4)$
symmetry to be only broken weakly, while chiral symmetry in the few-nucleon
sector has no such strong protection, simply because it is not a
characteristic symmetry of the FP. This allows for the possibility that chiral
symmetry is broken and the Wigner-breaking tensor part $V_T$ is consequently
suppressed. That Wigner symmetry (but not chiral symmetry) characterises the FP
is now particularly powerful. It virtually guarantees that it survives
renormalisation as long as the renormalisation scheme itself respects the FP
symmetries. Farther away from the FP, chiral symmetry may then become as
important as Wigner-$\SU(4)$ symmetry, and eventually dominate for large
enough momenta -- mandating similar strengths for the \OPE's tensor and
central pieces. The scale $\LambdaNN$ at which \OPE becomes nonperturbative is
an obvious candidate for this inversion. Since the zero- and one-nucleon
sectors of \ChiEFT are perturbative, \ie~the projection of the FP onto these
is Gau\3ian, their chiral counting is unaffected.  The argument is not
implausible but needs to be studied quantitatively.

\subsection{Unitarity, Wigner-$\SU(4)$ Symmetry and Entanglement}
\label{sec:entanglement}

Beane \etal~(BKKS)~\cite{Beane:2018oxh} pointed to a link between entanglement
suppression and Wigner-$\SU(4)$ symmetry, which in turn extends to an
$\SU(16)$ symmetry for three quark flavours that had first been discovered in
lattice computations~\cite{Wagman:2017tmp}; see also~\cite{Bennett:1996gf,
  Beane:2020wjl, Low:2021ufv, Liu:2022grf, Miller:2023ujx} for discussions and
relations to $S$-matrix properties\footnote{We thank one referee and M.~Savage
  for encouraging us to study this point, and N.~Klco for pointing out our
  initial misconceptions.}. In Quantum Information Theory, entanglement aims
to quantify the extent to which a state is ``non-classical'' at its core. It
asks about the a-causal/ethereal influence of measuring quantum numbers of one
particle on others at mutually space-like separated points (``QM
non-locality''). If the total state of the system is a tensor product of
isolated particles, then the system is quasi-classical and its entanglement is
zero. If measuring one particle instantaneously and uniquely defines the
quantum states of all others (``total collapse of the wave function at a
distance''), then entanglement is maximal and the system is fundamentally
quantum-mechanical. The most famous example of the latter is given in the
Einstein-Podolsky-Rosen Gedankenexperiment~\cite{Einstein:1935rr}, inspiring
both Bell's inequality~\cite{Bell:1964} and his comparison to Bertlmann's
socks~\cite{Bell:1980wg}. Entanglement (or entangling) power, in turn, is the
extent to which an operator is a direct product of single-particle
operators. So, entanglement power is the capacity of an operator to create
entanglement, namely the average entanglement of final states an operator
generates from a set of initial product (un-entangled) states drawn with some
given probability distribution~\cite{Zanardi:2001zza}. An operator which is a
direct product has zero entanglement power. Like entanglement, entanglement
power is nonlinear. An operator can map product states into product states
(zero entanglement power) and yet itself not contain any direct product
(maximal entanglement). The $\SWsigma$ operator discussed momentarily is an
example. However, a generally-agreed operator-entanglement measure
does not yet seem to exist.

In the $\N\N$ system, the operator of interest is the scattering matrix.  As
we wish to explore relations between Wigner-$\SU(4)$ symmetry and entanglement
inside the Unitarity Window, we consider the \oneS-\threeS system only but are
aware of Miller's critique that higher partial waves are not negligible at
higher momenta~\cite{Miller:2023ujx}.  The
$S$ matrix is then
\begin{equation}
  S=\begin{pmatrix}\e^{2\ii\delta[\threeS]}&0\\0&\e^{2\ii\delta[\oneS]}\end{pmatrix}=
  \e^{2\ii\delta[\threeS]}\;P[\threeS]+\e^{2\ii\delta[\oneS]}\;P[\oneS]\;\;.
\end{equation}
The partial-wave projectors $P$ lead to the ``SWAP'' operator
$\SWsigma:=\frac{1}{2}\;(\mathbbm{1}+\vect{\sigma}_1\cdot\vect{\sigma}_2)$~\cite{Low:2021ufv,
  Liu:2022grf}.  As it interchanges the spins of nucleon $1$ and $2$, it has
eigenvalue $+1$ for the spin-symmetric (\threeS) state, and $-1$ for the
spin-antisymmetric (\oneS).  Introducing the average and difference of the two
phase shifts as $\Sigma:=\frac{1}{2}(\delta[\threeS]+\delta[\oneS])$ and
$\Delta:=\delta[\threeS]-\delta[\oneS]$, one finds
\begin{equation}
  S=\e^{2\ii\Sigma}\Big[\mathbbm{1}\;\cos\Delta+
  \ii\;\SWsigma\,\sin\Delta\Big]
\end{equation}

One can now study the \OPE contribution to entanglement. The operator of the
central (Wigner-symmetric) term is
$(\vect{\sigma}_1\cdot\vect{\sigma}_2)(\vect{\tau}_1\cdot\vect{\tau}_2)=
-3\,\mathbbm{1}$ in both channels and hence impacts only the $\mathbbm{1}$
part of the $S$ matrix. On the other hand, iterations of the Wigner-breaking
tensor term
$ [3\left(\vect{\sigma}_1\cdot\ev_q\right)
\left(\vect{\sigma}_2\cdot\ev_q\right)
-\left(\vect{\sigma}_1\cdot\vect{\sigma}_2\right)]
\left(\vect{\tau}_1\cdot\vect{\tau}_2\right)$ are zero in \oneS but nonzero in
\threeS, and hence differentiate between the $\SWsigma$ and $\mathbbm{1}$
terms, and they mix Wigner super-multiplets of different $l$. Therefore, they
do entangle spins (and isospins) quite differently between \oneS and \threeS,
leading to a phase-shift difference $\Delta$. To quantify that, one needs to
find their impact on entanglement.

BKKS defined an operator's entanglement power\footnote{According to
  Klco~\cite{Klco:PrivComm}, this is strictly speaking the ``local purity'',
  related to entanglement power.} as a state-independent average over initial
direct-product (un-entangled) states which vanishes when out-states remain
such a tensor product, using the second R\'enyi entropy of the one-particle
density matrix:
\begin{equation}
  \label{eq:BKKS}
  \calE_\mathrm{BKKS}(\Sigma,\Delta):=\frac{\sin^22\Delta}{6}\in[0;\frac{1}{6}]
\end{equation}
Based on the binary von Neumann entropy of overlaps to Bell states of
spin-$\half$ particles~\cite{Bennett:1996gf},
\begin{equation}
  \label{eq:entropy}
  H[f(\Sigma,\Delta)]:=-x\;\ln x-(1-x)\ln(1-x)\in[0;\ln2]\;\;
  \mbox{ with }
  x:=\frac{1+\sqrt{f(\Sigma,\Delta)}}{2}\;\;,
\end{equation}
Miller~\cite{Miller:2023ujx} proposed a different measure\footnote{We thank
  G.~Miller for pointing out the typographical error that the denominator of
  eq.~(19) in ref.~\cite{Miller:2023ujx} should be squared.}:
\begin{equation}
  \label{eq:Miller}
  \calE_\mathrm{Miller}(\Sigma,\Delta):=H[
  \frac{\cos^2\Delta\,(\cos\Delta-\cos2\Sigma)^2}
  {(1-\cos\Delta\,\cos2\Sigma)^2}]\;\;.
\end{equation}

To isolate the effect of $\SWsigma$ from that of $\mathbbm{1}$, one may also
define a third criterion, an \emph{entanglement strength/nonlocality} of the
$S$ matrix as magnitude-square of the coefficient of $\SWsigma$, relative to
the total $S$-matrix strength, $\cos^2\Delta+\sin^2\Delta=1$. This concept is
meaningful as $S,\;\mathbbm{1}$ and $\SWsigma$ all have eigenvalues of
magnitude one. To some degree, it measures the nonlocality of the operator
$S$, and hence its potential to entangle. It thus sits between between the
\emph{entanglement} of an operator, \ie~the degree to which it can be written
as product, and its \emph{entanglement power}, \ie~the degree to which it
entangles initial product states. The former is zero for $\mathbbm{1}$ but
maximal for $\SWsigma$, while the latter is zero for both (mapping product
states into product states). Only their combination has nonzero entanglement
power, \ie~maps product states into entangled states, and only because the
$\SWsigma$ operator is nonlocal as it cannot be written as product state
(while $\mathbbm{1}$ is obviously not). One then converts that to a binary
entanglement entropy using eq.~\eqref{eq:entropy}\footnote{$\sin^2\Delta$
  without eq.~\eqref{eq:entropy} falls off slightly faster than $\calS$, with
  no marked change of the analysis below.}:
\begin{equation}
  \label{eq:adhoc}
  \calS(\Delta):=H[\sin^2\Delta]
\end{equation}
This definition does not depend on the mean phase shift, only on the
difference. It serves as \emph{relative, simplistic and ad-hoc} measure if one
thinks of $\mathbbm{1}$ as not adding entanglement to the incident, already
fully-entangled \oneS and \threeS states. No ``additional'' entanglement
enters in the Wigner-$\SU(4)$ symmetric case, $\calS(\Delta=0)=0$. On the
other hand, for a phase-shift difference of $90^\circ$,
$S(\Delta=\frac{\pi}{2})=
\ii\,\e^{2\ii\Sigma}\;\SWsigma=\pm\ii\,\e^{2\ii\Sigma}\propto\SWsigma$ is
maximally nonlocal, with upper (lower) sign for the spin-symmetric
(antisymmetric) channel, \threeS (\oneS). The spin wave functions of the final
state continue then to be maximally entangled, just like the initial
ones. Both become eigenstates to the $S$ matrix, and
$\calS(\Delta=\frac{\pi}{2})=\ln2$ is maximal.

Like $\calE_\mathrm{BKKS}$ and unlike $\calE_\mathrm{Miller}$, the relative
measure $\calS$ does not depend on the mean, $\Sigma$. However, it is like
$\calE_\mathrm{Miller}$ maximal for $\Delta=(2n+1)\frac{\pi}{2}$ and zero for
$\Delta=n\frac{\pi}{2}$ ($n\in\mathbbm{Z}$), while $\calE_\mathrm{BKKS}$
depends on $2\Delta$ and hence assigns zero where both $\calE_\mathrm{Miller}$
and $\calS$ predict a maximal measure of entanglement. Conversely,
$\calE_\mathrm{BKKS}$ ascribes maximal entanglement power to
$\Delta=(2n+1)\frac{\pi}{4}$, where neither $\calS\approx0.41\dots$ nor
$\calE_\mathrm{Miller}$ are maximal.  Miller's and the \emph{ad-hoc}
definition agree for $\Sigma=\pm\frac{\pi}{4}$, \ie~at the brink of the
Unitarity Window.

\begin{figure}[!b]
    \begin{center}
      \includegraphics[width=0.56\linewidth]
      {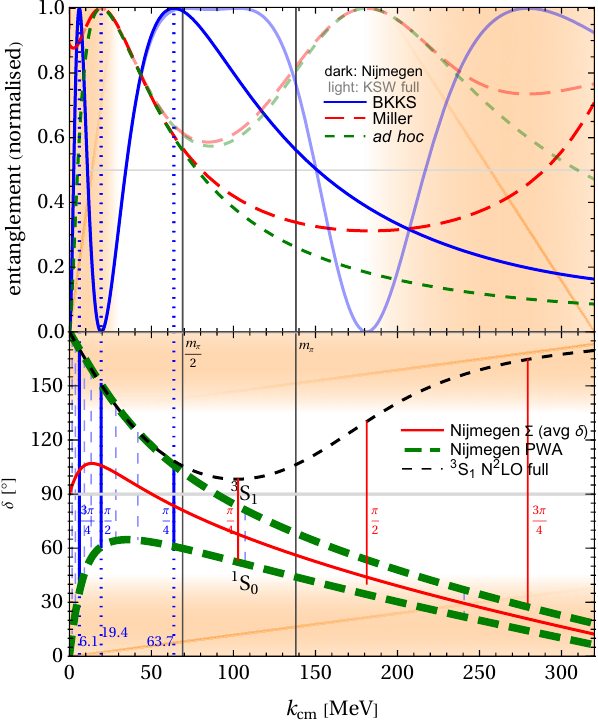}
      \caption{(Colour on-line) Top: Entanglement measures of the 
        phase shifts in the \oneS-\threeS system in the BKKS (blue solid),
        Miller (red long-dashed) and \emph{ad-hoc} measure (green
        short-dashed), each normalised to its maximum.
        Darker lines: Nijmegen PWA; lighter: \NXLO{2} with Wigner-breaking
        terms. 
        Bottom: Nijmegen PWA phase shifts (green thick dashed) and their mean $\Sigma$
        (red solid); the difference $\Delta$ is the width of the corridor
        between the \threeS and \oneS phase shifts, with blue lines every
        $\frac{\pi}{12}=15^\circ$, and $k$ shown at multiples of
        $\frac{\pi}{4}=45^\circ$.
        Black dashed: \threeS at \NXLO{2} with Wigner-breaking; red
        vertical lines above $k=\mpi$: \NXLO{2} \threeS-\oneS difference at
        multiples of $\frac{\pi}{4}=45^\circ$.
        Born Corridors and Unitarity Window as before.}
    \label{fig:entanglement}
  \end{center}
\end{figure}
 
We do not attempt to resolve the discrepancies between these measures for
operator entanglement.
Neither do we account for the fact that each initial state \oneS and \threeS
on which the $S$ matrix acts is separately already maximally entangled and
thus not a product state. Hence, the standard ideas described above of
measuring entanglement power by applying the operator to product states might
not quite apply to the $\N\N$ states themselves.
Rather, we study the degree of entanglement each measure assigns to the
$\rmS$-wave phase shift operator. Each reports zero entanglement for both
Unitarity and perfect Wigner-$\SU(4)$ symmetry ($\Delta=0$). If that drives
indeed tensor-\OPE suppression in the real-world Unitarity Window, some
measure of entanglement should be so small there as to serve as an efficient
expansion parameter. Ideally, it should be large(r) in the Born Corridors
where the Unitarity Expansion fails by design.

Figure~\ref{fig:entanglement} uses the Nijmegen PWA for results of each
variant (top, normalised to its respective maximum) and the corresponding
phase shifts with their average $\Sigma$ and difference $\Delta$
(bottom). Results for \NXLO{2} \ChiEFTPPUE are very similar since the \oneS
and \threeS phase shifts agree very well with their PWA counterparts.
The BKKS entanglement power is maximal well inside the Unitarity Window,
namely at $\Delta(k\approx63.7\;\MeV)=\frac{\pi}{4}$, and there always
$\gtrsim0.4$. It drops further in the high-momentum Born Corridor and thus
shows no correlation between its size and the Unitarity Window. It does not
provide a likely expansion parameter. On the other hand, both Miller's and the
\emph{ad-hoc} version are maximal just outside the lower end of the Window,
$\Delta(k\approx19.4\;\MeV)=\frac{\pi}{2}$. Both decrease monotonically
inside, with $\calE_\mathrm{Miller}\approx0.4$ ($\calS\approx0.35$) in its
centre ($k\approx100\;\MeV$) and $\calE_\mathrm{Miller}\approx0.3$
($\calS\approx0.17$) at its upper end. Miller's version increases again for
$k\ge200\;\MeV$, hinting indeed that entanglement thus defined may be
characteristic exclusively of the Unitarity Window, and not of the Born
Corridors. The \emph{ad-hoc} measure may provide a smaller, and hence
better-converging, expansion parameter, but it simply continues to decrease in
the Born Corridor, potentially indicating little correlation with the
Unitarity Window.

The top panel also contains in lighter colours the entanglement measures for
\ChiEFTPPUE at \NXLO{2} when the Wigner-breaking terms in \threeS are added,
and on the bottom its \threeS phase shift with markers where its difference to
\NXLO{2} \oneS is a multiple of $45^\circ$. All entanglement measures follow
in that case that of the Nijmegen PWA up to about $70\;\MeV$, where PWA
and \NXLO{2} phase shifts start to disagree. The BKKS variant is then
near-constant and saturated in the centre of the Unitarity Window, with a
rapid dip to zero entanglement at its upper border, followed by another rapid
rise inside the Born Corridor to the maximum. Miller's and the \emph{ad-hoc}
version are $\ge0.6$ everywhere and rapidly rise at the boundary. All variants
agree that the Wigner-breaking part of \OPE causes large entanglement inside
the Unitarity Window. This may indicate that $V_T$ lifts indeed the
entanglement suppression of the Unitarity Limit and its Wigner-$\SU(4)$
symmetry.

In conclusion, while there is a clear link between entanglement,
Wigner-$\SU(4)$ symmetry and the Unitarity Limit itself, the situation is
rather muddy in the real world where symmetries are not exact, phase-shift
differences not tiny, and measures of entanglement (power) not that small
inside the Unitarity Window. None of the definitions provides a clear path to
quantifying Wigner-symmetry and its breaking inside the Unitarity Window. It
is also a major deficiency that we found each entanglement measure only by
actually computing the phase shifts both with and without tensor-\OPE. Without
an \emph{a-priori} assessment of the relative entanglement measure of the $V_C$
and $V_T$ operators, entanglement as a concept is useless for power-counting
-- at least for now.  These findings clearly need further study.

\subsection{Unitarity, Wigner-$\SU(4)$ Symmetry and Large-$N_c$}
\label{sec:largeNc}

In contradistinction and returning to a point raised in the Introduction, the
typical sizes of contributions can be estimated in the large-$N_C$ expansion
of QCD.  Recently, Richardson \etal~studied the relation between large-$N_C$,
Unitarity and entanglement in two-baryon systems with $\Delta(1232)$ and
strangeness~\cite{Richardson:2024zln}. The Unitarity Expansion may actually
resolve apparent conflict between large-$N_C$ and Wigner symmetry. In
large-$N_C$, the leading non-tensor part of the $\N\N$ interaction is
automatically Wigner-$\SU(4)$ symmetric in partial waves with even orbital
angular momentum~\cite{Kaplan:1995yg, Kaplan:1996rk}. However, this only holds
if the tensor/Wigner-breaking interaction is neglected altogether, and does
not apply to odd partial waves (to which the Unitarity Expansion does not
apply). And yet, that Wigner symmetry is only weakly broken contradicts the
large-$N_C$ finding that tensor (Wigner-breaking) $\N\N$ interactions should
be leading contributions. Calle Cord\'on and Ruiz Arriola concisely summarised
this quandary and proposed as solution that Wigner symmetry is largely intact
in intermediate- and long-range tensor \OPE, but strongly broken at short
distances~\cite{CalleCordon:2008cz}.

We, too, see evidence for such a splitting. Recall from
sect.~\ref{sec:NLOamplitudes} that the ``pion contribution'' is actually only
the pionic long-distance (non-analytic) part in the chosen renormalisation
scheme and condition. By fitting to $a$ and $r$, Wigner-breaking at low
momenta is locked into the short-distance Counter Terms. At very low momenta,
that guarantees vanishing pion contributions and thus good agreement with
\EFTNoPion and PWAs. By the Uncertainty Principle
$\Delta\!x\Delta\!p\ge\frac{\hbar}{2}$, this is associated with very large
distance scales (very poor resolution) over which the potential is
averaged. As resolution increases, more detailed structures in the potential
are probed. While we found that the Wigner-breaking (tensor) interaction is
insignificant at low momenta (long range/averaged over large distance scales), it becomes strong in \threeS as
$k\gtrsim80\;\MeV$, namely when structures $\delta\!x\sim2\;\fm$ are resolved
which should be classified as ``short'' in a theory with perturbative
pions. We saw that this leads to unacceptable convergence problems. We hence
discard again the offending term by invoking that the Unitarity Expansion's
Wigner symmetry is broken weakly.

In this interpretation, the large-$N_C$ argument of dominant tensor
interaction only holds if its leading coefficient is of natural size. We
propose that an additional symmetry not accounted for by only applying
large-$N_C$ is imposed, namely Wigner-$\SU(4)$ close to the Unitarity FP. This
forces the coefficient to be zero, resolving the apparent conflict. By the
argument of sect.~\ref{sec:fixedpoint}, this exact zero survives
renormalisation because Wigner's is a FP symmetry. Clearly, this mechanism is
as of yet mere speculation.

In our approach, imposing Wigner symmetry as principle is only legitimate when
phase shifts are close to the Unitarity FP, \ie~for $\rmS$ waves
only. Therefore, no contradiction exists between Wigner symmetry and
large-$N_C$ for any higher partial waves, simply because a Wigner symmetry
borne out of the Unitarity Limit does not apply there. This also cures the
problem that this symmetry is realised at best extremely poorly for
$l\ge1$
~\cite{CalleCordon:2008cz}.

\section{Conclusions}
\label{sec:conclusions}

\subsection{Summary}
\label{sec:summary}

We formulated an EFT with pionic degrees of freedom and a natural expansion
scheme about Unitarity in the \oneS and \threeSD channels of the $\N\N$
system: Chiral Effective Field Theory with Perturbative/``KSW'' Pions in the
Unitarity Expansion (\ChiEFTPPUE). Its analytic next-to-next-to leading order
(\NXLO{2}) amplitudes are given in sect.~\ref{sec:amplitudesKSW}. We
interpreted and contextualised the results of sect.~\ref{sec:results} in
sect.~\ref{sec:interpretation}, leading to the Conjecture of the
Introduction. It resolves for $k\gtrsim\mpi$ the apparent conflict of
conformal and Wigner-$\SU(4)$ symmetries around Unitarity with \ChiEFT's
chiral symmetry dictated by the Goldstone mechanism, in favour of the
symmetries of the nontrivial fixed point. We reiterate that in the $\N\N$
system, it only applies to the $\rmS$ waves as only their phase shifts are
close to Unitarity.

In this formulation, LO is Unitarity itself, \ie~the nontrivial fixed point
(FP) of the renormalisation group flow. Its only scale is the relative momentum $k$, and its natural renormalisation
scale/fit point for Low-Energy Coefficients (LECs) is the momentum $\kfit=0$
at which no scale exists.  Around Unitarity ($\kcotdelta=0$,
$\delta=\frac{\pi}{2}$) lies the Unitarity Window of momenta roughly in the
r\'egime $35\;\MeV\lesssim k\lesssim200\;\MeV$ with phase shifts
$45^\circ\lesssim\delta(k)\lesssim135^\circ$ ($|\cotdelta|\lesssim1$). In the
$\N\N$ system, it is only relevant for $\rmS$-waves, but these dominate many
low-energy properties of nuclei. Unitarity implies Universality: Details of
the interactions are less important than that their impact is so big as to
saturate the bounds set by probability conservation. What separates different
universality classes is the symmetries imposed at Unitarity. Besides
nonrelativistic conformal invariance (which is automatic at the FP), that is
Wigner's combined $\SU(4)$ spin-isospin transformations acting on nucleons.
Indeed, we find in sects.~\ref{sec:results1S0} and~\ref{sec:results3S1} that
\ChiEFTPPUE is in this window very close to ``pionless'' EFT -- explicit
pionic degrees of freedom appear to be of little consequence.

This observation nicely aligns with the viewpoint of Information
Theory; \cf~the Introduction: The high degree of symmetry at Unitarity implies
that an anomalously large scattering length is important, but not its precise
value. In \ChiEFTPPUE, that value is indeed demoted: Its impact is only
comparable to that of interaction details parametrised by the effective range
and non-iterated \OPE. These enter at NLO and are suppressed by $Q\approx0.5$
relative to LO. The once-iterated central part of \OPE is the only information
which enters one order in $Q$ higher still, with $\approx25\%$ relevance, but
is completely determined by short-range information (\ie~LECs) set already at
NLO. Tensor \OPE is relegated to higher orders. All this emphasises a central
EFT promise to optimally encode the unresolved short-distance information: The
high degree of symmetry at Unitarity leads to a better-optimised compression,
namely to a reduction of the number and relative importance of independent
LECs which encode the information in $\N\N$ scattering at a given
accuracy. Important is that both \EFTNoPion and \ChiEFTPPUE classify as most
relevant the mere existence of anomalously small intrinsic scales, not how
their different explicit degrees of freedoms produce them. Unitarity makes
their results agree very well -- the rest is detail.

In \ChiEFTPPUE, both Wigner-$\SU(4)$ and conformal invariance are weakly
broken in a systematic expansion in small, dimensionless parameters:
$Q\sim\frac{1}{ak}\lesssim1$ defines the lower bound of the Unitarity Window;
and $Q\sim\frac{\ptyp}{\LambdaNN}\lesssim1$ the upper one of a chiral
expansion, with $\ptyp\sim k,\mpi$ typical low-momentum scales. We find that
\ChiEFTPPUE at \NXLO{2} describes the Physics inside the Unitarity Window
which includes pionic degrees of freedom and is both conceptually fully
consistent and convergent. Comparison to pole parameters
(sect.~\ref{sec:resultspoles}) and Partial Wave Analyses (PWAs;
sects.~\ref{sec:results1S0} and \ref{sec:results3S1}), assessments of theory
uncertainties (sect.~\ref{sec:consistency}) including a Bayesian analysis of
order-by-order convergence and variations of extraction methods and of the
renormalisation point (app.~\ref{app:fitpoints}), all show that the expansion
parameter is consistent with the above \emph{a-priori} estimates, and that its
empirical breakdown scale is consistent with
$\LambdaNN=\frac{16\pi\fpi^2}{\gA^2 M}\approx300\;\MeV$. This is the scale
where \ChiEFT with Perturbative Pions is expected to become inapplicable
because iterations of \OPE are not suppressed. Thus, \ChiEFTPPUE appears to
also apply somewhat into the Born Corridor $|\cotdelta|\gtrsim1$. That is not
surprising as the perturbative (Born) approximation for $\cotdelta\gtrsim1$
and \ChiEFTPPUE for $\cotdelta\lesssim1$ must match in the transition region,
$\cotdelta\approx1$. In contradistinction to the Unitarity Expansion in
\EFTNoPion, \ChiEFTPPUE naturally embeds the scales $\frac{\mpi}{2}$ and
$\mpi$ associated with the branch points of non- and once-iterated \OPE.

Thus, the following picture at nuclear scales might emerge.  The $\N\N$ $\rmS$
waves are treated in perturbation about Unitarity with suppressed tensor-\OPE
in \ChiEFT -- and for some processes even in \EFTNoPion. Either version
considerably reduces computational complexity without loss of information in
the momentum region most relevant for low-energy nuclear systems. The $\rmS$
waves set the dominant patterns of nuclear systems, while the other partial
waves are of course relevant for more detailed and complex phenomena. High
partial waves are treated in Born approximation (with perturbative
\OPE)~\cite{Kaiser:1997mw, Beane:2001bc, Birse:2005um, Birse:2009my}; and
intermediate partial waves (likely some $\rmP$ and possibly $\rmD$ waves) may
need resumming an infinite number of interactions at LO, including potentially
all of \OPE, but not about Unitarity~\cite{Birse:2005um, Birse:2009my,
  Wu:2018lai, Kaplan:2019znu, Peng:2020nyz}. Such a three-pronged approach is
computationally much less intensive than the traditional one of a
nonperturbative LO \OPE potential with a high number of partial waves.

\subsection{Outlook}
\label{sec:outlook}

Further studies are clearly warranted. If the Conjecture of
sect.~\ref{sec:inferences} is to be successful, then a systematic expansion
guided by Wigner-$\SU(4)$ symmetry and Unitarity must be found. Such an
ordering scheme must provide an \emph{a-priori} classification of the relative
importance of Wigner-symmetric and breaking contributions induced by chiral
symmetry. It might also differentiate between Unitarity Window and Born
Corridors, and illuminate at which distances from the FP Wigner and chiral
symmetry dominate, respectively (sect.~\ref{sec:fixedpoint}). We discussed two
candidates, both somewhat problematic: quantum-mechanical operator
entanglement (sect.~\ref{sec:entanglement}), and QCD's large-$N_C$ limit
(sect.~\ref{sec:largeNc}).

The Conjecture should also be tested in systems with at least $3$ nucleons and
with external probes.  The extant NLO work in \ChiEFT with Perturbative Pions
in $2\N$ and $3\N$ systems with external probes of~\eg~\cite{Kaplan:1998sz,
  Chen:1998vi, Chen:1998ie, Savage:1999cm, Bedaque:1999vb, Borasoy:2003gf} can
be expanded about Unitarity and extended by at least one order.  While we see
that \EFTNoPion and \ChiEFTPPUE agree in scattering even at higher momenta,
external probes most likely differentiate the two even at quite low energies,
\eg~in Compton scattering~\cite[sect.~5.4.1]{Griesshammer:2012we}. Because of
the issues with chiral symmetry, pion photoproduction and pion scattering may
be of particular interest. All these are well-known to reveal the considerable
strength of pion-exchange currents.  Fortunately, computations become actually
simpler in the Unitarity Expansion, as seen by comparing our \NXLO{2}
amplitudes to those for $\frac{1}{a}\ne0$ at the same
order~\cite{Rupak:1999aa, Fleming:1999bs, Fleming:1999ee}.

Finally, the Unitarity Expansion for $\N\N$ $\rmS$-waves system should also be
explored in the \ChiEFT version in which pions are nonperturbative. First LO
results are in preparation~\cite{future2} and reported in~\cite{pres1, pres2,
  talk3, talk4,Griesshammer:2025els}. Both for perturbative and
nonperturbative pions, further clarifications are needed. Can imposing a
preference for Unitarity as a highly symmetric state about which to expand
provide a quantitative answer to the questions of the Introduction: Why is
fine-tuning preferred? How does varying $\mpi$ affect it?

We see our Conjecture merely as a first step to merge two highly successful
but at first glance incompatible concepts of Nuclear Theory, namely the
expansion about Unitarity and \ChiEFT, for mutual benefit and a better
understanding of why fine-tuning emerges in low-energy Nuclear Physics.

\section*{Author Contributions}

The authors shared equally in the tasks related to amplitudes and results
described in sects.~\ref{sec:amplitudes} and~\ref{sec:results}, which are
based on Teng's MSc thesis~\cite{TengMSthesis} supervised by
Grie\3hammer. Grie\3hammer took the lead in the interpretations at the core of
sect.~\ref{sec:interpretation}.

\section*{Data Availability Statement}

All data underlying this work is available in full from the corresponding author upon request.

\section*{Code Availability Statement}

\textsc{Mathematica} notebooks are available in full from the
corresponding author upon request.

\section*{Declarations of Competing Interests}

The authors have no financial or non-financial competing interests deriving
from personal or financial relationships with people or organisations
which may cause them embarrassment.


\section*{Acknowledgements}

This project was conceived in 2019 with the warm hospitality and financial
support of Beihang University (Beijing, China) and of the workshop
\textsc{Effective Field Theories and Ab Initio Calculations of Nuclei} in
Nanjing (China). We gratefully acknowledge the organisers and participants of
the ECT* workshops \textsc{The Nuclear Interaction: Post-Modern Developments}
and \textsc{Universality in Strongly Interacting Systems: From QCD to Atoms},
the \textsc{11th International Workshop on Chiral Dynamics (CD2024)}, the
programme and workshop \textsc{INT-24-3 Quantum Few- and Many-Body Systems in
  Universal Regimes} and the workshop \textsc{INT-25-92W Chiral EFT: New
  Perspectives} at the Institute for Nuclear Theory of the University of
Washington (INT-PUB-24-052 and INT-PUB-25-012) for critically evaluating these
ideas, and for highly stimulating discussions, especially with Bingwei Long
and Dean Lee. We are indebted to Ubirajara van Kolck for encouragement early
on and extensive discussions in the final stages. Mario S\'anchez S\'anchez
contributed initially. Ian Stewart clarified typographical errors in the KSW
and FMS publications~\cite{Kaplan:1998we, Fleming:1999bs,
  Fleming:1999ee}. Some are fortunately not in the arXiv versions. Daniel
R.~Phillips' and Roxanne P.~Springer's suggestions and pointers to pertinent
publications about the large-$N_C$ limit as well as Mike Birse's concise
comments dramatically improved the first arXiv version. Martin J.~Savage and
Gerald A.~Miller pointed out links to quantum entanglement and discussed its
use; Silas R.~Beane clarified fine points; Natalie M.~Klco eradicated a number
of misconceptions.
We owe a debt of gratitude to the two anonymous referees, Daniel R.~Phillips
and Martin J.~Savage for strenuously insisting on more clarity and
conciseness, and to the editors for generous deadline extensions to help with
that goal.
This work was supported in part by the US Department of Energy under contract
DE-SC0015393, by the Fondazione Bruno Kessler via the ECT*, by a travel
stipend of the Europe-U.S.~Theory Institute for Physics with Exotic Nuclei
(\textsc{Eustipen}),
by the Istituto Nazionale di Fisica Nucleare (INFN, Italy), by
\textsc{Eurolabs}, by the Universit\`a degli Studi di Milano (Italy),
and by George Washington University: by the Office of the Vice President for
Research and the Dean of the Columbian College of Arts and Sciences; and by an
Enhanced Faculty Travel Award of the Columbian College of Arts and
Sciences. HWG's research was conducted in part in GW's Campus in the Closet.


\appendix
\section{Appendix to Section \ref{sec:amplitudes}}
\label{app:amplitudes}


\subsection{From Amplitudes to Phase Shifts}
\label{app:amplitudestophaseshifts}

In a perturbative calculation, converting an amplitude into observables is
dictated by the theorems of Mathematical Perturbation Theory~\cite{Bellman,
  BenderOrszag, Murdock, Holmes}. Only an order-by-order expansion in both
amplitude and observable is guaranteed to preserve the symmetries (including
$S$-matrix unitarity) at each order independently. Different paths to
observables must at a given order agree within uncertainties set by higher
orders. Therefore, we will in app.~\ref{app:uncertainties-delta} use different
extractions to estimate truncation uncertainties and order-by-order
convergence.

We first follow the original KSW/FMS prescription which starts from the $S$
matrix in the Stapp-Ypsilanti-Metropolis (SYM/''bar'') parametrisation of the
coupled \threeSD channel~\cite{SYM}:
\begin{equation}
\label{eq:Smatrix}
  S=\mathbbm{1}+\frac{\ii kM}{2\pi}\begin{pmatrix}
  A^{(\rmS\rmS)}&A^{(\rmS\rmD)}\\A^{(\rmS\rmD)}&A^{(\rmD\rmD)}\end{pmatrix} = \begin{pmatrix}
\e^{2\ii\delta_0}\cos2\varepsilon&\ii\e^{\ii(\delta_0+\delta_2)}\sin2\varepsilon\\
\ii\e^{\ii(\delta_0+\delta_2)}\sin2\varepsilon&\e^{2\ii\delta_2}\cos2\varepsilon
\end{pmatrix}\;\;.
\end{equation} 
Throughout, the \oneS case is retrieved by setting $\delta_2=\varepsilon=0$, $A^{(\rmS\rmD)}=A^{(\rmD\rmD)}=0$. 

Like the amplitudes $A=A_{-1}+A_0+A_1+A_2\dots$, each phase shift and the mixing angle $\varepsilon$ is expanded in powers of the small, dimensionless parameter: 
\begin{equation}
    \delta_0=\delta_{0,-1}+\delta_{0,0}+\delta_{0,1}+\delta_{0,2}+\dots \mbox{ and likewise for $\delta_2,\varepsilon_1$}\;\;,
\end{equation}
where $\delta_{0,-1}\gg\delta_{0,0}\gg\delta_{0,1}\gg\delta_{0,2}\dots$ and
the second subscript denotes again the order in $Q$. In \ChiEFT with
Perturbative Pions, only the $\rmS$ wave phase shift is nonzero at LO, while
$\delta_{2,-1}=\varepsilon_{2,-1}=0$ vanish. Expanding and matching
order-by-order, we quote the result from KSW/FMS~\cite{Kaplan:1998tg,
  Kaplan:1998we,Fleming:1999bs, Fleming:1999ee}. The only nonzero LO
contribution is
\begin{equation}
    \label{eq:LOphaseshift}
    \delta_{0,-1}=-\frac{\ii}{2}\ln\left[1+\frac{\ii kM}{2\pi}A_{-1}^{(\rmS\rmS)}\right]\;\;.
\end{equation} 
At NLO and \NXLO{2}, one adds for the $\rmS$ wave phase shifts
\begin{align}
  \label{eq:NLOphaseshiftSS}
  \delta_{0,0}&=\frac{kM}{4\pi}\;\frac{A_0^{(\rmS\rmS)}}{1+\frac{\ii
                kM}{2\pi}A_{-1}^{(\rmS\rmS)}}\;\;,\;\;
  \\
  \label{eq:N2LOphaseshiftSS}
  \delta_{0,1}&=\frac{kM}{4\pi}\;\frac{A_1^{(\rmS\rmS)}-\ii\;
                \frac{kM}{4\pi}[A_0^{(\rmS\rmD)}]^2}
  {1+\frac{\ii kM}{2\pi}A_{-1}^{(\rmS\rmS)}}
  -\ii\;(\delta_{0,0})^2\;\;.
\end{align}
While the $\rmS\rmD$-mixing NLO amplitude $A_0^{(\rmS\rmD)}$ enters at
\NXLO{2}, sect.~\ref{sec:N2LOamplitudes} shows that it cancels in the \threeS
channel against terms in $A_1^{(\rmS\rmS)}$. Formulae for $\delta_2$ and
$\varepsilon$ can be found in~\cite{Kaplan:1998we, Fleming:1999bs,
  Fleming:1999ee, TengMSthesis}.

In the Unitarity limit, the term in the denominators simplifies to $1+\frac{\ii kM}{2\pi}A_{-1}^{(\rmS\rmS)}|_\mathrm{Uni}=-1$ and the LO phase shift is in both $\rmS$ waves as expected identically
\begin{equation}
  \label{eq:LOdelta}
  \delta_{0,-1}\equiv\frac{\pi}{2}=90^\circ\;\;.
\end{equation}
However, from the pionless amplitudes of eq.~\eqref{eq:amplitudes-pionless},
the NLO correction,
\begin{equation}
  \label{eq:delta-divergence}
    \delta_{0,0}=\frac{1}{ka}-\frac{kr}{2}\;\;,
\end{equation}
suffers an apparent divergence for $1\gg ka\to0$. The same holds for \ChiEFTPPUE.  Strictly speaking, this is not an issue since such a
limit is outside the window $\frac{1}{ka}\lesssim1$ in which the Unitarity
Expansion converges. Still, the divergence may obscure features close to $k=0$
which can be of interest. Ultimately, its origin is a mismatch between
defining the low-energy parameters in the ERE of $\kcotdelta$,
eq.~\eqref{eq:effrange}, and the constraint that $\delta(k\to0)\to0$ or
$180^\circ$ at NLO and in physical systems, while the value is $90^\circ$ at
Unitarity.

We therefore use another definition which expands not the phase shifts but the
``physical'' part of the amplitude, \ie~$\kcotdelta$ as in
eq.~\eqref{eq:unitarity-amp}, extending the approach by RS~\cite{Rupak:1999aa}
to the coupled \threeSD system. Replace phase shifts ($l=0,2$) by
\begin{equation}
    \label{eq:kcotdelta}
    \e^{2\ii\delta_l}=1+\frac{2\ii k}{\kcotdelta(k)-\ii k}\mbox{ with }
    \kcotdelta=
    \kcotdelta_{l,-1}+\kcotdelta_{l,0}+\kcotdelta_{l,1}+\kcotdelta_{l,2}+\dots,
\end{equation}
expand the mixing parameter $\varepsilon$ as before, use that
$\kcotdelta_{2,-1}=0$ vanishes at LO to find
\begin{align}
    \label{eq:LOkcotdeltaSS}
    \kcotdelta_{0,-1}&=\ii k+\frac{4\pi}{M}\;\frac{1}{A_{-1}^{(\rmS\rmS)}}\\
    \label{eq:NLOkcotdeltaSS}
  \kcotdelta_{0,0}&=-\frac{4\pi}{M}\;\frac{A_{0}^{(\rmS\rmS)}}
                    {[A_{-1}^{(\rmS\rmS)}]^2}\\
    \label{eq:N2LOkcotdeltaSS}
  \kcotdelta_{0,1}&=-\frac{4\pi}{M}\left(\frac{A_{1}^{(\rmS\rmS)}}
                    {[A_{-1}^{(\rmS\rmS)}]^2}
    -\frac{[A_{0}^{(\rmS\rmS)}]^2}{[A_{-1}^{(\rmS\rmS)}]^3}\right)+\ii k
    \frac{[A_{0}^{(\rmS\rmD)}]^2}{[A_{-1}^{(\rmS\rmS)}]^2}
\end{align}
Deriving the corresponding formulae for $\kcotdelta_2$ and $\varepsilon$ is
again straightforward~\cite{TengMSthesis}.

The phase shifts are then simply found by inverting $\kcotdelta_0$ at the appropriate order. In the Unitarity limit, the LO expression is as expected 
\begin{equation}
    \kcotdelta_{0,-1}\big|_\mathrm{Uni}=0 \;\;,
\end{equation}
\ie~$\delta_{0,-1}(k)=\arccot[\frac{0}{k}]\equiv\frac{\pi}{2}=90^\circ$. Now, one finds at NLO in pionless EFT from eq.~\eqref{eq:amplitudes-pionless} by construction the effective-range result:
\begin{equation}
  \label{eq:kcotdelta-finite}
   \kcotdelta_{0,0}\big|_\mathrm{Uni}=-\frac{1}{a}+\frac{r}{2}k^2\;\;
   \Folgt\;\;\delta_{0}=\arccot[\frac{0-\frac{1}{a}+\frac{r}{2}k^2}{k}]
   \stackrel{k\to0}{=}\arctan0\;\;.
\end{equation}
This is finite ($0^\circ$ for $a<0$, $180^\circ$ for $a>0$) and reproduces
at NLO the first two terms of ERE and \EFTNoPion~\eqref{eq:effrange}. The
$\calO(Q^{n})$ (\NXLO{n+1}) terms are (also beyond Unitarity Expansion)
\begin{equation}
    \label{eq:EREcoeffs}
    \kcotdelta_{0,n}=\left\{
    \begin{array}{ll}v_{\frac{n+2}{2}} k^{n+2}&\mbox{ for $n\ge2$ even}\\ 0&\mbox{ for $n\ge1$ odd}\end{array}\right.\;\;.
\end{equation}
Since the \NXLO{2} contribution vanishes, we refer to such results as
``\EFTNoPion NLO$=$\NXLO{2}''.

In the following, we will mostly use the $\kcotdelta$ form, cognisant of the
fact that phase shifts at $\frac{1}{ka}\gtrsim1$ are outside the Unitarity
Window. In both variants to extract phase shifts, as in all expansions which
are consistent within Mathematical Perturbation Theory, all phase shifts
$\delta_0,\delta_2,\varepsilon$ as well as $\kcotdelta$ are at each order real
below the pion-production threshold, \ie~$S$-matrix unitarity is manifest at
each order. In sect.~\ref{sec:consistency} and app.~\ref{app:uncertainties},
we will compare both versions to assess truncation uncertainties and radius of
convergence.

\subsection{From Amplitudes to Pole Positions}
\label{app:amplitudestopoles}

Poles of the $S$ matrix, and hence of the amplitudes, at complex momentum
$\ii\gamma$ correspond to real bound states for $\Re[\gamma]>0,\Im[\gamma]=0$;
to virtual states for $\Re[\gamma]<0,\Im[\gamma]=0$, and to resonances for
$\Re[\gamma]<0,\Im[\gamma]\ne0$. Unitarity is at
$\gamma=0$. Nonrelativistically, the corresponding (real or virtual) binding
energies are $B=\frac{\gamma^2}{M}>0$ when bound.

Extracting bound state properties from amplitudes follows again Mathematical
Perturbation Theory~\cite{Bellman, BenderOrszag, Murdock, Holmes}; see
also~\cite{Rupak:1999aa, Bedaque:1999vb} and~\cite{hgrienotes}. Poles are
zeroes of the inverse $S$ matrix (or amplitude) and appear in all coupled
waves concurrently. It suffices to find the zeroes of the denominator in
eq.~\eqref{eq:kcotdelta}. Expand both $\kcotdelta(k)$ and the pole position,
\ie~its argument:
\begin{equation}
  k_\mathrm{pole}=\ii(\gamma_{-1}+\gamma_0+\gamma_1+\dots)\;\;.
\end{equation}
Likewise, the (rescaled) residue at the pole is found by expanding
\begin{equation}
    Z^{-1}=\frac{4\pi\ii}{M}\left.\frac{\dd}{\dd k}\frac{1}{A(k)}\right|_{k=\ii\gamma}
    =\ii\left.\frac{\dd}{\dd k}\left(\kcotdelta_0(k)-\ii k\right)\right|_{k=\ii\gamma}\;\;.
\end{equation}
For ease of notation, we also define the $m$th derivative of the $n$th
($\calO(Q^n)$, \NXLO{n+1}) contribution to $\kcotdelta_0$ at
$k=\ii\gamma_{-1}$ as
$\kcotdelta_{0,n}^{(m)}(\ii\gamma_{-1}):=\frac{\dd^m\kcotdelta_{0,n}(k)}{\dd
  k^m}|_{k=\ii\gamma_{-1}}$.  The leading-order pole position is determined
implicitly, and the residue is fixed by the Unitarity part, $\ii k$:
\begin{equation}
    \label{eq:LOpole}
    \kcotdelta_{0,-1}(\ii\gamma_{-1})+\gamma_{-1}\stackrel{!}{=}0\;\;,\;\;
    (Z_{-1})^{-1}=1+\ii\;\kcotdelta_{0,-1}^{(1)}(\ii\gamma_{-1})\;\;.
\end{equation}
The other orders are again solved by iteration:
\begin{align}
    \label{eq:NLOpole}
  \ii\gamma_0&=\frac{\kcotdelta_{0,0}(\ii\gamma_{-1})}
               {\ii-\kcotdelta_{0,-1}^{(1)}(\ii\gamma_{-1})}\\
    \label{eq:N2LOpole}
  \ii\gamma_1&=\frac{\ii\gamma_0}{\kcotdelta_{0,0}(\ii\gamma_{-1})}
               \left(\kcotdelta_{0,1}(\ii\gamma_{-1})+
               \ii\gamma_0\;\kcotdelta_{0,0}^{(1)}(\ii\gamma_{-1})
        +\frac{(\ii\gamma_0)^2}{2}\;\kcotdelta_{0,-1}^{(2)}(\ii\gamma_{-1})\right)
\end{align}
 For expressions via amplitudes, one starts from $[A_{-1}(k_\mathrm{pole})+A_0(k_\mathrm{pole})+\dots]^{-1}\stackrel{!}{=}0$ or inserts eqs.~\eqref{eq:LOkcotdeltaSS} to~\eqref{eq:N2LOkcotdeltaSS}.
Usually,  $\kcotdelta$ is analytic in $k^2$ around $\gamma_{-1}$, so that
$\kcotdelta_{0,n}^{(m)}(\ii\gamma_{-1})=0$ for odd $n\ge1$ from
eq.~\eqref{eq:EREcoeffs}. In the Unitarity Expansion ($\gamma_{-1}=0$), $\kcotdelta_{0,-1}\equiv0\equiv\kcotdelta_{-1\p}^{(m)}$ vanishes with all its derivatives, and the last term in parenthesis is zero. 

Expanding the residue is straightforward; we only quote the next nontrivial
term:
\begin{equation}
  Z_0=-\ii(Z_{-1})^2\left(\kcotdelta_{0,0}^{(1)}(\ii\gamma_{-1})+
    \ii\gamma_0\;\kcotdelta_{0,-1}^{(2)}(\ii\gamma_{-1})\right)\;\;.
\end{equation}
  
That both the Unitarity limit and ERE/\EFTNoPion expand about the same point
$k=0$ has a curious consequence which we present in some detail as
sect.~\ref{sec:resultspoles} demonstrates that the same postdiction emerges in
\ChiEFTPPUE at renormalisation point $k=0$. Since $\kcotdelta$ is there
analytic in $k^2$, one can use directly eq.~\eqref{eq:effrange}. All odd
orders of $\gamma_n$ and all even orders of $Z_n$ are zero. The first nonzero
terms are:
\begin{align}
        \ii\gamma_0&=\frac{\ii}{a}\;\;,\;\;\ii\gamma_2=\frac{\ii r}{2a^2}\;\;,\;\;\ii\gamma_4=\frac{\ii r^2}{2a^3}
        \;\;,\;\;\ii\gamma_6=\frac{\ii}{2a^4}\left(\frac{5r^3}{4}-2v_2\right)\\
        Z_{-1}&=1\;\;,\;\;Z_1=\frac{r}{a}\;\;,\;\;Z_3=\frac{3r^2}{2a^2}
                \;\;,\;\;Z_5=\frac{1}{2a^3}\left(5r^3-8v_2\right)
                \label{eq:preresidue}
\end{align}
[The first term in each line and $\ii\gamma_1=0=Z_0$ also follow from inserting the pionless amplitudes of eq.~\eqref{eq:amplitudes-pionless} into eqs.~\eqref{eq:NLOpole} and \eqref{eq:N2LOpole}.] 
These series converge since $\frac{r}{a}\ll1$ inside the Unitarity Window and
we assume $v_n\sim\mpi^{-2n+1}$ are natural as well.

While the shape parameter $v_2$ enters at $\calO(Q^2)$ (\NXLO{3}) in
ERE/\EFTNoPion, it shifts the pole only at a very high $\calO(Q^6)$ (\NXLO{7})
when the parameters of the Unitarity Expansion are determined at $k=0$. The
reason is that $\kcotdelta_{0,n}(k\to0)\propto k^{2n+2}$ beyond NLO, so that
at zero momentum $\kcotdelta_{0,n}^{(m)}(0)=0$ -- unless $m=2n+2$ so that
$\kcotdelta_{0,n}^{(2n+2)}(0)=(2n+2)!\;v_\frac{n+2}{2}$. This is
inconsequential as long as all derivatives and functions in the pole expansion
must be taken at a LO pole position $\gamma_{-1}\ne0$. But Unitarity mandates
$\ii\gamma_{-1}=0$, so nearly all contributions to the pole momentum
disappear. As a consequence, the binding momentum and residue can be
determined to quite high order in the Unitarity Expansion, even if the ERE is
given only to \NXLO{2} (expanding in $\sqrt{\frac{r}{a}}$ would induce
non-analyticities in $r,a$):
\begin{align}
  \label{eq:pole}
  \ii\gamma&=\frac{\ii}{a}\left(1+\frac{r}{2a}+
             \frac{r^2}{2a^2}+\calO(\frac{r^3}{a^3})\right)\\
  \label{eq:residue}
    Z&=1+\frac{r}{a}+\frac{3r^2}{2a^2}+\calO(\frac{r^3}{a^3})
\end{align}
This result is accurate only up to and including \NXLO{2} in $\frac{r}{a}$. It
coincides with the outcome of ERE/\EFTNoPion without Unitarity Expansion at
\NXLO{2}. When the scattering length is included at LO (instead of NLO as in
the Unitarity Expansion), terms with $n$ powers of $\frac{1}{a}\sim Q^{-1}$
are merely reshuffled to $n$ orders earlier, leading to
$\ii\gamma_{-1}^{\mathrm{trad}}=\ii\gamma_0$,
$\ii\gamma_{0}^{\mathrm{trad}}=\ii\gamma_2$, $Z_{0}^{\mathrm{trad}}=Z_1$ \etc

\subsection{The Zero-Momentum Limit}
\label{app:limits}

Reporting the $k\to0$ limit of the \ChiEFTPPUE amplitudes serves a dual
purpose. It confirms that the amplitudes reduce to those of \EFTNoPion; and it
allows further insight into the stark discrepancy between Wigner-$\SU(4)$
symmetric and breaking pion contributions.

At NLO, the impact of $A_{0}^{(\rmS)}(k)$ reduces for low momenta to
contributions analytic in $\mpi^2$, and hence in the quark mass, which are
eventually absorbed into higher-order \ChiEFT LECs:
\begin{equation}
  \label{eq:NLOlimit}
  \lim\limits_{k\to0}\kcotdelta_{0,0}(k)=
  \left(-\frac{1}{a}+\frac{r}{2}k^2\right)-\frac{2}{\mpi^2\,\LambdaNN}
  k^4\left(1+\calO(\frac{k^2}{\mpi^2})\right)\;\;.
\end{equation}
The Wigner-$\SU(4)$ symmetric \NXLO{2} amplitude
$A_{1}^{(\oneS)}(k)= A_{1\mathrm{sym}}^{(\rmS)}(k)$ reduces to
\begin{equation}
  \begin{split}
  \label{eq:N2LOlimitWigner}
  \lim\limits_{k\to0}&\left.\kcotdelta_{1,0}(k)\right|_\mathrm{sym}=\\[-3ex]
  &\frac{4}{\mpi\,\LambdaNN}\bigg[
    \left(\frac{8}{5a\mpi^2}+\frac{r}{3}\right)
    -\frac{4}{\mpi\,\LambdaNN}
    \overbrace{\frac{(384\ln2-227)}{960}}^{=0.040800\dots}\mpi\bigg]k^4
  \left(1+\calO(\frac{k^2}{\mpi^2})\right)
  \end{split}
\end{equation}
Finally, the symmetry-breaking \NXLO{2} amplitude
$A_{1\mathrm{break}}^{(\rmS)}(k)$ reduces at low momenta to\footnote{Since $A_{0}^{(\rmS)}$ has no
  Wigner-$\SU(4)$ breaking contribution, $A_0^{(\rmS\rmS)}=0$ in
  eq.~\eqref{eq:N2LOkcotdeltaSS}.}
\begin{equation}
  \label{eq:N2LOlimitbreak}
  \lim\limits_{k\to0}\left.\kcotdelta_{1,0}(k)\right|_\mathrm{break}=
  -\left(\frac{4}{\mpi\,\LambdaNN}\right)^2
  \overbrace{\frac{384\ln2-187}{280}}^{=0.28274\dots}\mpi k^4
  \left(1+\calO(\frac{k^2}{\mpi^2})\right)\;\;.
\end{equation}
Each OPE iteration adds one power of $\frac{4}{\LambdaNN\mpi}$.  Neither
\NXLO{2} contribution is analytic in the quark mass ($\mpi^2$). That is
relevant for sect.~\ref{sec:difference-sym-vs-break}.

\section{Appendix to Section \ref{sec:results}}
\label{app:results}

\subsection{Assessing Uncertainties}
\label{app:uncertainties}

To quantify the observations of sects.~\ref{sec:results1S0}
and~\ref{sec:results3S1}, we estimate theory uncertainties: \emph{a-priori}
order-by-order convergence; \emph{a-priori} estimates from the assumption to
be inside the Unitarity Window; \emph{data-driven} convergence to PWAs;
\emph{a-posteriori} order-by-order convergence via Bayesian statistics; and
\emph{a-posteriori} comparison of different ways to extract phase shifts.

\subsubsection{\emph{A-Priori}: Order-By-Order Estimates}
\label{app:uncertainties-LambdaNN}

The typical low-momentum scales are $\frac{1}{a}\to0$ in the Unitarity
Expansion, $r\sim\frac{1}{\mpi}$, $k$ and $\mpi$.  As discussed in
sects.~\ref{sec:NLOamplitudes} and~\ref{sec:N2LOamplitudes}, $n$ inverse
powers of the \emph{a-priori} breakdown scale $\LambdaNN\approx300\;\MeV$
parametrise the relative strength by which $(n-1)$-times-iterated \OPE is
suppressed against LO (no \OPE). Therefore, $\frac{\mpi}{\LambdaNN}\approx0.5$
and for $Q\lesssim1$ from eq.~\eqref{eq:Qunitarity}, the upper limit of
\ChiEFT with Perturbative Pions is expected to be
$k\lesssim\LambdaNN\approx300\;\MeV$.

Since $Q\approx0.5$ at $k\approx\mpi$ and \NXLO{2} is complete, corrections in
$\delta(k\sim\mpi)$ should be of order $Q^3\approx\frac{1}{8}\approx10\%$
relative to $\delta_{0,-1}=\frac{\pi}{2}$, if coefficients follow the
Naturalness Assumption~\cite{tHooft:1979rat, NDA,NDA2,Weinberg:1989dx,
  Georgi:1992dw, Griesshammer:2005ga, Hammer:2019poc, vanKolckSaclay,
  Griesshammer:2020fwr}. As $k\to\LambdaNN$, NLO and \NXLO{2} corrections
should become comparable to LO. That happens at a slightly lower
$k\approx150\;\MeV$ in the \oneS channel (\cf~fig.~\ref{fig:results1S0}). In
the Wigner-symmetrised \threeS wave, \NXLO{2} corrections are consistently
somewhat larger than NLO, as allowed by Naturalness, but they exceed thrice
NLO at $k\approx170\;\MeV$; fig.~\ref{fig:results3S1Wigner}.

\subsubsection{\emph{A-Priori}: Inside the Unitarity Window}
\label{app:uncertainties-Born}

According to eq.~\eqref{eq:Qunitarity}, the Unitarity Window $\frac{1}{ka}<1$
with upper limit $|\cotdelta|<1$ imposes $8\;\MeV\lesssim k\lesssim160\;\MeV$
in \oneS, and $35\;\MeV\lesssim k\lesssim230\;\MeV$ in \threeS. In both,
ratios of small scales are not inconsistent with Naturalness:
$|\frac{1}{a\mpi}|<0.1$ in \oneS and $\approx0.25$ in \threeS about Unitarity;
$\frac{r\mpi}{2}\approx0.9$ in \oneS and $\approx0.6$ in \threeS. Therefore,
changes from \NXLO{3} may at $k\approx\mpi$ be $0.6^3\approx0.2$ in \threeS,
and $0.9^3\approx0.7$ in \oneS~-- assuming again Naturalness. That appears
large, but actual NLO-to-\NXLO{2} corrections are much smaller.

\subsubsection{\emph{Data-Driven}: Convergence to the PWA}
\label{app:uncertainties-Lepage}

\begin{figure}[!b]
\begin{center}
  \includegraphics[height=0.56\textheight]
  {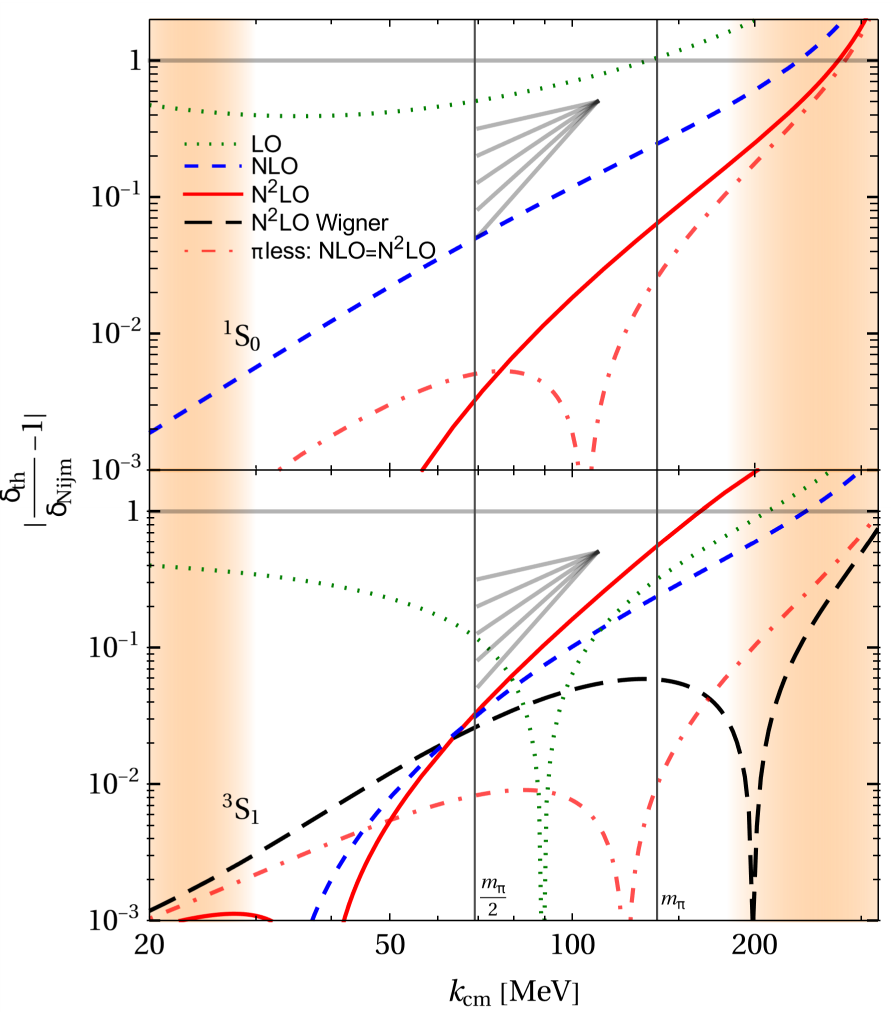}
  \caption{(Colour on-line) Double-logarithmic plots of the relative deviation
    between our results and the Nijmegen PWA in the \oneS (top) and \threeS
    (bottom) channels. The gray lines represent slopes of
    $k^{1,2,3,4,5}$. Colour-coding as in fig.~\ref{fig:results1S0}; shaded:
    Born Corridors.}
\label{fig:lepageplots}
\end{center}
\end{figure}

Comparison to an empirical PWA as next-best approximation to data can quantify
how accurately the EFT reproduces experimental
information~\cite[chap.~2.3]{Landau}~\cite{Lepage:1997cs}. The
double-logarithmic plots of fig.~\ref{fig:lepageplots} contain information
about expansion parameter and breakdown scale. First, deviations between
calculation and data should be of order $1$ for all orders when
$Q\approx1$. That determines the empirical breakdown scale
$\overline{\Lambda}_\mathrm{emp}$. Second, since each new order adds one power
of $Q=\frac{k,\mpi}{\overline{\Lambda}_\mathrm{emp}}$ in the expansion of
observables, slopes should increase by one power of $Q$ from one order to the
next for $k\gg\mpi$ where the momentum scale dominates.  Furthermore, the
plots help quantify the assertions in sects.~\ref{sec:results1S0}
and~\ref{sec:results3S1} that the Wigner-symmetric forms agree well with PWAs.

Excellent agreement at small $k$ is no surprise since the low-$k$ phase shifts
are fitted to the ERE which is determined from a PWA.
For the \oneS channel, NLO, \NXLO{2} and pionless corrections all approach
unity around $\overline{\Lambda}_\mathrm{emp}(\oneS)\approx270\;\MeV$ which is
close to the expected breakdown scale $\LambdaNN\approx300\;\MeV$. We believe
that is still sufficiently far away to be unaffected by the point
$k\approx350\;\MeV$ where the \ChiEFTPPUE result is nonzero but the Nijmegen
PWA crosses zero; \cf~\cite{SanchezSanchez:2017tws}. That induces an
artificial divergence which cannot be remedied by plotting $\cotdelta$
instead.  Slopes seem indeed to increase by roughly one from LO to NLO, and
from NLO to \NXLO{2}, and \NXLO{2} is consistently more aligned with data than
NLO. The \EFTNoPion result might have an even better slope, but it suffers
from an artificial zero since it crosses the empirical phase shift at an
already relatively high $k\approx100\;\MeV$. We estimate an empirical
expansion parameter from scaling at momenta where reproducing the ERE has worn
off: $Q(k\approx200\;\MeV)\approx0.6$ at NLO, and
$Q^2(k\approx200\;\MeV)\approx0.25$ at \NXLO{2}. That is not inconsistent with
the \emph{a-priori} estimate
$Q=\frac{k\approx200\;\MeV}{\LambdaNN}\approx0.7$.

In the \threeS channel, the failure of the \NXLO{2} result including
Wigner-breaking terms (red solid line) is obvious: It breaks down at
$k\approx150\;\MeV$, and \NXLO{2} corrections are already significantly larger
than NLO ones for $k\gtrsim100\;\MeV$. In contradistinction, the
Wigner-symmetric version at \NXLO{2} (black long-dashed line) has the expected
slope increase. For $k\gtrsim170\;\MeV$, it demonstrates consistent
improvement over NLO, towards the PWA, and, most remarkably, over
\EFTNoPion. This adds to the evidence that \ChiEFTPPUE has a radius of
convergence which is not only formally larger than that of a pionless version,
but also empirically.  The extractions of an empirical breakdown scale at NLO
($\approx250\;\MeV$) and \NXLO{2} ($\approx350\;\MeV$) differ but are both
well compatible with the expected breakdown scale
$\LambdaNN\approx300\;\MeV$. The issue may be that \NXLO{2} and PWA happen to
agree at $k=200\;\MeV$, inducing an artificial zero. That also makes it
difficult to read off estimates of the expansion parameter for $k\gg\mpi$. We
find instead $Q(k\approx120\;\MeV)\approx0.2$ from NLO, and
$Q^2(k\approx120\;\MeV)\approx0.06$ from \NXLO{2}, which is quite a bit
smaller than an \emph{a-priori} estimate
$Q=\frac{(k\approx120\;\MeV+\mpi)/2}{\LambdaNN}\approx0.4$ in which $\mpi$
cannot be neglected as typical low scale.

\subsubsection{\emph{A-Posteriori}: Bayesian Order-By-Order Convergence}
\label{app:uncertainties-Bayes}

The \emph{a-priori} estimates of apps.~\ref{app:uncertainties-LambdaNN}
and~\ref{app:uncertainties-Born} can account only qualitatively for the
Naturalness of coefficients~\cite{tHooft:1979rat, NDA,NDA2,Weinberg:1989dx,
  Georgi:1992dw, Griesshammer:2005ga, Hammer:2019poc, vanKolckSaclay,
  Griesshammer:2020fwr}. Factors of $2$ or $3$ can change $\LambdaNN$ and
order-by-order convergence substantially. A statistical interpretation via
Bayesian analysis of the information on Naturalness that is available from the
known orders quantifies theory uncertainties from truncation at a given order;
see \eg~\cite{JPhysG,JPhysG2, Phillips:2020dmw} and references therein.

We first analyse truncation uncertainties for the pole position,
eqs.~\eqref{eq:pole} and table~\ref{eq:polevalues}, following the simple approach
of refs.~\cite{Cacciari:2011ze, Furnstahl:2015rha,
  Griesshammer:2015ahu}. Expand
\begin{equation}
  \ii\gamma=\ii\left[\gamma_{-1}+\gamma_{0}+\sum\limits_{n=1}\gamma_{n}\right]=
  \ii\gamma_0\left[1+\sum\limits_{m=1}c_m\epsilon^n\right]
\end{equation}
to define dimensionless coefficients $c_m$ assumed to be of natural size with
expansion parameter $\epsilon$.  Truncated after $n$ known coefficients,
$|c_{\max,n+1}|\approx\max\limits_{m\le n}\{|c_m|\}$ reasonably estimates
higher orders and $\pm|c_{\max,n+1}|\;\epsilon^{n+1}$ is a reasonable
truncation uncertainty~\cite[sect.~4.4]{Griesshammer:2012we}.

The \EFTNoPion and \ChiEFTPPUE expressions for the pole parameters agree, and
eq.~\eqref{eq:pole} suggests $\epsilon=\frac{r}{a}$ with $c_1=c_2=\half$ for
$\gamma$. Since $\ii\gamma_{-1}=0$, only $\ii\gamma_0=\frac{\ii}{a}$ sets a
scale. If one would likewise choose $\epsilon=\frac{r}{a}$ for the residue,
eq.~\eqref{eq:residue}, coefficients
$c_1=\half,c_2=\frac{3}{2},c_3=\frac{5}{2},c_4=\frac{35}{8}\approx4.4,\dots$
quickly outgrow the Naturalness Assumption
$c_n\lesssim\frac{1}{\epsilon}$. The reason, well-explored in \EFTNoPion, is
its high sensitivity to the pole position~\cite{Phillips:1999hh}. From
eq.~\eqref{eq:preresidue} or the ERE residue with no expansion in
$\frac{r}{a}$ and $v_n=0$, $Z^{-1}=\sqrt{1-\frac{2r}{a}}$ proceeds in powers
of $\epsilon=\frac{2r}{a}$ and
$c_1=\half,c_2=\frac{3}{8},c_3=\frac{5}{16},c_4=\frac{35}{128}\approx0.3,\dots$
are indeed natural. Thus, the expansion parameter for $Z$ is twice that of
$\ii\gamma$. Ignoring such additional information in a Bayesian prior leads to
inconsistent and wildly under-estimated uncertainties in particular in \threeS
where $\frac{2r}{a}\approx0.6$ is not so small and dominates against
corrections from higher ERE parameters. In \oneS, corrections from
$\frac{2r}{a}\approx-0.2$ are similar in size to ERE corrections.

The \textsc{Buqeye} collaboration derived probability distributions and
degree-of-belief (DoB) intervals to statistically interpret
$\pm|c_{n+1}|\;\epsilon^{n+1}$~\cite{Furnstahl:2015rha}. A simple choice of
priors is a uniform (flat) distribution for all known and unknown $|c_m|$ up
to some (unknown but existing) maximum $\bar{c}$ which follows a
$\mathrm{log}$-uniform distribution. With $n=2$ known terms,
$\pm|c_{\max,3}|\;\epsilon^{3}$ encompasses roughly a $68\%$ DoB
interval. With only $n=1$ term known, $\pm|c_{\max,2}|\;\epsilon^{2}$ sets a
$50\%$ DoB interval since fewer information leads to greater uncertainty. For
these priors, the $68\%$ DoB for $n=1$ known term is actually at about
$\pm1.6\;|c_{\max,2}|\;\epsilon^{2}$, and the $95\%$ interval for $n=2$ known
terms at about $\pm2.7\;|c_{\max,3}|\;\epsilon^{3}$. This is not simply a
factor $2$ from the $68\%$ DoB width for a Gau\3ian distribution because the
posteriors fall off much slower, namely with inverse powers. Reasonable
variations of the priors lead for $n\ge2$ known drawings to variations by
$\lesssim20\%$ in these posteriors.  Thus, we set the $68\%$ DoB intervals as
$\pm\frac{\ii}{a}\frac{r^3}{2a^3}$ for the pole position and
$\pm\frac{4r^3}{a^3}$ for its residue, each multiplied by about $2.7$ for
$95\%$ DoBs.

\begin{figure}[!t]
\begin{center}
  \includegraphics[height=0.56\textheight]
  {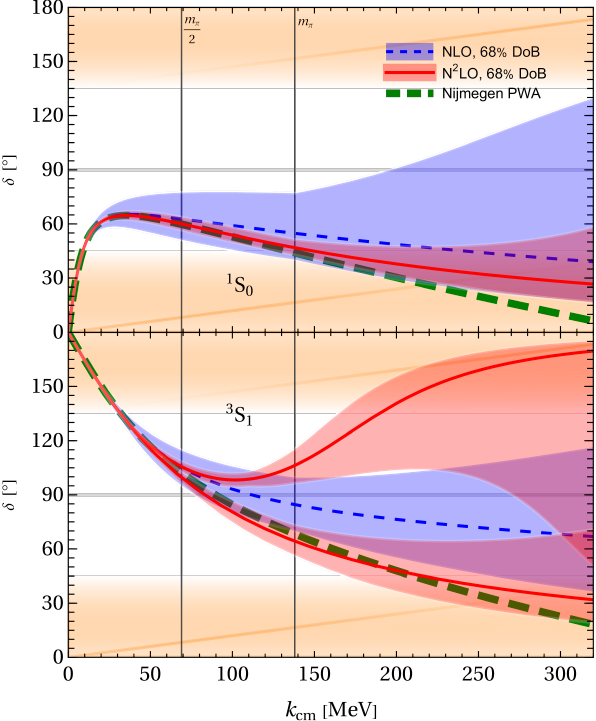}
  \caption{(Colour on-line) The $68\%$ DoB uncertainties of phase shifts at
    \NXLO{2} (red) and NLO (blue, using the rescaling factor from $50\%$
    described in the text) in the \oneS channel (top), and in the \threeS
    channel with and without Wigner-breaking terms (bottom), under the
    assumptions of priors described in the text. Details as in
    fig.~\ref{fig:results1S0}.}
\label{fig:NLODoB68N2LODoB68}
\end{center}
\end{figure}

We apply the same method to $\kcotdelta(k)$ at each $k$ and extract the
corresponding uncertainties for phase shifts. This assumes that uncertainties
at different $k$ are very strongly correlated, which appears reasonable given
that $\cotdelta(k)$ is near-linear for $k\gtrsim100\;\MeV$;
\cf~figs.~\ref{fig:results1S0} and \ref{fig:results3S1Wigner}. The LO
coefficient is again zero, $\kcotdelta_{0,-1}=0$. Around Unitarity, the
Unitarity-ensuring factor $-\ii k$ of eq.~\eqref{eq:amplitude} sets a natural
scale. Since the theory is renormalised at $k=0$ to the ERE, we
ensure that uncertainty bands vanish as $k\to0$ by setting
\begin{equation}
  c_0=\frac{\kcotdelta_{0,0}(k)-\kcotdelta_{0,0}(0)}{k\;Q}
  \;\;,\;\;c_1=\frac{\kcotdelta_{0,1}(k)}{k\;Q^2}\;\;.
\end{equation}
This choice is aligned with $\cotdelta$ as prime variable. We simply
accommodate the various scales which make up the numerator of the expansion
parameter in eq.~\eqref{eq:Qunitarity} as
\begin{equation}
  Q\approx\frac{\max\{k;\mpi\}}{\LambdaNN}\;\;,
\end{equation}
noting that this choice of breakdown scale is consistent with the numbers in
app.~\ref{app:uncertainties-Lepage}.

The \NXLO{2} uncertainties are shown in figs.~\ref{fig:results1S0}
to~\ref{fig:results3S1Wigner}. Figure~\ref{fig:NLODoB68N2LODoB68} includes the
NLO bands (rescaled as above to $68\%$ DoBs). They overlap very well in the
\oneS channel and reasonably in the Wigner-symmetric \threeS version. For
both, the \NXLO{2} band is throughout smaller than the NLO one. In the \threeS
wave with Wigner-breaking terms, NLO and \NXLO{2} uncertainties still
overlap. However, the width of the \NXLO{2} band approaches rapidly that of
NLO and exceeds it for $k\gtrsim200\;\MeV$, which is further proof that this
version does not converge well.

\subsubsection{\emph{A-Posteriori}: Different Ways to Extract Phase Shifts}
\label{app:uncertainties-delta}

\begin{figure}[!t]
\begin{center}
  \includegraphics[height=0.56\textheight]
  {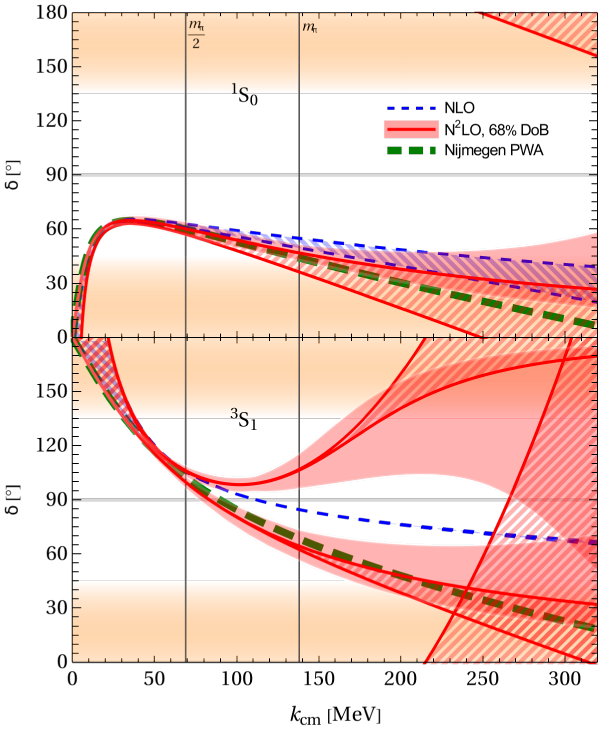}
  \caption{(Colour on-line) Phase shifts for the two variants to extract the
    \oneS (top) and \threeS (bottom) phase shifts of
    app.~\ref{app:amplitudestophaseshifts}: directly (pure lines), and via
    $\kcotdelta$ (lines with Bayesian uncertainty bands), at \NXLO{2} (red)
    and NLO (blue). The differences are marked by the hatched areas (blue:
    NLO; red: \NXLO{2}). Details as in fig.~\ref{fig:results1S0}.}
\label{fig:NLON2LOextractionsband}
\end{center}
\end{figure}

In app.~\ref{app:amplitudestophaseshifts}, we discuss two variants to extract
phase shifts from amplitudes: directly, or via $\kcotdelta$. As noted there,
when both are preformed at the same order in $Q$, these two must agree up to
higher-order corrections where their convergence ranges overlap. If the
results differ vastly, the expansion breaks down. This is a particularly
interesting tool to measure the size of the Unitarity Window. When one expands
about it centre at $\cotdelta=0$,
\begin{equation}
  \cotdelta(k)=\left(\frac{\pi}{2}-\delta(k)\right)
  +\frac{1}{3}\left(\frac{\pi}{2}-\delta(k)\right)^3
  +\calO(\left(\frac{\pi}{2}-\delta(k)\right)^5)
\end{equation}
proceeds in odd powers of $90^\circ-\delta(k)$. The first correction is
smaller than $\frac{1}{4}$ of the leading term for approximately
$\delta\in[40^\circ;140^\circ]$, \ie~close to the points
$\delta=45^\circ,135^\circ$ at which the Unitarity Expansion of
eq.~\eqref{eq:unitarity-amp} is expected to fail because $|\cotdelta|=1$.

The very good agreement of extractions both at NLO and \NXLO{2} in all
\ChiEFTPPUE variants, fig.~\ref{fig:NLON2LOextractionsband}, is therefore no
surprise. At the low end of the Unitarity Window, they diverge starkly for the
reason discussed in app.~\ref{app:amplitudestophaseshifts}: According to
eq.~\eqref{eq:kcotdelta-finite}, using $\kcotdelta(k\to0)$ leads to
$\delta(k\to0)=0,180^\circ$ at Unitarity from NLO on, while the phase-shift
extraction encounters a divergence, $\delta(k\to0)\to\frac{1}{ka}\to\infty$,
eq.~\eqref{eq:delta-divergence}. Thus, discrepancies are expected for
$k\lesssim\frac{1}{|a|}=8\;\MeV$ in \oneS and $\lesssim35\;\MeV$ in \threeS.
Interestingly, for higher momenta pushing into the Born Corridors, the induced
bands are smaller than the Bayesian estimates, and phase shifts from the
direct methods tend to lie slightly below the $\kcotdelta$ variant.

\subsection{Varying the Renormalisation Point}
\label{app:fitpoints}

\begin{table}[!b]
  \centering\footnotesize\setlength{\tabcolsep}{4.5pt}
  \begin{tabular}{|l@{\hspace*{1ex}}l||lll||lll|}
    \hline
    \rule[-1.5ex]{0ex}{4ex}&&\multicolumn{3}{c||}{\oneS}
    &\multicolumn{3}{c|}{\threeS}\\[-0.5ex]
    $\kfit$&&$a\;[\fm]$&$r\;[\fm]$&$(\gamma\;[\MeV],Z)$
    &$a\;[\fm]$&$r\;[\fm]$&$(\gamma\;[\MeV],Z)$\\
    \hline
    \hline
    \multicolumn{2}{|l||}{empirical}
    &$-23.735(6)*$&$2.673(9)*$&\multirow{2}{*}{$(-7.892,0.9034)$}
    &$5.435(2)*$&$1.852(2)*$&\multirow{2}{*}{$(+47.7023\whitey\whit,1.689)*$}\\
    \multicolumn{2}{|l||}{pole}
    &$-23.7104$&$2.7783$&&$5.6128$&$2.3682$&\\
    \hline
    \multirow{3}{*}{$\frac{\mpi}{2}$}
    &NLO&$-38.988$&$3.3270$&$(-4.86\whitey,0.925)$
    &$4.9310$&$2.4966$&$(+55.\whitey\whitey\whitey\whitey\whitey\whit,1.9)$\\
    &\NXLO{2}&$-25.428$&$2.7281$&$(-7.38\whitey,0.910(5))$
    &$4.7768$&$2.4492$&$(+57(3).\whitey\whitey\whitey,1.9(5))$\\
    &~~~sym.&&&&$5.4625$&$1.6124$&$(+43.0(5)\whitey\whitey,1.4(1))$\\
    \hline
    \multirow{3}{*}{$\mpi$}
    &NLO&$+\whitey9.2856$&$4.2285$&$(+28.\whitey\whitey\whit,1.8)$
    &$3.3442$\contradict&$3.1886$\contradict
    &$(+114.\whitey\whitey\whitey\whitey\whit,3.)$\contradict\\
    &\NXLO{2}&$+34.3335$&$2.8956$&$(+6.01\whitey\whit,1.10)$
    &$1.8376$\contradict&$3.3741$\contradict&$(+387(330).,8(25).)$\contradict\\
    &~~~sym.&&&&$4.5344$&$1.7006$&$(+55(1).\whitey\whitey\whitey,1.6(2))$\\
    \hline
  \end{tabular}
  \caption{Values of $a,r$ interpreted as LECs to reproduce $\kcotdelta$ and
    its derivative at a certain $\kfit$. 
    The empirical values are found from the Granada
    values~\cite{RuizArriola:2019nnv} as described in
    app.~\ref{app:amplitudestopoles}. An asterisk $*$ is input; a lightning
    bolt \contradict~indicates the result cannot converge because
    $r\gtrsim a$.  For $a,r$, spurious precision is used because of some
    fine-tuning. \NXLO{2} uncertainties for pole position $\gamma$ and residue
    $Z$ are based on the uncertainty estimate of
    app.~\ref{app:amplitudestopoles}. If none is given, it is smaller than
    the last quoted significant figure.}
  \label{tab:fitpoints}
\end{table}

So far, we used the ERE around $k=0$ because that is the natural scale at
Unitarity. Now, we treat the symbols $a$ and $r$ as if they were not
scattering length and effective range but (finite pieces of) LECs whose values
reproduce the Nijmegen values of $\kcotdelta$ and its derivative at a
particular $\kfit$.  Any choice is legitimate, as long as
$\kfit\ll\LambdaNN$~\cite{Griesshammer:2021zzz}.

\begin{figure}[!t]
\begin{center}
  \includegraphics[height=0.64\textheight]
  {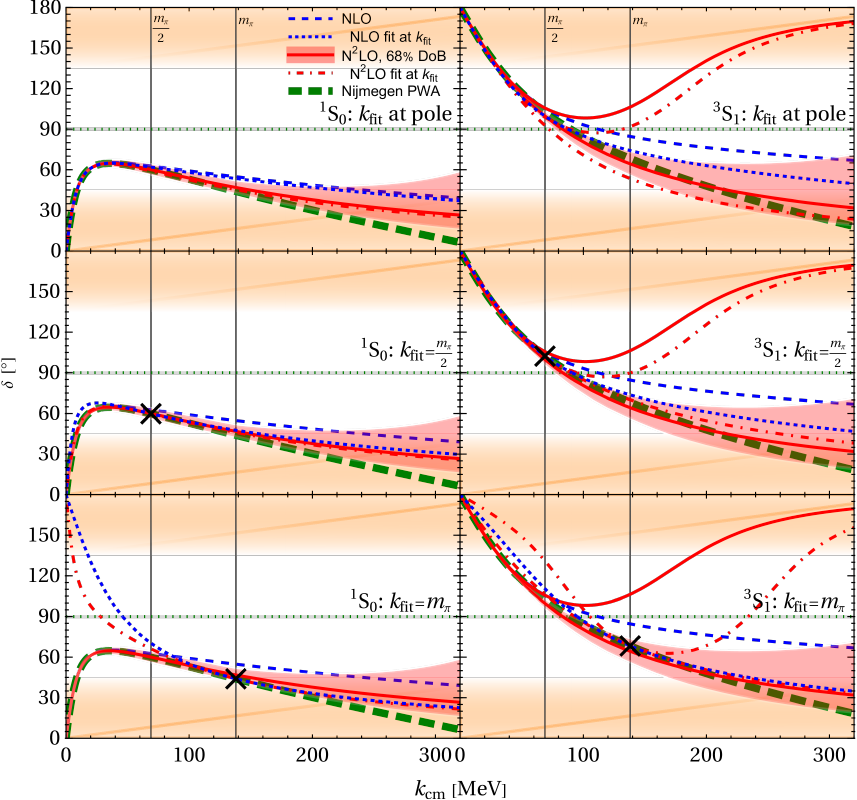}
  \caption{(Colour on-line) Phase shifts in the \oneS (left) and \threeS
    (right) channels, fitted to the pole (top),
    $\kfit=\frac{\mpi}{2}$ (centre) and $\mpi$ (bottom). NLO: blue
    short-dashed. \NXLO{2} and Wigner-invariant \NXLO{2} are both red
    dot-dashed but simple to differentiate.  Crosses mark fit points. The
    results for $\kfit=0$ with Bayesian \NXLO{2} (in \threeS for the
    symmetric form only) bands are colour-coded as in
    fig.~\ref{fig:results1S0}.}
\label{fig:phaseshifts-different-fits}
\end{center}
\end{figure}

RS and FMS used the pole position of the amplitude as
input~\cite{Rupak:1999aa, Fleming:1999ee}. This natural choice for bound-state
properties avoids universal (and thus trivial) correlations between binding
energies and observables like the charge radius. One can find such $a$ and $r$
from the empirical pole position and residue of table~\ref{eq:polevalues} up
to $\calO(\frac{r^3}{a^3})$ by inverting
eqs.~(\ref{eq:pole}/\ref{eq:residue}). No pion contributions enter. The
results in table~\ref{tab:fitpoints} differ only slightly from the Granada
group's best values for physical scattering length and effective range at
$k=0$. On the other hand, since the leading nonzero contribution (NLO) is
$\ii\gamma_0=\frac{\ii}{a}$ from eq.~\eqref{eq:NLOpole},
$\kfit a=\gamma a\approx1$ lies by design just at the brink of the Unitarity
Window. That is dangerous.

We thus also explore two natural renormalisation points of \ChiEFTPPUE inside
the Unitarity Window: the branch-point scales $\kfit=\frac{\mpi}{2}$ and
$\mpi$ of non- and once-iterated \OPE. 

\begin{figure}[!t]
\begin{center}
  \includegraphics[height=0.64\textheight]
  {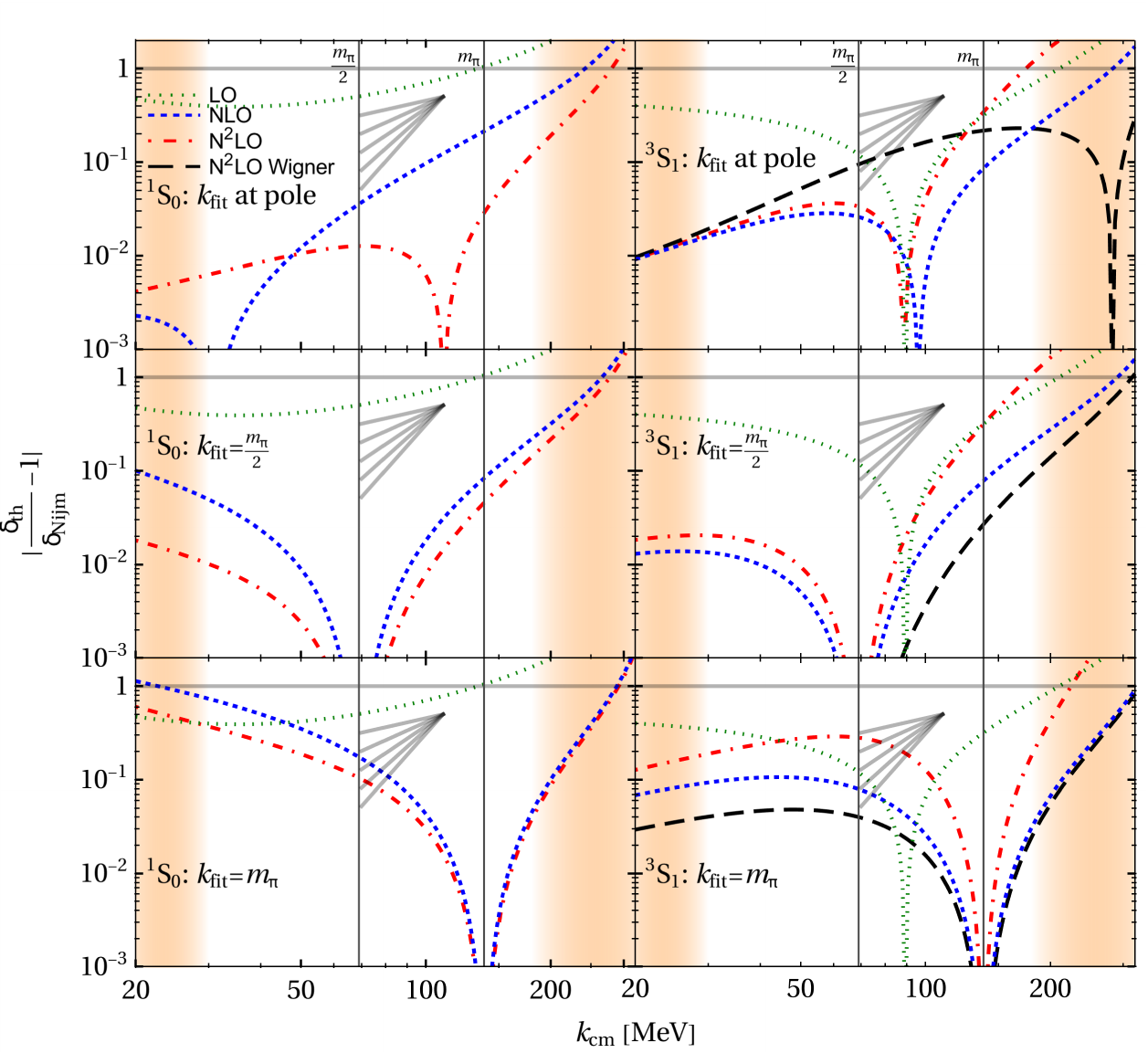}
  \caption{(Colour on-line) Double-logarithmic plots in the \oneS (left) and
    \threeS (right) channels of the relative deviation of results fitted to
    the pole (top), $\kfit=\frac{\mpi}{2}$ (centre) and $\mpi$
    (bottom), to the Nijmegen PWA. LO: green dotted; NLO: blue short-dashed;
    \NXLO{2}: red dot-dashed; Wigner-symmetric \NXLO{2}: black dashed; see
    also fig.~\ref{fig:lepageplots}.}
\label{fig:lepageplots-different-fits}
\end{center}
\end{figure}

Figure~\ref{fig:phaseshifts-different-fits} shows the phase shifts for the
three choices, and fig.~\ref{fig:lepageplots-different-fits} the
double-logarithmic plots of convergence to the PWA. The results of both fits
to the pole and $\frac{\mpi}{2}$ are well within the \NXLO{2} bands of $68\%$
DoB uncertainties from Bayesian truncation uncertainties with
$\kfit=0$. According to table~\ref{tab:fitpoints}, the pole position and
residue is still captured adequately, though discrepancies increase and
order-by-order convergence gets poorer for $\kfit=\frac{\mpi}{2}$. The nominal
\oneS scattering length changes sign for $\kfit=\mpi$ (bottom panels), and the
extracted pole parameters and their uncertainties are meaningless since the
phase shifts clearly diverge from the PWA well inside the Unitarity Window for
$k\lesssim60\;\MeV$. The \threeS results at NLO and Wigner-symmetric \NXLO{2}
have $\frac{r}{a}\gtrsim1$ (marked by lightning bolts \contradict). This does not only violate the expansion of pole and residue in powers
of $\frac{r}{a}$ (eqs.~(\ref{eq:pole}/\ref{eq:residue})), but also the
Unitarity-Window constraint $\frac{r}{2a}\ll1$.  Apparently, the ``leverage
arm'' from the fit point to the lower bound of the Unitarity Window is too
large.

Interestingly, NLO is for all fit points at least as good, and sometimes even
better, than the $\kfit=0$ answers. Some of this may be because the phase
shift is simply required to go through the PWA result. However, the \NXLO{2}
curves with Wigner-$\SU(4)$ symmetry do not show such a uniform trend. The
convergence plots of fig.~\ref{fig:lepageplots-different-fits} are very
similar to the $\kfit=0$ case for the pole fit. Except for the pole fit, they
are dominated by the enforced identity with the phase shifts and derivatives
of the PWA at the fit points. This makes the slopes seem steeper. However, the
empirical breakdown scale of all fits is unchanged and in the same range of
around $300\;\MeV$ as before. We therefore consider this to be a robust
result.

The only exception is, of course, the \threeS channel with Wigner-$\SU(4)$
breaking interactions (red dot-dashed lines). Even when the fit forces
matching at $\kfit=\mpi$, does it diverge quickly at higher $k$. We see no
scenario in which these amplitudes can be made to reasonably agree with the
PWA, or with the pole position and residue (table~\ref{tab:fitpoints}).

Different but reasonable choices of the renormalisation point at \NXLO{2} do
therefore not change phase shifts or empirical breakdown scales in a
statistically significant way.

\end{document}